\documentclass[twocolumn,english,aps,pra,showpacs,sor]{revtex4}
\usepackage{amssymb}
\usepackage{amsmath}
\usepackage{colordvi}
\usepackage{amsthm}
\usepackage{epsfig}
\usepackage{graphicx}
\usepackage{subfigure}
\usepackage{color}
\usepackage{mathrsfs}
\usepackage{bbm}
\usepackage{slashed}
\usepackage{feyn}
\usepackage{feynmf}
\usepackage{bm}
\usepackage{simplewick}

\allowdisplaybreaks[4]

\def\avg(#1){\langle#1\rangle}

\def\be{\begin{equation}}
\def\ee{\end{equation}}
\def\bea{\begin{eqnarray}}
\def\eea{\end{eqnarray}}

\makeatletter
\newcommand{\rmnum}[1]{\romannumeral #1}
\newcommand{\Rmnum}[1]{\expandafter\@slowromancap\romannumeral #1@}
\makeatother

\expandafter\ifx\csname useextended\endcsname\relax
  \let\ff\feyn
  \let\dd\Diagram
\else
  \let\ff\FEYN
  \let\dd\DIAGRAM
\fi

\begin{document}

\title{Background Field Method in Thermo Field Dynamics for Wave Propagation in Unmagnetized Spinor QED Plasmas}
\author{Shan Wu}
\email{wushan@jhun.edu.cn}
\author{Ji-ying Zhang}
\affiliation{Institute for Interdisciplinary Research, Jianghan University, Wuhan 430056, People's Republic of China}

\begin{abstract}
   In this paper, we propose a relativistic quantum many-body theory for the collective modes in spinor quantum electrodynamic plasma. Different from the usual quantization scheme, we use the self-consistency nontrivial background field method in the framework of thermo field dynamics, in which the resulting quanta are temperature-dependent particles instead of the observable ones such as electron, positron and photon. The theory provides a general scheme for many-body physics which overcomes the disadvantages of random phase, Hartree-Fock, or other equivalent meanfield approximations. The essential point for our theory is to exactly evaluate the background fields. In this paper, we propose a general and efficient method to determine them, which we name as ``classical limit method'' for convenience. To demonstrate how to apply the theory, we discuss the collective modes in unmagnetized electron-positron plasma, in both the low-energy and high-energy limits. It yields the well-known dispersion relations of longitudinal and transverse modes for non-relativistic degenerate plasmas, at zero and nonzero temperature. Furthermore, it gives the additional relativistic and vacuum fluctuation corrections, including mass increasing, effective charge decreasing, finite light velocity influence on the dispersion relation, and virtual charges redistribution. The last effect is reported for the first time.
\end{abstract}

 \maketitle
\section{Introduction}

Quantum plasma is an important topic in astrophysics including giant planet \cite{Militzer Hubbard Vorberger Tamblyn Bonev}\cite{Redmer Mattsson Nettelmann French}, white dwarf star \cite{Chabrier}\cite{Chabrier Brassard Fontaine Saumon}, outer crust of neutron star \cite{Haensel Potekhin Yakovlev}, gamma-ray bursts \cite{Uzdensky Rightley} and black hole \cite{Ruffini Vereshchagin Xue}. It also plays an important role in laser physics \cite{Piazza Muller Hatsagortsyan Keitel}, where a particularly exciting field is inertial confinement fusion \cite{Hurricane}. Besides, many condensed matters exhibit quantum plasma properties \cite{Manfredi}.

Traditionally, to determine if the quantum effect is taken into account or not, the characteristic parameter $\theta=k_BT/E_F$ is used in literature \cite{Bonitz Moldabekov Ramazanov}, where $k_B$ is the Boltzmann constant, $T$ is the temperature and $E_F=\hbar ^2(3\pi ^2n)^{\frac{2}{3}}/2m\gtrsim k_BT$ is the Fermi energy with the particle mass $m$, and the Plank constant $\hbar$ (Here we ignore the external field and internal structure of plasma particles). However, it just gives the coherent degree of plasma particles. If the energy is high enough, particle creation and annihilation effect appear, in this case the plasma is also considered as the quantum one. Besides, the low-energy relativistic quantum effect such as vacuum polarization should be taken into account either. Especially, when interparticle distance $\sim 1/m$, charges begin to penetrate each others' vacuum polarization clouds \cite{Peskin Schroeder}, the interaction will be strongly modified. To character this effect, we define another parameter
\begin{eqnarray}
  \kappa=\frac{a}{2\lambda_c},
 \label{(Eq.1a)}
\end{eqnarray}
where $a=(3/4\pi n)^\frac{1}{3}$ is the Wigner-Seitz radius ($n$ is the particle number density), and $\lambda_c=1/m$ is the Compton wavelength. $\kappa\lesssim O(1)$ obviously affects the properties of plasmas.

A general theory should be a relativistic quantum many-body one which include all the quantum effects mentioned in the last paragraph. For collective modes in plasmas, Yuan Shi, et al. \cite{Shi Fisch Qin} demonstrate that aside from their nontrivial background method, the existing relativistic quantum plasma theoretic schemes are not capable of showing that all linear wave modes well known in classical plasma theories can be recovered when taking the classical limit in relativistic quantum results. It strongly suggests further potential study of this method in QED plasmas. However, their theory is limited, and its defects and weaknesses include: (1) They focus on Boson plasmas. Their ground state wave function method in evaluating the background field is difficult to extend to the Fermi case. (2) It is almost impossible to extend their theory to the non-ideal plasmas since it involves solving the wave function equation for the many-body system containing interaction terms.  (3) The temperature effect is not taken into account. (4) Their theory is incapable of explaining the vacuum polarization behavior when $\kappa\lesssim O(1)$.

In this paper, we are going to extend their theory to overcome these difficulties, to develop a general perturbative spinor QED plasma scheme for collective modes. To explain how to apply the theory, we focus on the unmagnetized spinor QED particle-antiparticle pair plasmas. It is a widely studied topic, useful for showing the validity of our theory. To overcome the above-mentioned points (1) and (2), we develop a new method, named as ``classical limit method'' for convenience, to exactly evaluate the nontrivial background field. To overcome the point (3), thermo field dynamics (TFD) \cite{Leplae Umezawa Mancini}-\cite{Ojima} is used. The point (4) is naturally solved in our scheme. One will find that the theory we are going to develop in this paper shows generality for many-body system, enabling one to go beyond the scope of the random phase, Hartree-Fock, and other equivalent mean field approximations. Furthermore, it potentially avoids the uniform system assumption.

The rest of this paper is organized as follows. In Sec. \Rmnum{2}, we review some points of TFD useful for our study. The relativistic spinor QED plasma model in TFD is summarized. In Sec. \Rmnum{3}, we discuss the quantization scheme in the presence of nontrivial background fields in TFD. The Feynman rules for our model are given. The expressions of temperature-dependent self-energy of the fluctuation vector boson are obtained. Besides, we present the way to obtain the collective modes from the temperature-dependent self-energy. In Sec. \Rmnum{4}, we give the general idea of ``classical limit method''. It helps us to calculate the self-energy in both the low-energy and high-energy limits. In Sec. \Rmnum{5}, a series of collective modes are obtained. Their low-energy limits show quite coincide with the well-known results in the classical plasmas, non-relativistic quantum plasmas, and relativistic QED plasmas theories. Furthermore, new vacuum fluctuation correction due to the virtual charge redistribution is found. The high-energy case is also analyzed. Summary and remarks are made in Sec. \Rmnum{6}, which includes the analysis of potential extension of our theory to the general many-body case. Effective action for temperature dependent fluctuation boson is given in Appendix A. The analysis of potential violation of Ward identity is presented in Appendix B.

We use the Heavisde-Lorentz units with the speed of light $c=1$ and Plank constant $\hbar=1$. Moreover, we set the Boltzmann constant $k_B=1$. The metric signature is ($+$,$-$,$-$,$-$).  Einstein's summation convention is used only for repeated space-time indices.

\section{\label{sec:level1}Review of TFD and its application to QED}

This work is restricted in the realm of spinor QED plasmas. In addition, particles such as muon, tauon, etc, other than those of electron, positron, and photon, are excluded. We use the QED electron-positron plasma model to develop our theory. It can be extended naturally to other spinor QED plasma systems. Our starting point is the standard Lagrangian density for spinor QED,
\begin{eqnarray}
  \mathcal{L}=\bar{\Psi}(i\slashed{D}-m)\Psi-\frac{1}{4}(F_{\mu\nu})^2,
\label{(Eq.11b)}
\end{eqnarray}
where $\Psi$ is the four components Dirac spinor field, $\bar{\Psi}=\Psi^\dag \gamma^0$, $\slashed{D}=\gamma^\mu D_\mu$ ($\gamma^\mu $ is the $4\times 4$ gamma matrices, $D_\mu =\partial_\mu+ieA_\mu $ is the covariant derivative operator with  the electron charge $e$ and the electromagnetic vector potential  $A_\mu$), and $F_{\mu\nu}=\partial_\mu A_\nu-\partial_\nu A_\mu$ is the electromagnetic field strength tensor. Below, we briefly review some ideas and results of TFD.

TFD is a very powerful method for constructing our relativistic quantum many body theory for plasmas. Different from the usual Green function theory, it has a theoretical structure almost parallel to the usual QFT, enabling one to absorb most techniques from QFT. If we consider a quantum system in a Fock space $\mathbb{F}$, the key point of TFD is to introduce an ancillary Fock space $\tilde{\mathbb{F}}$ which has the same structure of $\mathbb{F}$. It enables one to construct the so called ``total space" $\mathbb{F}\otimes\tilde{\mathbb{F}}$ in TFD. In addition to the physical set $\mathscr{U}=\{A\}$ corresponding to $\mathbb{F}$, a fictitious operator set $\tilde{\mathscr{U}}=\{\tilde{A}\}$ corresponding to $\tilde{\mathbb{F}}$ is introduced. In TFD, the operators are governed by the following axions \cite{Matsumoto}:
\begin{eqnarray}
  \tilde{\tilde{A}}=A,
\label{(Eq.3a)}
\end{eqnarray}
\begin{eqnarray}
  (AB)^{\sim}=\tilde{A}\tilde{B},
\label{(Eq.4a)}
\end{eqnarray}
\begin{eqnarray}
  (c_{1}A+c_{2}B)^{\sim}=c_{1}^{*}\tilde{A}+c_{2}^{*}\tilde{B},
\label{(Eq.5a)}
\end{eqnarray}
for any $c_{1}$,$c_{2}\in \mathbb{C}$. The equal time commutation and anticommutation relations for operators of boson and fermion are
\begin{eqnarray}
  [A,\tilde{B}]=0,
\label{(Eq.6a)}
\end{eqnarray}
\begin{eqnarray}
  \{A,\tilde{B}\}=0,
\label{(Eq.7a)}
\end{eqnarray}
respectively. In the statistical equilibrium state, the thermal average of an arbitrary operator $A$ can be expressed as the expectation value concerning the thermal vacuum $|\Omega (\beta)\rangle\ $:
\begin{eqnarray}
  \langle \Omega (\beta) |A| \Omega (\beta) \rangle=\frac{\mathrm{Tr}(Ae^{-\beta (H-\mu N)})}{\mathrm{Tr}(e^{-\beta (H-\mu N)})} ,
\label{(Eq.8a)}
\end{eqnarray}
where $H$ is the total Hamiltonian of the system, $\mu$ is the chemical potential, $N$ is the particle number operator, and $| \Omega(\beta)\rangle $ is constructed with the vectors in $\mathbb{F}\otimes\tilde{\mathbb{F}}$ \cite{Takahashi Umezawa}. It is invariant under the tilde conjugation
\begin{eqnarray}
  |\Omega(\beta)\rangle=|\tilde{\Omega}(\beta)\rangle  ,
\label{(Eq.8b)}
\end{eqnarray}
and satisfies the thermal state condition \cite{Matsumoto Nakano Umezawa}.
The evolution of system is governed by the Heisenberg equation, i.e., $\dot{A}=i[H,A]$. In TFD, one can unify it and its tilde form $\dot{\tilde{A}}=-i[\tilde{H},\tilde{A}]$ into $\dot{R}=i[\hat{H},R]$. Here $R$ is an arbitrary operator corresponding to $\mathbb{F}\otimes\tilde{\mathbb{F}}$, and
\begin{eqnarray}
	\hat{H}=H\otimes\mathbbm{1}-\mathbbm{1}\otimes\tilde{H}
  \label{(Eq.10a)}
\end{eqnarray}
is often referred to as ``total Hamiltonian". It implies that many body system is governed by $\hat{H}$. For the case of relativistic field theory, the starting point is the total Lagrangian density, i.e., $\mathcal{\hat{L}}=\mathcal{L}\otimes\mathbbm{1}-\mathbbm{1}\otimes\mathcal{\tilde{L}}$.

For QED,
\begin{eqnarray}
  \mathcal{\hat{L}}=\sum_{\alpha=1,2}\varepsilon_\alpha P_\alpha [\bar{\bm{\Psi}}^{(\alpha)}(i\slashed{\bm{D}}^{(\alpha)}-m)\mathbf{\Psi}^{(\alpha)}-\frac{1}{4}(\bm{F}_{\mu\nu}^{(\alpha)})^2],
\label{(Eq.11a)}
\end{eqnarray}
where $P_\alpha$ is the ordering operator which is defined as \cite{Umezawa Mastsumot Tachiki}
\begin{eqnarray}
P_\alpha (A^{(\alpha)} B^{(\alpha)}\cdot\cdot\cdot C^{(\alpha)})=
	\begin{cases}
		A^{(\alpha)} B^{(\alpha)}\cdot\cdot\cdot C^{(\alpha)} &  \alpha=1,\\
		C^{(\alpha)}\cdot\cdot\cdot  B^{(\alpha)}A^{(\alpha)} &  \alpha=2,
	\end{cases}\nonumber\\
\label{(Eq.12a)}
\end{eqnarray}
for arbitrary operators $A$, $B$, $C$, etc. If $\hat{\mathcal{L}}$ is a c-number, $P_\alpha$ is understood as acting on the coefficients of Fourier expansions of the $\bm{\Psi}^{(\alpha)}$. Besides,
\begin{eqnarray}
\varepsilon_\alpha=
		\begin{cases}
		  1 & \quad \mbox{for}\quad\alpha=1,\ \\
		  -1 & \quad \mbox{for}\quad \alpha=2.
	\end{cases}
\label{(Eq.13a)}
\end{eqnarray}
In Eq.({\ref{(Eq.11a)}}), for convenience, the thermal doublets are introduced as follows
\begin{eqnarray}
  \mathbf{\Psi}(x)=
    \left(
      \begin{array}{c}
        \mathbf{\Psi}^{(1)}(x)\\
        \mathbf{\Psi}^{(2)}(x)
      \end{array}
    \right)\;=
    \left(
      \begin{array}{c}
        \Psi(x)\\
        i[\tilde{\Psi}(x)^\dag]^\mathrm{T}
      \end{array}
    \right)\;,
  \label{(Eq.14a)}
\end{eqnarray}
\begin{eqnarray}
  \bm{A}_\mu(x)=
    \left(
      \begin{array}{c}
        \bm{A}_\mu^{(1)}(x)\\
        \bm{A}_\mu^{(2)}(x)
      \end{array}
    \right)\;=
    \left(
      \begin{array}{c}
        A_\mu(x)\\
        \tilde{A}_\mu(x)
      \end{array}
    \right)\;,
  \label{(Eq.15a)}
\end{eqnarray}
\begin{eqnarray}
  \bm{F}_{\mu\nu}(x)=
    \left(
      \begin{array}{c}
        \bm{F}_{\mu\nu}^{(1)}(x)\\
        \bm{F}_{\mu\nu}^{(2)}(x)
      \end{array}
    \right)\;=
    \left(
      \begin{array}{c}
        F_{\mu\nu}(x)\\
        \tilde{F}_{\mu\nu}(x)
      \end{array}
    \right)\;
  \label{(Eq.15b)},
\end{eqnarray}
where $\mathrm{T}$ denotes the matrix transposition, $i$ in Eq.(\ref{(Eq.14a)}) is introduced in order to satisfy Eq.(\ref{(Eq.3a)}), and $\alpha$ specifies a component of thermal doublets such that $\mathbf{\Psi}^{(1)}(x)=\Psi(x)$, $\mathbf{\Psi}^{(2)}(x)=i[\tilde{\Psi}^\dag(x)]^T$, $\bm{A}_\mu^{(1)}(x)=A_\mu(x)$, $\bm{A}_\mu^{(2)}(x)=\tilde{A}_\mu(x)$, $\bm{F}_{\mu\nu}^{(1)}(x)=F_{\mu\nu}(x)$, and $\bm{F}_{\mu\nu}^{(2)}(x)=\tilde{F}_{\mu\nu}(x)$. Besides, $\slashed{\bm{D}}^{(1)}=\slashed{\partial}+ie\slashed{A}(x)$ and $\slashed{\bm{D}}^{(2)}=\slashed{\partial}+ie\tilde{\slashed{A}}(x)$. In canonical quantization, one can expand the thermal doublets $\bm{A}$, $\bm{\Psi}$, and $\bar{\bm{\Psi}}$ as follows
\begin{eqnarray}
   \bm{A}_\mu(x)&=&\int \frac{d^3p}{(2\pi)^3}\frac{1}{\sqrt{2E_p}}\sum_{r=1,2}[\left(                 % 左括号
      \begin{array}{c}   %该矩阵一共2列，每一列都居中放置
        a^r_\textbf{p} \\  %第一行元素
        \tilde{a}^{r\dag}_\textbf{p} \\  % 第二行元素
      \end{array}
   \right)\varepsilon^r_\mu(\textbf{p})e^{-ip\cdot x}\nonumber\\
            &~&~~~~~~~~~~~~+\left(                 %左括号
      \begin{array}{c}   %该矩阵一共2列，每一列都居中放置
        a^{r\dag}_\textbf{p} \\  %第一行元素
        \tilde{a}^r_\textbf{p} \\  %第二行元素
      \end{array}
   \right)\varepsilon_\mu^r(\textbf{p})e^{ip\cdot x}],
   \label{(Eq.15i)}
\end{eqnarray}
\begin{eqnarray}
   \bm{\Psi}(x)&=&\int \frac{d^3p}{(2\pi)^3}\frac{1}{\sqrt{2E_p}}\sum_{s=1,2}[\left(                 % 左括号
      \begin{array}{c}   %该矩阵一共2列，每一列都居中放置
        c^s_\textbf{p} \\  %第一行元素
        i\tilde{c}^{s\dag}_\textbf{p} \\  % 第二行元素
      \end{array}
   \right)u^s(\textbf{p})e^{-ip\cdot x}\nonumber\\
            &~&~~~~~~~~~~~~+\left(                 %左括号
      \begin{array}{c}   %该矩阵一共2列，每一列都居中放置
        d^{s\dag}_\textbf{p} \\  %第一行元素
        i\tilde{d}^s_\textbf{p} \\  %第二行元素
      \end{array}
   \right)v^s(\textbf{p})e^{ip\cdot x}],
   \label{(Eq.15c)}
\end{eqnarray}
\begin{eqnarray}
   \bar{\bm{\Psi}}(x)&=&\int \frac{d^3p}{(2\pi)^3}\frac{1}{\sqrt{2E_p}}\sum_{s=1,2}[\left(                 % 左括号
      \begin{array}{c}   %该矩阵一共2列，每一列都居中放置
        c^{s\dag}_\textbf{p} \\  %第一行元素
        -i\tilde{c}^{s}_\textbf{p} \\  % 第二行元素
      \end{array}
   \right)^T
   \bar{u}^s(\textbf{p})e^{ip\cdot x}\nonumber\\
            &~&~~~~~~~~~~~~+\left(                 %左括号
      \begin{array}{c}   %该矩阵一共2列，每一列都居中放置
        d^{s}_\textbf{p} \\  %第一行元素
        -i\tilde{d}^{s\dag}_\textbf{p} \\  %第二行元素
      \end{array}
   \right)^T
   \bar{v}^s(\textbf{p})e^{ip\cdot x}],
   \label{(Eq.15f)}
\end{eqnarray}
with $E_{\bm{p}}=\sqrt{\bm{p}^2+m^2}$ and $p^\mu=(p^0,\bm{p})$. Hereafter, for the sake of notational convenience, we denote the creation and annihilation operators
\begin{eqnarray}
  \begin{split}
    &~&a^r_{\bm{p}}\otimes \mathbbm{1}\Rightarrow a^r_{\bm{p}},~~~~~~a^{r\dag}_{\bm{p}}\otimes \mathbbm{1}\Rightarrow a^{r\dag}_{\bm{p}},\\
    &~&\mathbbm{1}\otimes \tilde{a}^r_{\bm{p}}\Rightarrow \tilde{a}^r_{\bm{p}},~~~~~~ \mathbbm{1}\otimes \tilde{a}^{r\dag}_{\bm{p}}\Rightarrow \tilde{a}^{r\dag}_{\bm{p}},\\
    &~&c^s_{\bm{p}}\otimes \mathbbm{1}\Rightarrow c^s_{\bm{p}},~~~~~~c^{s\dag}_{\bm{p}}\otimes \mathbbm{1}\Rightarrow c^{s\dag}_{\bm{p}},\\
    &~&\mathbbm{1}\otimes \tilde{c}^s_{\bm{p}}\Rightarrow \tilde{c}^s_{\bm{p}},~~~~~~ \mathbbm{1}\otimes \tilde{c}^{s\dag}_{\bm{p}}\Rightarrow \tilde{c}^{s\dag}_{\bm{p}},\\
    &~&d^s_{\bm{p}}\otimes \mathbbm{1}\Rightarrow d^s_{\bm{p}},~~~~~~d^{s\dag}_{\bm{p}}\otimes \mathbbm{1}\Rightarrow b^{s\dag}_{\bm{p}},\\
    &~&\mathbbm{1}\otimes \tilde{d}^s_{\bm{p}}\Rightarrow \tilde{d}^s_{\bm{p}},~~~~~~ \mathbbm{1}\otimes \tilde{d}^{s\dag}_{\bm{p}}\Rightarrow \tilde{d}^{s\dag}_{\bm{p}},
  \end{split}
\label{Eq.1a}
\end{eqnarray}
and they obey the usual commutation and anticommutation rules
\begin{eqnarray}
  [a_\textbf{p}^r,a_\textbf{q}^{s\dag}]=(2\pi)^3\delta^{rs}\delta^{(3)}(\textbf{p}-\textbf{q}),
\label{(Eq.15g)}
\end{eqnarray}
\begin{eqnarray}
  [\tilde{a}_\textbf{p}^r,\tilde{a}_\textbf{q}^{s\dag}]=(2\pi)^3\delta^{rs}\delta^{(3)}(\textbf{p}-\textbf{q}),
\label{(Eq.15h)}
\end{eqnarray}
\begin{eqnarray}
  \{a_\textbf{p}^s,a_\textbf{q}^{r\dag}\}=\{b_\textbf{p}^s,b_\textbf{q}^{r\dag}\}=(2\pi)^3\delta^{rs}\delta^{(3)}(\textbf{p}-\textbf{q}),
\label{(Eq.15d)}
\end{eqnarray}
\begin{eqnarray}
  \{\tilde{a}_\textbf{p}^s,\tilde{a}_\textbf{q}^{r\dag}\}=\{\tilde{b}_\textbf{p}^s,\tilde{b}_\textbf{q}^{r\dag}\}=(2\pi)^3\delta^{rs}\delta^{(3)}(\textbf{p}-\textbf{q}).
\label{(Eq.15e)}
\end{eqnarray}
Besides, $\varepsilon^r_\mu(\bm{p})$ is the physical polarization vector for vector particle, and  $u^s(\textbf{p})$ and $v^s(\textbf{p})$ are two momentum dependent Dirac spinors
\begin{eqnarray}
     u^s(\textbf{p})=
     \left(                 %左括号
      \begin{array}{c}   %该矩阵一共2列，每一列都居中放置
        \sqrt{p\cdot\sigma}\xi^s \\  %第一行元素
        \sqrt{p\cdot\bar{\sigma}}\xi^s \\  %第二行元素
      \end{array}
   \right),
   \label{(Eq.65a)}
\end{eqnarray}
\begin{eqnarray}
     v^s(\textbf{p})=
     \left(                 %左括号
      \begin{array}{c}   %该矩阵一共2列，每一列都居中放置
        \sqrt{p\cdot\sigma}\eta^s \\  %第一行元素
        -\sqrt{p\cdot\bar{\sigma}}\eta^s \\  %第二行元素
      \end{array}
   \right)\,
   \label{(Eq.66a)}
\end{eqnarray}
with $\sigma=(\mathbbm{1}_{2\times2}, \boldsymbol{\sigma})$. Here $\bm{\sigma}$ is the Pauli matrix, and $\xi^s$ and $\eta^s$ are two two-component spinors depending on $\textbf{p}$. What's more, $\bar{u}^s(\bm{p})=u^s(\bm{p})^\dag\gamma^0$ and $\bar{v}^s(\bm{p})=v^s(\bm{p})^\dag\gamma^0$.

We note that $\tilde{\mathscr{U}}$ and $\tilde{\mathbb{F}}$ are purely mathematical objects, which are introduced for obtaining the formulas for many body system parallel to the ones in the usual QFT. However, one can still give them the simple physical picture. For example, $\langle \Omega (\beta) |a^\dag a| \Omega (\beta) \rangle$ represents the normal type particle number while we interprets $\langle \Omega (\beta) |\tilde{a}^\dag \tilde{a}| \Omega (\beta) \rangle$ as the tilde type one.

\section{\label{sec:level1}background field method in TFD for spinor QED plasma model}

For simplicity, we assume that fermions are in the equilibrium state, while the electromagnetic field is constituted with a group of noninteracting photons exhibiting no thermal properties. It is equivalent to the usual assumptions in the classical plasmas studies.

\subsection{\label{sec:level2}Nontrivial background field method in TFD}

The core idea of the nontrivial background field method has two folds. First, it is the formalism known as ``Furry picture'' in whose attitude the background field is considered as classical \cite{Furry}. Intuitively speaking, the complex arises from the infinite particle interaction is reduced by incorporating it into the background fields. From Raicher et. al's perspective \cite{Raicher Eliezer Zigler}, it is proper for decomposing both of the Dirac field $\Psi$ and the vector potential field $A_{\mu}$ into classical background fields and quantum fluctuations
\begin{eqnarray}
  \mathbf{\Psi}=\bm{\psi}_0+\bm{\mathbf{\psi}},~~~~\bf{\bm{A}}_\mu= \bar{\mathbf{\bm{A}}}_\mu+\bm{\bm{A}}_\mu,
\label{(Eq.16a)}
\end{eqnarray}
where $\bm{\psi}_0$ and $\bar{\mathbf{A}}_\mu$ are thermal doublets for the classical Dirac field and the vector potential field, respectively. Meanwhile, $\bm{\mathbf{\psi}}$ and $\bm{\mathbf{\mathcal{A}}}_\mu$ denote the corresponding quantum fluctuations. Second, following Yuan Shi et. al's idea \cite{Shi Fisch Qin}, it is better for the classical fields to satisfy the self-consistency classical equations. For our model, by using the variational method, we obtain the Maxwell-Dirac equations for the background thermal doublets as follows
\begin{eqnarray}
  P_\alpha[(i\bar{\bm{\slashed{D}}}^{(\alpha)}-m)\bm{\psi}_0^{(\alpha)}]=0,
\label{(Eq.17a)}
\end{eqnarray}
\begin{eqnarray}
  P_\alpha(\partial_\mu\bar{\bm{F}}^{(\alpha)\mu\nu}-e\bar{\bm{\psi}}_0^{(\alpha)}\gamma^{\nu}\bm{\psi}_0^{(\alpha)})=0.
\label{(Eq.18a)}
\end{eqnarray}
Here, $\bar{\bm{F}}^{(\alpha)}_{\mu\nu}=\partial_\mu \bar{\bm{A}}^{(\alpha)}_\nu-\partial_\nu \bar{\bm{A}}^{(\alpha)}_\mu$ is the thermal doublet for the background electromagnetic field strength tensor, and $\bm{\slashed{\bm{D}}}^{(\alpha)}=\slashed{\partial}+ie\bar{\slashed{\bm{A}}}^{(\alpha)}(x)$.  In this subsection, we first derive the total effective Lagrangian for $\bm{\psi}$, and $\bm{\mathcal{A}}_\mu$. Then we show the gauge invariance of the new model. Lastly, we explain why we can perform the perturbation calculation for our theory, even when a large number of particles participate in.

Inserting Eq.(\ref{(Eq.16a)})-Eq.(\ref{(Eq.18a)}) into Eq.(\ref{(Eq.11a)}), we get the total action of quantum fluctuation fields, up to a useless constant $\frac{1}{4}(\bar{\bm{F}}_{\mu\nu}^{(\alpha)})^2$, as follows
\begin{eqnarray}
  \hat{S}_{\mathrm{eff}}&=&\int d^4x\hat{\mathcal{L}}_{\mathrm{e}\mathrm{f}\mathrm{f}}[\bar{\bm{\psi}},\bm{\psi},\bm{\mathcal{A}}]\nonumber\\
  &=&\sum_{\alpha=1,2}\varepsilon_\alpha P_\alpha\int d^4x[\bar{\bm{\psi}}^{(\alpha)}(i\bar{\slashed{\bm{D}}}^{(\alpha)}-m)\bm{\psi}^{(\alpha)}\nonumber\\
  &~&-\frac{1}{4}(\bm{\mathcal{F}}^{(\alpha)}_{\mu\nu})^2
  -e\bar{\bm{\psi}}^{(\alpha)}_0\bm{\slashed{\mathcal{A}}}^{(\alpha)}\bm{\psi}^{(\alpha)}\nonumber\\
  &~&-e\bar{\bm{\psi}}^{(\alpha)}\bm{\slashed{\mathcal{A}}}^{(\alpha)}\bm{\psi}^{(\alpha)}_0
  -e\bar{\bm{\psi}}^{(\alpha)}\bm{\slashed{\mathcal{A}}}^{(\alpha)}\bm{\psi}^{(\alpha)}],
\label{(Eq.19a)}
\end{eqnarray}
where $\bar{\slashed{\bm{D}}}=\mathrm{diag}(\slashed{\partial}+ie\bar{\slashed{A}},\slashed{\partial}+ie\bar{\slashed{A}})$, and $\bm{\mathcal{F}}^{(\alpha)}_{\mu\nu}=\partial_\mu \bm{\mathcal{A}}_\nu^{(\alpha)}-\partial_\nu \bm{\mathcal{A}}_\mu^{(\alpha)}$ is the thermal doublet for the fluctuation tensor field. The first two terms in the second line correspond to the free Lagrangian density, while the last three terms describe the interactions. $\hat{\mathcal{L}}_{\mathrm{eff}}$ gives a new description for QED plasmas, which focus on $\psi$ and $\mathcal{A}$ instead of electrons, positrons, and photons. For convenience, we denote the corresponding quanta as $e_{\psi}$ and $\gamma_{_\mathcal{A}}$, respectively.

It is worth to show that $\hat{S}_{\mathrm{eff}}$ is invariant under the following local transformations,
\begin{eqnarray}
  \bm{\psi}^{(\alpha)}(x)\rightarrow (e^{i\chi_\alpha(x)}-1)\bm{\psi}_0^{(\alpha)}(x)+e^{i\chi_\alpha(x)}\bm{\psi}^{(\alpha)}(x),
\label{(Eq.20a)}
\end{eqnarray}
\begin{eqnarray}
  \bm{\mathcal{A}}_\mu^{(\alpha)}(x)\rightarrow \bm{\mathcal{A}}_\mu^{(\alpha)} (x)-\frac{1}{e}\partial_\mu\chi_\alpha(x),
\label{(Eq.21a)}
\end{eqnarray}
where $\chi_\alpha(x)$ is any scalar function of $x$. This symmetry is just the natural result of local $U(1)$ symmetry of the standard spinor QED Lagrangian density, i.e., the expression (\ref{(Eq.11b)}). It implies a conservative current $\bm{\mathcal{J}}^{\mu}(x)$ satisfying
\begin{eqnarray}
  \partial_\mu\bm{\mathcal{J}}^{(\alpha)\mu}=0.
\label{(Eq.23a)}
\end{eqnarray}
where
\begin{eqnarray}
  \bm{\mathcal{J}}^{(\alpha)\mu}= P_\alpha[\bar{\bm{\psi}}^{(\alpha)}\gamma^{\mu}\bm{\psi}^{(\alpha)}+\bar{\bm{\psi}}_0^{(\alpha)}\gamma^{\mu}\bm{\psi}^{(\alpha)}
  +\bar{\bm{\psi}}^{(\alpha)}\gamma^{\mu}\bm{\psi}_0^{(\alpha)}]\nonumber\\
\label{(Eq.22a)}
\end{eqnarray}
If we define the vacuum current $\bm{\mathcal{J}}_{\mathrm{v}\mathrm{a}\mathrm{c}}^{(\alpha)\mu}$ and the background current $\bm{\mathcal{J}}_{\mathrm{b}\mathrm{a}\mathrm{c}}^{(\alpha)\mu}$ as
\begin{eqnarray}
  \bm{\mathcal{J}}_{\mathrm{v}\mathrm{a}\mathrm{c}}^{(\alpha)\mu}=P_\alpha(\bar{\bm{\psi}}^{(\alpha)}\gamma^{\mu}\bm{\psi}^{(\alpha)})
\label{(Eq.24a)}
\end{eqnarray}
and
\begin{eqnarray}
  \bm{\mathcal{J}}^{(\alpha)\mu}_{\mathrm{b}\mathrm{a}\mathrm{c}}=P_\alpha(\bar{\bm{\psi}}_0^{(\alpha)}
  \gamma^{\mu}\bm{\psi}^{(\alpha)}+\bar{\bm{\psi}}^{(\alpha)}\gamma^{\mu}\bm{\psi}_0^{(\alpha)}),
\label{(Eq.25a)}
\end{eqnarray}
respectively, the interaction Lagrangian density can be written as
\begin{eqnarray}
  \hat{\mathcal{L}}_{\mathrm{i}\mathrm{n}\mathrm{t}}&=&\sum_{\alpha=1,2}\varepsilon_\alpha P_\alpha[\bm{\mathcal{J}}^{(\alpha) \mu}\bm{\mathcal{A}}^{(\alpha)}_{\mu}]\nonumber\\
  &=&\sum_{\alpha=1,2}\varepsilon_\alpha P_\alpha[(\bm{\mathcal{J}}^{(\alpha) \mu}_{\mathrm{v}\mathrm{a}\mathrm{c}}+\bm{\mathcal{J}}^{(\alpha) \mu}_{\mathrm{b}\mathrm{a}\mathrm{c}})\bm{\mathcal{A}}^{(\alpha)}_{\mu}],
\label{(Eq.26a)}
\end{eqnarray}
which involves the fluctuation vector field $\mathcal{A}_\mu$ coupling to both the background field $\psi_0$ and the fluctuation Dirac spinor filed $\psi$.

When enormous states corresponding to a large number of particles participate in, the calculation of $e^n$ order correction will involve an enormous number of Feynman diagrams, the contribution may even exceed that of the $e^{n-1}$ order, the standard perturbative theory is of no use. It is more clear in the perspective of the path integral. The evolution matrix element can be written as
\begin{eqnarray}
  \langle \bar{\Psi}(\!\!\!\!&t&\!\!\!\!\!,\mathbf{x})\Psi(T,\mathbf{x})A_{\mu}(T,\mathbf{x})|e^{-iHT}|\bar{\Psi}(0,\mathbf{x})\Psi(0,\mathbf{x})A_{\mu}(0,\mathbf{x})\rangle \nonumber\\
  &=&\int D\bar{\Psi} D\Psi DA \exp(i\int_{0}^{T}d^4x\mathcal{L}[\bar{\Psi},\Psi,A]),
\label{(Eq.19b)}
\end{eqnarray}
where $\mathcal{L}[\bar{\Psi},\Psi,A]$ is denoted by Eq. ({\ref{(Eq.11b)}}). The functions $\bar{\Psi}$, $\Psi$, and $A$ over which we integrate are constrained to the specific configurations with $\bar{\Psi}(0,\mathbf{x})$, $\Psi(0,\mathbf{x})$, $A_{\mu}(0,\mathbf{x})$ at $x^0=0$ and $\bar{\Psi}(T,\mathbf{x})$, $\Psi(T,\mathbf{x})$, $A_{\mu}(T,\mathbf{x})$ at $x^0=T$. Only the fields in the function space around the classical trajectory contribute to the integral significantly. As will be shown in Sec. \Rmnum{3}, for the ideal gas, $\Psi_0(x)\sim \sqrt{n_0(t,\mathbf{x})}$, where $n_0(t,\mathbf{x})$ is the number density. The classical electromagnetic potential $A_\mu$ is also a macroscopic physical quantity. It makes the interaction action $S_{\mathrm{i}\mathrm{n}\mathrm{t}}=-e\int d^4x\bar{\Psi}\slashed{A}\Psi$ big enough that one can not simply ignore the higher order contribution. Perturbative calculation in terms of the interaction Hamiltonian $H_{\mathrm{i}\mathrm{n}\mathrm{t}}=-\int d^4x\mathcal{L}_{\mathrm{i}\mathrm{n}\mathrm{t}}$ is inapplicable. Alternatively, in the nontrivial background field scheme, one focus on the evolution element
\begin{eqnarray}
  \langle \bar{\psi}(\!\!\!\!&T&\!\!\!\!\!,\mathbf{x})\psi(T,\mathbf{x})\mathcal{A}_{\mu}(T,\mathbf{x})|e^{-iH_{\mathrm{e}\mathrm{f}\mathrm{f}}T}
  |\bar{\psi}(0,\mathbf{x})\psi(0,\mathbf{x})\mathcal{A}_{\mu}(0,\mathbf{x})\rangle \nonumber\\
  &=&\int D\bar{\psi} D\psi D\mathcal{A} \exp(i\int_{0}^{T}d^4x\mathcal{L}_{\mathrm{e}\mathrm{f}\mathrm{f}}[\bar{\psi},\psi,\mathcal{A}])
\label{(Eq.19c)},
\end{eqnarray}
where $H_{\mathrm{e}\mathrm{f}\mathrm{f}}$ is the Hamiltonian for effective field system corresponding to $\mathcal{L}_{\mathrm{e}\mathrm{f}\mathrm{f}}$. Subsequently, the decomposition of fields, i.e., Eq.(\ref{(Eq.16a)}), makes the contribution of integral mainly from the fluctuation fields around zero. The perturbation theory is acceptable in this case.%In this work, we just focus on the small amplitude wave propagation. The nonlinear QED effects is excluded.

In TFD, propagator can be derived from both of the canonical quantization and the path integral quantization method. Gell-Mann-Low formula and Wick's theorem are also available. It shows a very similar way of constructing the perturbation theory to the usual QFT. We summarize the new Feynman rules in the subsection C. The free thermal doublet propagator and self-energy in the presence of the background fields are discussed in subsection D and E, respectively.

\subsection{\label{sec:level2}Feynman rules}

According to Eq.({\ref{(Eq.19a)}}), the Feynman rules in momentum space for this effective field theory in TFD are listed as follows
\begin{eqnarray}
  \ff{\vertexlabel^{(\alpha)}\vertexlabel_a fs!{fA}pfsx\vertexlabel^{A}f \vertexlabel^{(\beta)}\vertexlabel_b}~~=~S_{TFab}^{(\alpha\beta)}(p;\bar{A}),
\label{(Eq.26c)}
\end{eqnarray}
\begin{eqnarray}
  \ff{\vertexlabel^{(\alpha)}\vertexlabel_a g!{gA}kg \vertexlabel^{(\beta)}\vertexlabel_b}~~=~D_{T\mu\nu}^{(\alpha\beta)}(k),
  \label{(Eq.26d)},
\end{eqnarray}
\begin{eqnarray}
\dd{\vertexlabel^{(\rho)} \\
       fdV \\
& g\vertexlabel_{(\sigma)}\vertexlabel^{\mu} \\
\vertexlabel_{(\tau)} fuA \\
}
~=~-ie\gamma^{\mu}\delta^{\rho\tau\sigma}
\label{(Eq.26e)},
\end{eqnarray}

\begin{eqnarray}
  \dd{        & ![ulft]{x} {\bar{\bm{\psi}}_0^{(\rho)}(p)} \\
                   &  ![ulft]{fvA}p  \\
          \vertexlabel^{(\tau)~~~~} \vertexlabel_{\mu~~~~} ![bot]{gV}k & ![bot]{fV}{k-p} \vertexlabel^{~~~(\sigma)}\\
        }
  ~~~~=~-ie\bar{\bm{\psi}}_0^{(\rho)}(p)\gamma^\mu\delta^{\rho\tau\sigma}
\label{(Eq.26g)},
\end{eqnarray}
\begin{eqnarray}
  \dd{        & ![ulft]{x} {\bm{\psi}_0^{(\rho)}(p)} \\
                   &  ![ulft]{fvV}p  \\
          \vertexlabel^{(\sigma)~~~~} \vertexlabel_{\mu~~~~} ![bot]{fV}{k+p} & ![bot]{gV}k \vertexlabel^{~~~(\tau)}\\
        }
  ~~~~=~-ie\gamma^\mu\bm{\psi}_0^{(\rho)}(p)\delta^{\rho\tau\sigma}
\label{(Eq.26h)},
\end{eqnarray}
where $S_{TFab}^{(\alpha\beta)}(p;\bar{A})$ is the temperature-dependent propagator for $\bm{\psi}$ with momentum $p$ in the presence of $\bar{A}$, and $D_{T\mu\nu}^{(\alpha\beta)}(k)$ is the one for $\bm{\mathcal{A}}$ with momentum $k$. Besides, Eq.({\ref{(Eq.26e)}}) is the ``thermal version" of usual the interaction in QED, the expressions ({\ref{(Eq.26g)}}) and ({\ref{(Eq.26h)}}) are two new interactions. Here, $\alpha$, $\beta$, $\rho$, $\sigma$, and $\tau$ denote the component of thermal doublet, $a$ and $b$ denote the component of Dirac spinor, and $\mu$ and $\nu$ denote the component of 4-vector. What's more,
\begin{eqnarray}
\delta^{\rho\tau\sigma}=
		\begin{cases}
		  1 & \quad \mbox{for}\quad\rho=\tau=\sigma=1,\ \\
		  0 & \quad \mbox{otherwise}.
	\end{cases}
\label{(Eq.26i)}
\end{eqnarray}

\subsection{\label{sec:level2}Propagator in the presence of background fields}

The temperature-dependent vacuum state vector is obtained, with the help of the unitary operator $e^{-iG}$, as follows \cite{Takahashi Umezawa}
\begin{eqnarray}
  | \Omega(\beta)\rangle=e^{-iG}| \Omega\rangle\otimes| \tilde{\Omega}\rangle
\label{(Eq.26b)},
\end{eqnarray}
where $G$ is the generator of this unitary transformation operator, $ | \Omega\rangle$ is the vacuum state corresponding to $\mathcal{L}$, and $ | \tilde{\Omega}\rangle$ is the tilde-conjugate one. One can introduce the temperature-dependent operator
\begin{eqnarray}
  R(\beta)=e^{-iG}Re^{iG}
\label{(Eq.26j)},
\end{eqnarray}
where $R$ is an arbitrary operator corresponding to the space $\mathbb{F}\otimes\tilde{\mathbb{F}}$. For the QED plasma model described by $\hat{\mathcal{L}}[\bar{\bm{\Psi}},\bm{\Psi},\bm{A}_\mu]$, all the physical operators are defined in terms of $\Psi(x)$ and $A_\mu(x)$. The corresponding temperature dependent thermal doublets are
\begin{eqnarray}
  \bm{\Psi}^{(\alpha)}(\beta;x)&=&e^{-iG}\bm{\Psi}^{(\alpha)}(x)e^{iG}\nonumber\\
  &=&\sum_{\tau=1,2} W^{(\alpha\tau)}_F(\beta,x) \bm{\Psi}^{(\tau)}(x)
\label{(Eq.28a)},
\end{eqnarray}
\begin{eqnarray}
  \bm{A}_\mu^{(\alpha)}(\beta;x)&=&e^{-iG}\bm{A}_\mu^{(\alpha)}(x)e^{iG}\nonumber\\
  &=&\sum_{\tau=1,2} W^{(\alpha\tau)}_B(\beta,x) \bm{A}_\mu^{(\tau)}(x)
\label{(Eq.28b)}.
\end{eqnarray}
Here, $W_F(\beta,x)$ and $W_B(\beta,x)$ are two $2\times2$ transformation matrices for fluctuation fermion and boson, respectively, and determined by the statistical properties of the system.

Then, for the effective field theory described by $\hat{\mathcal{L}}_{\mathrm{eff}}[\bar{\bm{\psi}},\bm{\psi},\bm{\mathcal{A}}_\mu]$, we use $W_F(\beta,x)$ to define a similar unitary transformation operator $e^{-i\mathcal{G}}$ as
\begin{eqnarray}
  \bm{\mathcal{\psi}}^{(\alpha)}(\beta;x)&=&e^{-i\mathcal{G}}\bm{\mathcal{\psi}}_\mu^{(\alpha)}(x)e^{i\mathcal{G}}\nonumber\\
                                         &=&W^{(\alpha\tau)}_F(\beta,x) \bm{\psi}^{(\tau)}(x)
\label{(Eq.26m)}.
\end{eqnarray}
With the help of $e^{-i\mathcal{G}}$, we define $|\Omega_{\mathrm{eff}}(\beta)\rangle$ as
\begin{eqnarray}
  |\Omega_{\mathrm{eff}}(\beta)\rangle=e^{-i\mathcal{G}}| \Omega_{\mathrm{eff}}\rangle\otimes| \widetilde{\Omega_{\mathrm{eff}}}\rangle,
\label{(Eq.26n)}
\end{eqnarray}
where $|\Omega_{\mathrm{eff}}\rangle$ is the vacuum state vector corresponding to $\mathcal{L}_{\mathrm{eff}}$, and $| \widetilde{\Omega_{\mathrm{eff}}}\rangle$ is the tilde conjugate one. The significance of the definition ({\ref{(Eq.26m)}}) is that the statistical distribution of the effective fluctuation fermions is the same as that of the real fermions. It is quite natural since they share the same mass, charge, and spin.
%In most cases, photons are not in equilibrium state. For simplicity, we consider the single frequency wave propagation case which is common seen in the studies of linear wave propagation in classical plasmas and non-relativistic degenerate plasmas. Then the state of incident light is denoted as $|k\rangle|k\rangle\cdot\cdot\cdot|k\rangle$. It indicates a group of single photons propagating in plasmas. Subsequently, the temperature dependent propagator of photon is similar to that in the usual QFT. The transformation matrix $U_B=1$.

Then, we define the free propagator for $\bm{\psi}$ as
\begin{eqnarray}
  S_{TF}(x-y;\bar{A})&=&\langle 0_{\mathrm{eff}}(\beta)|T\bm{\psi}(x)\bar{\bm{\psi}}(y)|0_{\mathrm{eff}}(\beta)\rangle\nonumber\\
             &=&\int \frac{d^4p}{(2\pi)^4}e^{-ip\cdot (x-y)}S_{TF}(p;\bar{A})
\label{(Eq.31a)},
\end{eqnarray}
where
\begin{eqnarray}
  |0_{\mathrm{eff}}(\beta)\rangle=e^{-i\mathcal{G}}|0_{\mathrm{eff}}\rangle\otimes| \widetilde{0_{\mathrm{eff}}}\rangle
\label{(Eq.26o)}.
\end{eqnarray}
$|0_{\mathrm{eff}}\rangle$ is the vacuum state vector for the free effective theory, and $|\widetilde{0_{\mathrm{eff}}}\rangle$ is the tilde conjugate one. Using the path integral method, we have
\begin{eqnarray}
  S_{TF}(x-y;\bar{A})\!&=&\!\frac{1}{Z_{0F}}W_F(\beta,x)\int D\bar{\bm{\psi}}D\bm{\psi}e^{i\int\! d^4x\hat{\mathcal{L}}_{0\bar{A}}[\bar{\bm{\psi}},\bm{\psi}]}\nonumber\\
                      &~&~~~~~\times\bm{\psi}(x)\bar{\bm{\psi}}(y)W_F^{-1}(\beta,y)\nonumber\\
                      &=&\!W_F(\beta,x)\!
   \left(\!\!                 %左括号
      \begin{array}{cc}   %该矩阵一共2列，每一列都居中放置
        S_F(x-y;\bar{A})\!\!\! &  0 \\  %第一行元素
        0  & \!\!\!\tilde{S}_F(x-y;\bar{A}) \\  %第二行元素
      \end{array}
   \!\!\right)\nonumber\\
   &~&~~~~~~~\times W_F^{-1}(\beta,y)
\label{(Eq.31b)},
\end{eqnarray}
where,
\begin{eqnarray}
\hat{\mathcal{L}}_{0\bar{A}}[\bar{\bm{\psi}},\bm{\psi}]=\sum_\alpha \varepsilon_\alpha P_\alpha[\bar{\bm{\psi}}^{(\alpha)}(i\bar{\slashed{D}}^{(\alpha)}-m)\bm{\psi}^{(\alpha)}]
\label{(Eq.31c)},
\end{eqnarray}
and
\begin{eqnarray}
  Z_{0F}=\int D\bar{\bm{\psi}}D\bm{\psi}e^{i\int d^4x\hat{\mathcal{L}}_{0\bar{A}}[\bar{\bm{\psi}},\bm{\psi}]}.
  \label{(Eq.44a)}
\end{eqnarray}.
\begin{eqnarray}
S_{F}(x-y;\bar{A})&=&\int \frac{d^4p}{(2\pi)^4}e^{-ip\cdot(x-y)}S_F(p;A)\nonumber\\
&=&\int \frac{d^4p}{(2\pi)^4}\frac{ie^{-ip\cdot(x-y)}}{\slashed{p}-e\bar{\slashed{A}}-m+i\epsilon}
\label{(Eq.31d)},
\end{eqnarray}
and
\begin{eqnarray}
\tilde{S}_{F}(x-y;\bar{A})&=&\int \frac{d^4p}{(2\pi)^4}e^{-ip\cdot(x-y)}\tilde{S}_F(p;A)\nonumber\\
&=&\int \frac{d^4p}{(2\pi)^4}\frac{ie^{-ip\cdot(x-y)}}{\slashed{p}-e\bar{\slashed{A}}-m-i\epsilon}
\label{(Eq.31e)}.
\end{eqnarray}
In momentum space,
\begin{eqnarray}
  S_{TF}(p;\bar{A})=
   U_F(\beta,p)\!
   \left(\!\!                 %左括号
      \begin{array}{cc}   %该矩阵一共2列，每一列都居中放置
        S_F(p;\bar{A}) &  0 \\  %第一行元素
        0  & \tilde{S}_F(p;\bar{A}) \\  %第二行元素
      \end{array}
   \!\!\right)
   \!U_F^{-1}(\beta,p),\nonumber\\               %右括号.
\label{(Eq.32a)}
\end{eqnarray}
where, $U_F(\beta,p)$ is a $2\times2$ matrix satisfying
\begin{eqnarray}
  U_F(\beta,i\partial_{x\mu})=W_F(\beta,x)
\label{(Eq.32b)}.
\end{eqnarray}
The corresponding inverse matrix $W^{-1}_F(\beta,x)$ satisfies
\begin{eqnarray}
  U_F^{-1}(\beta,i\overleftarrow{\partial}_{y\mu})=W_F^{-1}(\beta,y)
\label{(Eq.32c)},
\end{eqnarray}
where $\leftarrow$ indicates differentiating to left.

If the electron-positron system obeys Fermi distribution, by direct computation, one can show that \cite{Ojima}\cite{Saito Maruyama Soutome}
\begin{eqnarray}
  U_F(\beta,p)=
    \left(                 %左括号
      \begin{array}{cc}   %该矩阵一共2列，每一列都居中放置
        \cos\theta_{p^0}(\beta) & \sin\theta_{p^0}(\beta) \\  %第一行元素
        -\sin\theta_{p^0}(\beta) & \cos\theta_{p^0}(\beta) \\  %第二行元素
      \end{array}
   \right).
\label{(Eq.33a)}
\end{eqnarray}
Here,
\begin{eqnarray}
  \cos\theta_{p^0}(\beta)=\frac{\theta (p^0)}{\sqrt{e^{\beta (\mu-p^0)}+1}}+\frac{\theta (-p^0)}{\sqrt{e^{\beta (\mu-p^0)}+1}},
\label{(Eq.34a)}
\end{eqnarray}
\begin{eqnarray}
  \sin\theta_{p^0}(\beta)=\frac{e^{-\frac{1}{2}\beta (\mu-p^0)}\theta (p^0)}{\sqrt{e^{-\beta (\mu-p^0)}+1}}+\frac{e^{\frac{1}{2}\beta (\mu-p^0)}\theta (p^0)}{\sqrt{e^{\beta (\mu-p^0)}+1}},
\label{(Eq.35a)}
\end{eqnarray}
where $\mu$ is the chemical potential of Dirac particle system and $\theta(p^0)$ is the step function
\begin{eqnarray}
  \theta(p^0)=
		\begin{cases}
		  1 & p^0\geqslant 0,\ \\
		  0 & p^0< 0.
	\end{cases}
  \label{(Eq.36a)},
\end{eqnarray}

\subsection{\label{sec:level2}Temperature-dependent self-energy in the presence of background fields}

We denote the temperature-dependent self-energy for $\gamma_{_\mathcal{A}}$ as $\Pi_T^{(\alpha\beta)\mu\nu}(x,y)$. It depends only on $x-y$ if the system is uniform. In this case, by using Fourier's transformation, we obtain
\begin{eqnarray}
   &~&\Pi_T^{(\alpha\beta)\mu\nu}(x,y)= \int \frac{d^4kd^4k'}{(2\pi)^8}e^{-ik\cdot x-ik\cdot y}\Pi_T^{(\alpha\beta)\mu\nu}(k,k')\nonumber\\
   &=&\Pi_T^{(\alpha\beta)\mu\nu}(x-y)=\int \frac{d^4kd^4k'}{(2\pi)^8}e^{-ik\cdot (x-y)}\Pi_T^{(\alpha\beta)\mu\nu}(k).\nonumber\\
   \label{(Eq.48a)}
\end{eqnarray}
Therefore
\begin{eqnarray}
   \Pi_T^{(\alpha\beta)\mu\nu}(k,k')=(2\pi)^4\delta^{(4)}(k+k')\Pi_T^{(\alpha\beta)\mu\nu}(k).
   \label{(Eq.48b)}
\end{eqnarray}

To the order of $e^2$, there are three connected diagrams contribute $\Pi_T^{\mu\nu}(k)$, as
\begin{eqnarray}
  \Pi_{T,2}^{(\alpha\beta)\mu\nu}(k)=\Pi^{(\alpha\beta)\mu\nu}_{T\mathrm{bac},2}(k)+\Pi^{(\alpha\beta)\mu\nu}_{T\mathrm{vac},2}(k),
  \label{(Eq.39b)}
\end{eqnarray}
with
\begin{eqnarray}
  &~&i\Pi^{(\alpha\beta)\mu\nu}_{T\mathrm{bac},2}(k)\nonumber\\
  \nonumber\\
  &=&\!\!\int\!\! \frac{d^4q}{(2\pi)^4}\Bigg(~\Diagram{ & ![ulft]{x} {\bar{\bm{\psi}}_0(q)} \\
                                           & ![ulft]{fvA}{q} \\
           \vertexlabel^{\mu}\vertexlabel_{(\alpha)}~ ![bot]{gA}k & ![bot]{fV}{q-k}xfs & ![lrt]{gA}k ~\vertexlabel^{\nu}\vertexlabel_{(\beta)} \\
                                           &                  & ![lrt]{fvA}{q} \\
                                           &                  & ![lrt]x {\bm{\psi}_0(q)} \\
   }
                               ~+~\Diagram{ & ![ulft] x {\bm{\psi}_0(q)} \\
                                            & ![ulft]{fvV}{q} \\
             \vertexlabel^{\mu}\vertexlabel_{(\alpha)}~![bot]{gA}k & fsx![bot]{fA}{q+k}  & ![lrt]{gA}k  ~\vertexlabel^{\nu}\vertexlabel_{(\beta)} \\
                                            &                  & ![lrt]{fvV}{q} \\
                                            &                  & ![lrt]x {\bar{\bm{\psi}}_0(q)} \\
   }~\Bigg).\nonumber\\
  \nonumber\\
  \label{(Eq.39c)}
\end{eqnarray}
and
\begin{eqnarray}
i\Pi_{T\mathrm{vac},2}^{(\alpha\beta)\mu\nu}(k)\!=\!\int\!\frac{d^4q}{(2\pi)^4}~~\ff{\vertexlabel^{\mu}\vertexlabel_{(\alpha)}g~~~~\momentum{flSA}{p+k}\momentum{flSuV}{p}~~~~g\vertexlabel^{\nu}\vertexlabel_{(\beta)}}
 \label{(Eq.39d)}~~.\nonumber\\
\end{eqnarray}
Here, $\Pi^{\mu\nu}_{T\mathrm{bac},2}(k)$ and $\Pi_{Tvac,2}^{\mu\nu}(k)$ are thermal background dependent and vacuum dependent self-energy, to the second order of $e^2$, respectively. We note that the fermion propagator in the expression ({\ref{(Eq.39d)}}) must be a thermal one, but not just $S_F$. Previous studies made misunderstanding on this point \cite{Shi Fisch Qin}\cite{Chin}.
%For later use, with the help of the relation ({\ref{(Eq.32a)}}), we rewrite these two polarization tensors as follows
%\begin{widetext}
%\begin{eqnarray}
%i\Pi_{T\mathrm{bac},2}^{(\alpha\beta)\mu\nu}(k)
%&=&\!-e^2\!\int \frac{d^4q}{(2\pi)^4}\sum_{\rho\sigma}[\bar{\bm{\psi}}^{(\alpha)}_0(q)\Gamma^\mu_{(\alpha\rho)} S_{TF}^{(\rho\sigma)}(q-k;\bar{A})\Gamma^\nu_{(\sigma\beta)}\bm{\psi}^{(\beta)}_0(q)+\bar{\bm{\psi}}^{(\alpha)}_0(q)\Gamma^\mu_{(\alpha\rho)} S_{TF}^{(\rho\sigma)}(q+k;\bar{A})\Gamma^\nu_{(\sigma\beta)}\bm{\psi}^{(\beta)}_0(q)]\nonumber\\
%&=&\!-e^2\!\int \frac{d^4q}{(2\pi)^4}\bigg[\mathcal{U}_F(\beta,p)
%    \left(                 %左括号
%      \begin{array}{cc}   %该矩阵一共2列，每一列都居中放置
%        \bar{\bm{\psi}}^{(1)}_0\gamma^\mu S_F(q-k;\bar{A})\gamma^\nu\bm{\psi}^{(1)}_0   &   0 \\  % 第一行元素
%        0   &  \bar{\bm{\psi}}^{(2)}_0\gamma^\mu \tilde{S}_F(q-k;\bar{A})\gamma^\nu\bm{\psi}^{(2)}_0 \\  % 第二行元素
%      \end{array}
%   \right)
%   \mathcal{U}_F^{-1}(\beta,p)\bigg]^{(\alpha\beta)},
% \label{(Eq.39f)}
%\end{eqnarray}
%and
%\begin{eqnarray}
%  i\Pi_{T\mathrm{vac},2}^{(\alpha\beta)\mu\nu}(k)=-e^2\sum_{\rho\sigma}\int\frac{d^4p}{(2\pi)^4}\mathrm{tr}[\Gamma_{(\alpha\rho)}^\mu S_{TF}^{(\rho\sigma)}(p)\Gamma_{(\rho\beta)}^\nu S_{TF}^{(\sigma\rho)}(p+k)]
% \label{(Eq.39g)}
%\end{eqnarray}
%where we use $\mathrm{tr}()$ to denote Dirac traces.
%\end{widetext}

To calculate these two self-energies, one should determine the terms containing background fields. New method is needed whose details are delayed to Sec. \Rmnum{4}. Here we just point out that although the theory is gauge invariance, it has no restriction on $\Pi^{(\alpha\beta)\mu\nu}_{T\mathrm{bac}}(k)$. More specifically, it is not necessary for $\Pi^{(\alpha\beta)\mu\nu}_{T\mathrm{bac}}(k)$ to satisfy Ward identity $k_\mu\Pi^{(\alpha\beta)\mu\nu}_{T\mathrm{bac}}(k)=0$, which indicates a possible different form comparing with $(k^2g^{\mu\nu}-k^\mu k^\nu)\Pi^{(\alpha\beta)}_T(k)$, where $k$ denotes the modulus of $k^\mu$ in Minkovsky space. To ensure the coherence of the paper, we delay the explanation to Appendix B.

\subsection{\label{sec:level2}Effective self-energy and collective mode}

In this subsection, an effective self-energy for fluctuation boson is defined. Then we are going to explain how to relate it with the thermal self-energy, which enables us to obtain the collective modes.

We consider a physical process that a group incident photons entering into plasma, then ejecting. From the effective field point of view, it is equivalent to the scattering of $\gamma_{_\mathcal{A}}$ by an alternative medium constituted with a huge number of $e_{\psi}$ in thermal equilibrium state. The interaction between $\gamma_{_\mathcal{A}}$ and the effective medium is described by
\begin{displaymath}
  \mathcal{L}_{\mathrm{eff}}^{\mathrm{int}}=-e\bar{\psi}_0\slashed{\mathcal{A}}\psi
  -e\bar{\psi}\slashed{\mathcal{A}}\psi_0-e\bar{\psi}\slashed{\mathcal{A}}\psi.
\end{displaymath}

Next, with the help of one-particle irreducible (1PI) diagram in usual QFT, we define a usual self-energy for $\gamma_{_\mathcal{A}}$ and name it as effective self-energy $\Pi_{\mathrm{eff}}^{\mu\nu}(k)$
\begin{eqnarray}
  i\Pi_{\mathrm{eff}}^{\mu\nu}(k)=~~~~\Diagram{
           \vertexlabel^\mu  &     &       &\vertexlabel^\nu  \\
                      gd     &  p  &  gu   &  \\
     \vertexlabel_{\mathrm{eff~med}}  ms  &     &  ms \vertexlabel_{\mathrm{eff~med}}
          }~~~
 \label{(Eq.50h)},
\end{eqnarray}
where ``$\feyn{ms}$ eff med'' represents the effective medium. The full propagator for $\gamma_{_\mathcal{A}}$ in the medium is
\begin{eqnarray}
  D_F(k)+D_F(k)(i\Pi(k))D_F(k)
 \label{(Eq.50d)},
\end{eqnarray}
with
\begin{eqnarray}
    &~&i\Pi^{\mu\nu}(k)\nonumber\\
    &=&i\Pi^{\mu\nu}_{\mathrm{eff}}(k)+(i\Pi_{\mathrm{eff}}(k)D_F(k)i\Pi_{\mathrm{eff}}(k))^{\mu\nu}\nonumber\\
    &~&+(i\Pi_{\mathrm{eff}}(k)D_F(k)i\Pi_{\mathrm{eff}}(k)D_F(k)i\Pi_{\mathrm{eff}}(k))^{\mu\nu}+\cdot\cdot\cdot\nonumber\\
    &=&\{i\Pi_{\mathrm{eff}}(k)[1-iD_F(k)\Pi_{\mathrm{eff}}(k)]^{-1}\}^{\mu\nu},
   \label{(Eq.50c)}
\end{eqnarray}
and
\begin{eqnarray}
  D^{\mu\nu}_F(k)=\frac{-i}{k^2+i\epsilon}[g^{\mu\nu}-(1-\frac{1}{\xi})\frac{k^\mu k^\nu}{k^2}]
 \label{(Eq.36e)},
\end{eqnarray}
where $\xi$ is any finite constant.
The collective modes can be obtained from the poles of the full propagator, i.e., the equation
\begin{eqnarray}
  \det[D^{-1}_F(k)-i\Pi_{\mathrm{eff}}(k)]=0
 \label{(Eq.50f)}.
\end{eqnarray}

The rest work is devoted into relating $\Pi_{\mathrm{eff}}^{\mu\nu}(k)$ to $\Pi_T^{(\alpha\beta)\mu\nu}(k)$.
%First, we note that the S-matrix for the forward scattering
%\begin{eqnarray}
%  \gamma_{_\mathcal{A}},\mathrm{effective~medium}\rightarrow \gamma_{_\mathcal{A}},\mathrm{effective~medium}
% \label{(Eq.50e)}
%\end{eqnarray}
%is
%\begin{eqnarray}
%    &~&\langle\bm{k}|\otimes\langle\mathrm{eff~med}|S|\mathrm{eff~ med}\rangle\otimes|\bm{k}\rangle\nonumber\\
%    &=&\!\mathbbm{1}\!+\!\!\!\!\!\!\sum_{\mathrm{polarizations}}\!\!\!\!\!\!\!\varepsilon_\mu(\bm{k})\varepsilon_\nu(\bm{k})^*
%    \Bigg(~~~~\Diagram{
%           \vertexlabel^\mu  &     &       &\vertexlabel^\nu  \\
%                      gd     &  p  &  gu   &  \\
%     \vertexlabel_{\mathrm{eff~med}}  ms  &     &  ms \vertexlabel_{\mathrm{eff~med}}
%          }~~~\Bigg)(2\pi)^4\delta^{(4)}(0),\nonumber\\
%   \label{(Eq.50b)}
%\end{eqnarray}
%where $|\bm{k}\rangle$ and $|\mathrm{eff~med}\rangle$ are the state vectors of $\gamma_{_\mathcal{A}}$ and effective medium, respectively.
From the expression ({\ref{(Eq.15i)}}), absorbing a $\gamma_{_\mathcal{A}}$ in usual QFT is equivalent to absorbing temperature-dependent fluctuation boson $\gamma_{_\mathcal{A}}(\beta)$ corresponding to $a_{\bm{k}}^\dag(\beta)$, or to emitting a tilde conjugate one $\tilde{\gamma}_{_\mathcal{A}}(\beta)$ corresponding to $\tilde{a}_{\bm{k}}(\beta)$ in TFD, since they share the same 4-momentum $k$ and polarization vector $\varepsilon^*(k)$. Analogously, emitting a $\gamma_{_\mathcal{A}}$ is equivalent to absorbing $\tilde{\gamma}_{_\mathcal{A}}(\beta)$, or to emitting a tilde conjugate one. It indicates that, for the thermal equilibrium state system, the process
\begin{eqnarray}
  \mathrm{effective~medium}+\gamma_{_\mathcal{A}}\rightarrow \mathrm{effective~medium}-\gamma_{_\mathcal{A}}
 \label{(Eq.50e)}
\end{eqnarray}
in usual QFT, gives rise to the following four and only four processes in TFD:
\begin{eqnarray}
  \begin{split}
   \mathrm{effective~medium}+\gamma_{_\mathcal{A}}(\beta)\rightarrow \mathrm{effective~medium}-\gamma_{_\mathcal{A}}(\beta),\\
   \mathrm{effective~medium}+\gamma_{_\mathcal{A}}(\beta)\rightarrow \mathrm{effective~medium}+\tilde{\gamma}_{_\mathcal{A}}(\beta),\\
   \mathrm{effective~medium}-\tilde{\gamma}_{_\mathcal{A}}(\beta)\rightarrow \mathrm{effective~medium}-\gamma_{_\mathcal{A}}(\beta),\\
   \mathrm{effective~medium}-\tilde{\gamma}_{_\mathcal{A}}(\beta)\rightarrow \mathrm{effective~medium}+\tilde{\gamma}_{_\mathcal{A}}(\beta),\\
  \end{split}\nonumber\\
 \label{(Eq.50l)}
\end{eqnarray}
where $+$ and $-$ are used to denote absorbing and emitting particles, respectively. In the processes ({\ref{(Eq.50e)}}) and ({\ref{(Eq.50l)}}), both on-shell and off-shell $\gamma_{_\mathcal{A}}$ and $\gamma_{_\mathcal{A}}(\beta)$ are acceptable.  We assert that these four processes contribute equally to the process ({\ref{(Eq.50e)}}). An intuitive explanation is that there is no reason for any of these four being special. A more rigorous argument relies on the following observations: From Eq. ({\ref{(Eq.8b)}}) and ({\ref{(Eq.10a)}}), any operator $F(a^r_{\bm{k}},a^{r\dagger}_{\bm{k}},c^s_{\bm{p}},c^{s\dagger}_{\bm{p}},d^t_{\bm{q}},d^{t\dagger}_{\bm{q}})$
satisfies
\begin{eqnarray}
  &~&\!\!\!\!\!\!\!\!\!\!\!\!\!\!\!\!\langle\Omega(\beta)|\exp(-i\hat{H}t)F\exp(i\hat{H}t)|\Omega(\beta)\rangle\nonumber\\
  &~&=\langle\Omega(\beta)|\exp[-i\tilde{\hat{H}}(-t)]\tilde{F}\exp[i\tilde{\hat{H}}(-t)]|\Omega(\beta)\rangle,
 \label{(Eq.50n)}
\end{eqnarray}
where
\begin{eqnarray}
  \tilde{F}=F(\tilde{a}^r_{\bm{k}},\tilde{a}^{r\dagger}_{\bm{k}},\tilde{c}^s_{\bm{p}},
  \tilde{c}^{s\dagger}_{\bm{p}},\tilde{d}^t_{\bm{q}},\tilde{d}^{t\dagger}_{\bm{q}}).
 \label{(Eq.50p)}
\end{eqnarray}
It indicates the symmetry: In the thermal equilibrium state, the system is invariant under the combination of the following transforms:

(a) particle and tilde conjugate particle interchange
\begin{eqnarray}
  \begin{split}
    &~&a^r_{\bm{k}}(\beta)\longleftrightarrow\tilde{a}^{r}_{\bm{k}}(\beta),
    ~~~~~~a^{r\dagger}_{\bm{k}}(\beta)\longleftrightarrow\tilde{a}^{r\dagger}_{\bm{k}}(\beta),\\
    &~&c^{s\dagger}_{\bm{p}}(\beta)\longleftrightarrow\tilde{c}^{s\dagger}_{\bm{p}}(\beta),
    ~~~~~~c^{s\dagger}_{\bm{p}}(\beta)\longleftrightarrow\tilde{c}^{s\dagger}_{\bm{p}}(\beta),\\
    &~&d^{s\dagger}_{\bm{q}}(\beta)\longleftrightarrow\tilde{d}^{t\dagger}_{\bm{q}}(\beta),
    ~~~~~~d^{t\dagger}_{\bm{q}}(\beta)\longleftrightarrow\tilde{d}^{t\dagger}_{\bm{q}}(\beta)
  \end{split}
 \label{(Eq.50m)}
\end{eqnarray}

(b) time inversion
\begin{eqnarray}
  t\longrightarrow -t.
 \label{(Eq.50o)}
\end{eqnarray}
After these operations, in the laboratory frame, we have the interchanges
\begin{eqnarray}
  \pm\gamma_{_\mathcal{A}}(\beta)\longleftrightarrow \mp\tilde{\gamma}_{_\mathcal{A}}(\beta)
  \label{(Eq.50p)}
\end{eqnarray}
in the expression ({\ref{(Eq.50l)}}). The effective medium is invariant under these operations because that the number of particle and conjugate particle is equal:
\begin{eqnarray}
  &~&\!\!\!\!\!\!\!\!\!\int\frac{d^3p}{(2\pi)^3}\sum_{r=1,2}\langle\Omega(\beta)|c^{r\dag}_{\bm{p}}c^r_{\bm{p}}|\Omega(\beta)\rangle\nonumber\\
  &~&~~~~~~~~~~~=\int\frac{d^3p}{(2\pi)^3}
  \sum_{r=1,2}\langle\Omega(\beta)|\tilde{c}^{r\dag}_{\bm{p}}\tilde{c}^r_{\bm{p}}|\Omega(\beta)|\rangle
 \label{(Eq.50m)},
\end{eqnarray}
\begin{eqnarray}
  &~&\!\!\!\!\!\!\!\!\!\int\frac{d^3q}{(2\pi)^3}\sum_{r=1,2}\langle\Omega(\beta)|d^{r\dag}_{\bm{q}}d^r_{\bm{q}}|\Omega(\beta)\rangle\nonumber\\
  &~&~~~~~~~~~~~=\int\frac{d^3q}{(2\pi)^3}
  \sum_{r=1,2}\langle\Omega(\beta)|\tilde{d}^{r\dag}_{\bm{q}}\tilde{d}^r_{\bm{q}}|\Omega(\beta)|\rangle
 \label{(Eq.50r)}.
\end{eqnarray}
Then, we conclude that the processes ({\ref{(Eq.50l)}}) equally contribute to $\Pi_{\mathrm{eff}}^{\mu\nu}(k)$.

Next, we insert $\Pi_{\mathrm{eff}}^{\mu\nu}(k)$ into the scattering process $e^+_\psi e^-_\psi\rightarrow e^+_\psi e^-_\psi$ whose invariant matrix $\mathcal{M}$ satisfies
\begin{eqnarray}
  i\mathcal{M}=-ie^2\bar{u}(p)\gamma_\mu v(p_+)\frac{-i}{k^2}(i\Pi_{\mathrm{eff}}^{\mu\nu}(k))\frac{-i}{k^2}\bar{v}(p_+)\gamma_\nu u(p).\nonumber\\
 \label{(Eq.50j)}
\end{eqnarray}
The corresponding diagram is
\begin{eqnarray}
  \Diagram{
      !{fdA}{p_+} & !{gA}{k} P !{gA}{k} & !{fuA}{p}  \\
      !{fuV}{p}   &                     & !{fdV}{p_+}
          }
 \label{(Eq.50k)}~~.
\end{eqnarray}
Here $\feyn{gPg}$ is evaluated with TFD, denoting the amplitude for effective medium absorbing a quanta with 4-momentum $k$, and then emitting one. Because $|\mathcal{M}|^2$ is proportional to the cross section for $e^+_\psi e^-_\psi\rightarrow e^+_\psi e^-_\psi$ and, from the optical theorem, $\mathrm{Im}\mathcal{M}$ is uniquely determined by the total cross section for $e^+_\psi e^-_\psi\rightarrow \mathrm{all~possible~indipendent~states}$, we conclude that $\mathcal{M}$ is uniquely determined no matter what calculation method is used. Subsequently, TFD method must give the same value of $\feyn{gPg}$ as usual QFT does. Considering that $\Pi_{T}^{(\rho\lambda)\mu\nu}(k)$ represents the amplitude for effective medium to absorb a $\rho$-type boson, and then to emit a $\lambda$-type one, and that these amplitudes equally contribute to $\feyn{gPg}$, we have
\begin{eqnarray}
  \Pi_{\mathrm{eff}}^{\mu\nu}(k)=\sum_{\rho\lambda=1,2}\Pi_{T}^{(\rho\lambda)\mu\nu}(k).
 \label{(Eq.39x)}
\end{eqnarray}
Substituting it into Eq. ({\ref{(Eq.50f)}}), we have
\begin{eqnarray}
  \det[D^{-1}_F(k)-i\!\sum_{\rho\lambda=1,2}\!\Pi_{T}^{(\rho\lambda)}(k)]=0
 \label{(Eq.39w)}.
\end{eqnarray}
It can be also obtained from the effective Lagrangian for $\mathcal{A}_\mu$, as shown in Appendix A. We must point out that, unlike random phase approximation (RPA) in which $\Pi_{\mathrm{eff}}^{\mu\nu}(k)$ is expressed as a circle diagram, our method gives a more general expression for it, as seen in Eq. ({\ref{(Eq.39c)}}), ({\ref{(Eq.39d)}}), and ({\ref{(Eq.39x)}}). The many-body effect is equipped in the background field, which is the core idea of ``Furry picture''. It gives more information about the system than the standard RPA does. For the ideal plasma, the method leads to a new vacuum fluctuation correction. Details are shown in Sec. \Rmnum{5}.

\section{\label{sec:level1}Calculation of polarization tensor}

Determining the terms containing $\psi_0$ is essential for calculating $\sum_{\rho,\lambda=1,2}\Pi^{(\rho\lambda)\mu\nu}_{T\mathrm{bac},2}(k)$. The classical limit method is developed to obtain these terms in the subsection A. The discussion of $\sum_{\rho,\lambda=1,2}\Pi^{(\rho\lambda)\mu\nu}_{Tvac,2}(k)$ is shown in subsection B.

\subsection{\label{sec:level2}Calculation of $\sum_{\rho,\lambda=1,2}\Pi^{(\rho\lambda)\mu\nu}_{Tbac,2}(k)$}

From the expression ({\ref{(Eq.39c)}}), we have
\begin{eqnarray}
  &~&i\sum_{\rho,\lambda=1,2}\Pi^{(\rho,\lambda)\mu\nu}_{T\mathrm{bac},2}(k)(2\pi)^4\delta^{(4)}(0)\nonumber\\
  &=&-e^2\!\!\sum_{\rho,\lambda,\sigma,\tau=1,2}\int \frac{d^4q}{(2\pi)^4}\bar{\bm{\psi}}_0^{(\rho)}(q)\nonumber\\
  &~&\times[(\Gamma_{(\rho\sigma)})^\mu S_{TF}^{(\sigma\tau)}(q-k;\bar{A})(\Gamma_{(\tau\lambda)})^\nu\nonumber\\
  &~&+(\Gamma_{(\rho\sigma)})^\nu S^{({\sigma\tau})}_{TF}(q+k;\bar{A})(\Gamma_{(\tau\lambda)})^\mu]\bm{\psi}^{(\lambda)}_0(q),
\label{(Eq.52a)}
\end{eqnarray}
where delta function $\delta^{(4)}(0)$ comes from the law of momentum conservation which is easily seen if one set $k=-k^\prime$ in Eq.({\ref{(Eq.48a)}}), and $\Gamma^\mu$ is introduced for notation convenience, as follows
\begin{eqnarray}
  \Gamma^\mu=
    \left(                 %左括号
      \begin{array}{cc}   %该矩阵一共2列，每一列都居中放置
        \gamma^\mu & 0 \\  %第一行元素
        0 & \gamma^\mu \\  %第二行元素
      \end{array}
   \right).
\label{(Eq.38a)}
\end{eqnarray}

For unmagnetized plasma, $\bar{A}^\mu=0$, $S_F$ and $\tilde{S}_F$ become
\begin{eqnarray}
  S_F(p)=\frac{i}{\slashed{p}-m+i\epsilon}
\label{(Eq.32d)},
\end{eqnarray}
\begin{eqnarray}
  \tilde{S}_F(p)=\frac{i}{\slashed{p}-m-i\epsilon}
\label{(Eq.32e)}.
\end{eqnarray}
To the order of $e^2$, by using the well-known gamma matrix properties of $[\gamma^\mu,\gamma^\nu]=\frac{i}{4}S^{\mu\nu}$ and $\{\gamma^\mu,\gamma^\nu\}=2g^{\mu\nu}$, inserting Eq. ({\ref{(Eq.32d)}}) and Eq. ({\ref{(Eq.32e)}}) into Eq. ({\ref{(Eq.52a)}}), we have
\begin{widetext}
\begin{eqnarray}
  &~&\sum_{\rho\lambda=1,2}i\Pi^{(\rho\lambda)\mu\nu}_{T\mathrm{bac},2}(k)(2\pi)^4\delta^{(4)}(0)\!\nonumber\\
  &~&~~~~~~~~~~~~=\!-2ie^2\int\frac{d^4q}{(2\pi)^4}\{\bar{\psi}_0(q)\{[(q-k)^2-m^2+i\epsilon][(q+k)^2-m^2+i\epsilon]\}^{-1}\{\gamma^\mu[(q^2+k^2-m^2)q^\nu-2q\cdot kk^\nu]\nonumber\\
  &~&~~~~~~~~~~~~~~~+2q\cdot k(\slashed{d}-\slashed{q})g^{\mu\nu}+(q^\mu-k^\mu)[(q+k)^2-m^2]\gamma^\nu+g^{\mu\nu}(m+\slashed{k}-\slashed{q})[(q+k)^2-m^2]\nonumber\\
  &~&~~~~~~~~~~~~~~~+4imq\cdot kS^{\mu\nu}+2iS^{\nu\rho}\gamma^\mu[(q^2+k^2-m^2)k_\rho+2q\cdot kq_\rho]\}\psi_0(q)\nonumber\\
  &~&~~~~~~~~~~~~~~~+\tilde{\psi}^T_0(q)\gamma^0\{[(q-k)^2-m^2+i\epsilon][(q+k)^2-m^2+i\epsilon]\}^{-1}\{\gamma^\mu[(q^2+k^2-m^2)q^\nu-2q\cdot kk^\nu]\nonumber\\
  &~&~~~~~~~~~~~~~~~+2q\cdot k(\slashed{d}-\slashed{q})g^{\mu\nu}+(q^\mu-k^\mu)[(q+k)^2-m^2]\gamma^\nu+g^{\mu\nu}(m+\slashed{k}-\slashed{q})[(q+k)^2-m^2]\nonumber\\
  &~&~~~~~~~~~~~~~~~+4imq\cdot kS^{\mu\nu}+2iS^{\nu\rho}\gamma^\mu[(q^2+k^2-m^2)k_\rho+2q\cdot kq_\rho]\}\tilde{\psi}^*_0(q)\nonumber\\
  &~&~~~~~~~~~~~~~~~-\bar{\psi}_0(q)[2\pi\delta[(q-k)^2-m^2]\gamma^\mu(\slashed{q}-\slashed{k}+m)\gamma^\nu\sin^2\theta_{q^0-k^0}(\beta) \nonumber\\ &~&~~~~~~~~~~~~~~~+2\pi\delta[(q+k)^2-m^2]\gamma^\mu(\slashed{q}+\slashed{k}+m)\gamma^\nu\sin^2\theta_{q^0+k^0}(\beta) ]\psi_0(q)\nonumber\\
  &~&~~~~~~~~~~~~~~~-\tilde{\psi}^T_0(q)\gamma^0[2\pi\delta[(q-k)^2-m^2]\gamma^\mu(\slashed{q}-\slashed{k}+m)\gamma^\nu \cos^2\theta_{q^0-k^0}(\beta)\nonumber\\
  &~&~~~~~~~~~~~~~~~+2\pi\delta[(q+k)^2-m^2]\gamma^\mu(\slashed{q}
  +\slashed{k}+m)\gamma^\nu \cos^2\theta_{q^0+k^0}(\beta) ]\tilde{\psi}^*_0(q)\}.\nonumber\\
\label{(Eq.53a)}
\end{eqnarray}
\end{widetext}

The calculation of this integral is divided into three steps. First, we develop a scheme, which we name as ``classical limit method" in this paper, to evaluate the terms containing the classical fields, i.e., the terms of $\psi_0$ and $\bar{A}_\mu$. The main idea is that the functions of these fields can be associated with some kind of classical macroscopic quantities, while terms of $\psi$ or $\mathcal{A}_\mu$ associate with fluctuations. Second, to simplify the calculation, we assume the system is ideal, static, and neutral. Lastly, the integral is calculated in the non-relativistic and relativistic limit. We will elaborate on each of them as detailed below.

\subsubsection{\label{sec:level2}Classical limit method}

We are going to develop the classical limit method to evaluate the following eight quantities $\bar{\psi}_0(q)\psi_0(q)$, $\bar{\psi}_0(q)\gamma^\mu\psi_0(q)$, $\bar{\psi}_0(q)S^{\mu\nu}\psi_0(q)$, $\bar{\psi}_0(q)S^{\nu\rho}\gamma^\mu\psi_0(q)$, $\tilde{\psi}^T_0(q)\gamma^0\tilde{\psi}^*_0(q)$, $\tilde{\psi}^T_0(q)\gamma^0\gamma^\mu\tilde{\psi}^*_0(q)$, $\tilde{\psi}^T_0(q)\gamma^0 S^{\mu\nu}\tilde{\psi}^*_0(q)$, and $\tilde{\psi}^T_0(q)\gamma^0 S^{\nu\rho}\gamma^\mu\tilde{\psi}^*_0(q)$ in Expression ({\ref{(Eq.53a)}}). For clarity, we temporarily put back the Plank constant $\hbar$ in the formulas ({\ref{(Eq.58b)}})-({\ref{(Eq.59f)}}) as follows
\begin{eqnarray}
   &~&\langle\Omega(\beta)|\bar{\Psi}(x)\Psi(x)|\Omega(\beta)\rangle\nonumber\\
   &=&W_F(\beta,x)\!\!\!\!\lim_{\substack{x'^{0}\rightarrow x^0+0^+\\
                            \mathbf{x}'\rightarrow \mathbf{x}}}\!\frac{1}{Z_\hbar}\!\int \! D\bar{\bm{\Psi}}D\bm{\Psi}D\bm{A}e^{\frac{i}{\hbar}\int                 d^4x\hat{\mathcal{L}}[\bar{\bm{\Psi}},\bm{\Psi},\bm{A}]}\nonumber\\
   &~&~~~~~~~\times[\bar{\psi}_0(x')+\bar{\psi}(x')][\psi_0(x)+\psi(x)]W_F^{-1}(\beta,x)\nonumber\\
   &=&\bar{\psi}_0(x)\psi_0(x)+\hbar\langle\Omega(\beta)|\bar{\psi}(x)\psi(x)|\Omega(\beta)\rangle
   \label{(Eq.58b)},
\end{eqnarray}
where
\begin{eqnarray}
   Z_\hbar=\int D\bar{\bm{\Psi}}D\bm{\Psi}D\bm{A} e^{\frac{i}{\hbar}\int d^4x\hat{\mathcal{L}}[\bm{\bar{\Psi}},\bm{\Psi},\bm{A}]}
   \label{(Eq.58c)},
\end{eqnarray}
and
\begin{eqnarray}
  &~&\langle\Omega(\beta)|{\psi}(x)\psi(x)|\Omega(\beta)\rangle\nonumber\\
  &=&W_F(\beta,x)\lim_{\substack{x'^{0}\rightarrow x^0+0^+\\
                            \mathbf{x}'\rightarrow \mathbf{x}}}\frac{1}{Z}\int D\bar{\bm{\Psi}}D\bm{\Psi}D\bm{A}\nonumber\\
                            &~&~~~~~\times e^{i\int d^4x\hat{\mathcal{L}}[\bm{\bar{\Psi}},\bm{\Psi},\bm{A}]}\bar{\psi}(x)\psi(x)W^{-1}_F(\beta,x).
   \label{(Eq.58d)}
\end{eqnarray}
with
\begin{eqnarray}
   Z=\int D\bar{\bm{\Psi}}D\bm{\Psi}D\bm{A} e^{i\int d^4x\hat{\mathcal{L}}[\bm{\bar{\Psi}},\bm{\Psi},\bm{A}]}
   \label{(Eq.58e)}.
\end{eqnarray}
Thus,
\begin{eqnarray}
   \langle\bar{\Psi}(x)\Psi(x)\rangle_T&=&\frac{\mathrm{Tr}(\bar{\Psi}(x)\Psi(x)e^{-\beta (H-\mu N)})}{\mathrm{Tr}(e^{-\beta (H-\mu N)})}\nonumber\\
   &\sim&\bar{\psi}_0(x)\psi_0(x),
   \label{(Eq.59a)}
\end{eqnarray}
where $\langle\rangle_T$ denotes the thermal average, and we use the symbol $\sim$ to mean that both sides are identical in the ``classical limit'', named $\hbar\rightarrow 0$. Then, we can consider $\langle\bar{\psi}(x)\psi(x)\rangle$ as a ``quantum fluctuation'' term. In general, it is easily seen that
\begin{eqnarray}
   \langle\mathcal{O}(\bar{\Psi}(x),\Psi(x))\rangle_T\sim\mathcal{O}(\bar{\psi}_0(x),\psi_0(x)),
   \label{(Eq.59c)}
\end{eqnarray}
where $\mathcal{O}(\bar{\Psi}(x),\Psi(x))$ is an arbitrary functional of $\bar{\Psi}(x)$ and $\Psi(x)$. Performing tilde conjugation on Eq. ({\ref{(Eq.58b)}}), with the help of Eq. ({\ref{(Eq.8b)}}), we have
\begin{eqnarray}
   \langle\Omega(\beta)|\bar{\Psi}(x)\Psi(x)|\Omega(\beta)\rangle\
   &=&\langle\Omega(\beta)|\bar{\tilde{\Psi}}(x)\tilde{\Psi}(x)|\Omega(\beta)\rangle\nonumber\\
   &\sim&\bar{\tilde{\psi}}_0(x)\tilde{\psi}_0(x)
   \label{(Eq.59d)},
\end{eqnarray}
It gives
\begin{eqnarray}
   \bar{\psi}_0(x)\psi_0(x)=[\bar{\psi}_0(x)\psi_0(x)]^\sim
   \label{(Eq.59e)}.
\end{eqnarray}
The following relation is immediately obtained
\begin{eqnarray}
   \mathcal{O}(\bar{\psi}_0(x)\psi_0(x))=\tilde{\mathcal{O}}(\bar{\psi}_0(x),\psi_0(x))
   \label{(Eq.59f)}.
\end{eqnarray}
The above discussion indicates the core idea of the classical limit method: Instead of calculating the background fields, one just need to determine the expectations of the corresponding operators in classical limit. In principle, it is not difficult to calculate these classical expectations since one can always relate them with some characteristic quantities, such as charge distribution, particle density, electric current, etc., which are usually easily obtained. Most of the rest effort is devoted to finding these relations.

First, we discuss the first four of the eight quantities mentioned in the beginning of this subsection.

\paragraph{$\bar{\psi}_0(q)\psi_0(q):$}

Let
\begin{eqnarray}
     \mathcal{O}(\bar{\Psi}(x),\Psi(x))=\bar{\Psi}(x)i\gamma^\mu\partial^\nu\Psi(x),
   \label{(Eq.70d)}
\end{eqnarray}
by using the relation ({\ref{(Eq.59c)}}), we have
\begin{eqnarray}
     \langle\bar{\Psi}(x)i\gamma^\mu\partial^\nu\Psi(x)\rangle_T\sim\bar{\psi}_0(x)\psi_0(x).
   \label{(Eq.70c)}
\end{eqnarray}
From the energy-momentum tensor $T^{\mu\nu}$ corresponding to the Lagrangian density $\mathcal{L}[\bar{\Psi}_0,\Psi_0,A]$, and the symmetric energy-momentum tensor $T_{\mathrm{EM}}^{\mu\nu}=F^{\mu\rho}F^\nu_{~\rho}\!-\frac{1}{4}(F_{\rho\sigma})^2g^{\mu\nu}$ for free electromagnetic field, we have
\begin{eqnarray}
     \langle\bar{\Psi}(x)i\gamma^\mu\partial^\nu\Psi(x)\rangle_T=T^{\mu\nu}(x)-T_{\mathrm{EM}}^{\mu\nu}(x).
   \label{(Eq.70b)}
\end{eqnarray}
With the help of Dirac equation, Eq. ({\ref{(Eq.70b)}}), and $\bar{A}_\mu=0$, we obtain
\begin{eqnarray}
     \!\!\!\!\!\!\!\!\bar{\psi}_0(x)\psi_0(x)&=&\frac{1}{m}\{\bar{T}^0_0(x)-(\bar{T}_{EM})^0_0(x)\nonumber\\
     &~&~~~~~~~~~~-[\bar{T}^i_i(x)-(\bar{T}_{EM})^i_i(x)]\}\nonumber\\
     &=&\frac{1}{m}\{\bar{E}(x)-\bar{E}_{EM}(x)\nonumber\\
     &~&~~~~~~~~~~-\sum_i[\bar{P}^i(x)-\bar{P}^i_{EM}(x)]\},
   \label{(Eq.71b)}
\end{eqnarray}
where $T^{\mu\nu}\sim \bar{T}^{\mu\nu}$, $T_{EM}^{\mu\nu}\sim \bar{T}_{EM}^{\mu\nu}$, $\bar{E}$ and $\bar{E}_{EM}$ are the energy of the whole system and electromagnetic field at $x$ in the classical limit, respectively, and $\bar{P}^i$ and $\bar{P}^i_{EM}$ are the corresponding momentums. Then we rewrite Eq. ({\ref{(Eq.71b)}}) in momentum space as follows
\begin{eqnarray}
     \!\!\!\int\!\frac{d^4p}{(4\pi)^4}\bar{\psi}_0(p)\psi_0(p)=\frac{1}{m}\int\! dt[\bar{E}_D(t)-\sum_i \bar{P}_D^i(t)].
   \label{(Eq.74a)}
\end{eqnarray}
Here,
\begin{eqnarray}
  \bar{E}_D(t)=\bar{E}(t)-\bar{E}_{EM}(t)
\label{(Eq.74b)}
\end{eqnarray}
and
\begin{eqnarray}
  \bar{P}_D^i(t)=\bar{P}^i(t)-\bar{P}^i_{EM}(t)
\label{(Eq.74c)}
\end{eqnarray}
where $\bar{E}_{EM}(t)$ and $\bar{P}_{EM}^i(t)$ are the energy and momentum of free propagating electromagnetic field at time $t$, respectively. In the center of mass frame of the charged particle system, $P_D^i(t)=0$. It gives
\begin{eqnarray}
     \int\!\frac{d^4p}{(4\pi)^4}\bar{\psi}_0(p)\psi_0(p)=\frac{1}{m}\int\! dtE_D(t).
   \label{(Eq.75a)}
\end{eqnarray}

\paragraph{$\bar{\psi}_0(q)\gamma^\mu\psi_0(q):$}

Let
\begin{eqnarray}
   \mathcal{O}(\bar{\bm{\Psi}}^{(1)}(x),\bm{\Psi}^{(1)}(x))=\bar{\Psi}(x)\gamma^\mu\Psi(x)
   \label{(Eq.70e)}
\end{eqnarray}
by using the relation ({\ref{(Eq.59c)}}), we have
\begin{eqnarray}
   \langle\bar{\Psi}(x)\gamma^\mu\Psi(x)\rangle_T\sim\bar{\psi}_0(x)\gamma^\mu\psi_0(x).
   \label{(Eq.59b)}
\end{eqnarray}
It gives
\begin{eqnarray}
     &~&\int d^4x\langle\Omega(\beta)|\bar{\Psi}(x)\gamma^\mu\Psi(x)|\Omega(\beta)\rangle\nonumber\\
     &=&\int \frac{d^4q}{(2\pi)^4}\langle\Omega(\beta)|\bar{\Psi}(q)\gamma^\mu\Psi(q)|\Omega(\beta)\rangle\nonumber\\
     &\sim & \int \frac{d^4q}{(2\pi)^4}\bar{\psi}_0(q)\gamma^\mu\psi_0(q)\nonumber\\
     &=&\int dt\bar{j}^\mu(t,\textbf{q}) ,
   \label{(Eq.76a)}
\end{eqnarray}

\paragraph{$\bar{\psi}_0(q)S^{\mu\nu}\psi_0(q):$}

The calculation will be included in Eq.(\ref{(Eq.103a)}). We just mention here that the quantity relates to the spin of the system if $\mu=i$ and $\nu=j$. It is easily seen by setting
\begin{eqnarray}
   \mathcal{O}(\bar{\bm{\Psi}}^{(1)}(x),\bm{\Psi}^{(1)}(x))=\bar{\Psi}(x)S^{\mu\nu}\Psi(x).
   \label{(Eq.60a)}
\end{eqnarray}
Then, we have
\begin{eqnarray}
      &~&\langle\bar{\Psi}(q)S^{ij}\Psi(q)\rangle_T=\frac{1}{2}\langle\bar{\Psi}(q)\varepsilon^{ijk}\Sigma^k\Psi(q)\rangle_T\nonumber\\
      &\sim&\bar{\psi}_0(q)S^{ij}\psi_0(q)=\frac{1}{2}\bar{\psi}_0(q)\varepsilon^{ijk}\Sigma^k\psi_0(q),
   \label{(Eq.90b)}
\end{eqnarray}
where
\begin{eqnarray}
      \Sigma^k=
    \left(                 %左括号
      \begin{array}{cc}   %该矩阵一共2列，每一列都居中放置
        \sigma^k & 0 \\  %第一行元素
        0 & \sigma^k \\  %第二行元素
      \end{array}
   \right)
   \label{(Eq.90c)}
\end{eqnarray}
with Pauli's matrix $\sigma^k$.

\paragraph{$\bar{\psi}_0(q)S^{\nu\rho}\gamma^\mu\psi_0(q):$}

Let
\begin{eqnarray}
   \mathcal{O}(\bar{\bm{\Psi}}^{(1)}(x),\bm{\Psi}^{(1)}(x))=\bar{\Psi}(x)S^{\nu\rho}\gamma^\mu\Psi(x),
   \label{(Eq.70f)}
\end{eqnarray}
with the help of the relation ({\ref{(Eq.59c)}}), we have
\begin{eqnarray}
   \langle\bar{\Psi}(x)S^{\nu\rho}\gamma^\mu\Psi(x)\rangle_T\sim\bar{\psi}_0(x)S^{\nu\rho}\gamma^\mu\psi_0(x)
   \label{(Eq.61a)}.
\end{eqnarray}
This quantity will be discussed in eight cases as follows.

(1) For the case of $\mu=0,\nu=0,\rho=0$,
\begin{eqnarray}
     &~&\int d^4x\bar{\psi}_0(x)S^{\nu\rho}\gamma^\mu\psi_0(x)\nonumber\\
     &=&\int\frac{d^4q}{(2\pi)^4}\bar{\psi}_0(q)S^{00}\gamma^0\psi_0(q)\nonumber\\
     &=&0.
   \label{(Eq.77a)}
\end{eqnarray}

(2) For the case of $\mu=0,\nu=0,\rho=i$,
\begin{eqnarray}
     &~&\int d^4x\langle\Omega(\beta)|\bar{\Psi}(x)S^{\nu\rho}\gamma^\mu\Psi(x)|\Omega(\beta)\rangle \nonumber\\
     &\sim &\int \frac{d^4q}{(2\pi)^4}\bar{\psi}_0(q)S^{0i}\gamma^0\psi_0(q)\nonumber\\
     &=&\int dt \int \frac{d^3q}{(2\pi)^3}\bar{\psi}_0(t,\textbf{q})S^{0i}\gamma^0\psi_0(t,\textbf{q})\nonumber\\
     &=&-\frac{i}{2}\int dt \int \frac{d^3q}{(2\pi)^3}\bar{j}^i(t,\textbf{q}).
   \label{(Eq.78a)}
\end{eqnarray}

(3) For the case of $\mu=0,\nu=j,\rho=0$,
\begin{eqnarray}
      &~&\int d^4x\langle\Omega(\beta)|\bar{\Psi}(x)S^{\nu\rho}\gamma^\mu\Psi(x)|\Omega(\beta)\rangle \nonumber\\
     &\sim &\int \frac{d^4q}{(2\pi)^4}\bar{\psi}_0(q)S^{j0}\gamma^0\psi_0(q)\nonumber\\
     &=&\int dt \int \frac{d^3q}{(2\pi)^3}\bar{\psi}_0(t,\textbf{q})S^{j0}\gamma^0\psi_0(t,\textbf{q})\nonumber\\
     &=&\frac{i}{2}\int dt \int \frac{d^3q}{(2\pi)^3}\bar{j}^i(t,\textbf{q}).
   \label{(Eq.79a)}
\end{eqnarray}

(4) For the case of $\mu=0,\nu=j,\rho=i$,
\begin{eqnarray}
      &~&\int d^4x\langle\Omega(\beta)|\bar{\Psi}(q)S^{\nu\rho}\gamma^\mu\Psi(x)|\Omega(\beta)\rangle \nonumber\\
     &\sim &\int \frac{d^4q}{(2\pi)^4}\bar{\psi}_0(q)S^{ji}\gamma^0\psi_0(q)\nonumber\\
     &=&\int dt \int \frac{d^3q}{(2\pi)^3}\bar{\psi}_0(t,\textbf{q})S^{ji}\gamma^0\psi_0(t,\textbf{q})\nonumber\\
     &=&\frac{1}{2}\int dt \int \frac{d^3q}{(2\pi)^3}\varepsilon^{ijk}[\bar{j}_{R}^k(t,\textbf{q})-\bar{j}_{L}^k(t,\textbf{q})].
   \label{(Eq.80a)}
\end{eqnarray}

(5) For the case of $\mu=l,\nu=0,\rho=0$,
\begin{eqnarray}
      &~&\int d^4x\bar{\psi}_0(x)S^{\nu\rho}\gamma^\mu\psi_0(x)\nonumber\\
     &=&\int \frac{d^4q}{(2\pi)^4}\bar{\psi}_0(q)S^{00}\gamma^l\psi_0(q)\nonumber\\
     &=&0.
   \label{(Eq.81a)}
\end{eqnarray}

(6) For the case of $\mu=l,\nu=0,\rho=i$,
\begin{eqnarray}
      &~&\int d^4x\langle\Omega(\beta)|\bar{\Psi}(x)S^{\nu\rho}\gamma^\mu\Psi(x)|\Omega(\beta)\rangle \nonumber\\
     &\sim &\int \frac{d^4q}{(2\pi)^4}\bar{\psi}_0(q)S^{0i}\gamma^l\psi_0(q)\nonumber\\
     &=&\int dt \int \frac{d^3q}{(2\pi)^3}\bar{\psi}_0(t,\textbf{q})S^{0i}\gamma^l\psi_0(t,\textbf{q})\nonumber\\
     &=&-\frac{i}{2}\int dt \int \frac{d^3q}{(2\pi)^3}\{\delta^{il}j_0^0(t,\textbf{q})\nonumber\\
     &~&~~~~~~~~+\varepsilon^{ilt}[\bar{j}_{R}^t(t,\textbf{q})-\bar{j}_{L}^t(t,\textbf{q})]\}.
   \label{(Eq.82a)}
\end{eqnarray}

(7) For the case of $\mu=l,\nu=j,\rho=0$,
\begin{eqnarray}
      &~&\int d^4x\langle\Omega(\beta)|\bar{\Psi}(x)S^{\nu\rho}\gamma^\mu\Psi(x)|\Omega(\beta)\rangle \nonumber\\
     &\sim &\int \frac{d^4q}{(2\pi)^4}\bar{\psi}_0(q)S^{j0}\gamma^l\psi_0(q)\nonumber\\
     &=&\int dt \int \frac{d^3q}{(2\pi)^3}\bar{\psi}_0(t,\textbf{q})S^{j0}\gamma^l\psi_0(t,\textbf{q})\nonumber\\
     &=&\frac{i}{2}\int dt \int \frac{d^3q}{(2\pi)^3}\{\delta^{jl}\bar{j}^0(t,\textbf{q})\nonumber\\
     &~&~~~~~~~~+\varepsilon^{jlt}[\bar{j}_{R}^t(t,\textbf{q})-\bar{j}_{L}^t(t,\textbf{q})]\}.
   \label{(Eq.83a)}
\end{eqnarray}

(8) For the case of $\mu=l,\nu=j,\rho=i$,
\begin{eqnarray}
      &~&\int d^4x\langle\Omega(\beta)|\bar{\Psi}(x)S^{\nu\rho}\gamma^\mu\Psi(x)|\Omega(\beta)\rangle \nonumber\\
     &\sim &\int \frac{d^4q}{(2\pi)^4}\bar{\psi}_0(q)S^{ji}\gamma^l\psi_0(q)\nonumber\\
     &=&\int dt \int \frac{d^3q}{(2\pi)^3}\bar{\psi}_0(t,\textbf{q})S^{ji}\gamma^l\psi_0(t,\textbf{q})\nonumber\\
     &=&\frac{1}{2}\int dt \int \frac{d^3q}{(2\pi)^3}\{\varepsilon^{jil}[\bar{N}_{R}(t,\textbf{q})-\bar{N}_{L}(t,\textbf{q})]\nonumber\\
     &~&~~~~~~~~+i\delta^{jl}[\bar{j}_{R}^i(t,\textbf{q})-\bar{j}_{L}^i(t,\textbf{q})]\nonumber\\
     &~&~~~~~~~~-i\delta^{il}[\bar{j}_{R}^j(t,\textbf{q})-\bar{j}_{L}^j(t,\textbf{q})]\}.
   \label{(Eq.84a)}
\end{eqnarray}
In the above eight expressions, $\bar{\textbf{j}}_{L}(t,\textbf{q})$ and $\bar{\textbf{j}}_{R}(t,\textbf{q})$ are the average left-handed and right-handed currents in momentum space at time $t$, respectively.
\begin{eqnarray}
      \bar{\textbf{j}}_{L}(t,\textbf{q})=\int d^3xe^{i\textbf{q}\cdot \textbf{x}}\bar{\psi}_0(x)\boldsymbol{\gamma}\frac{1-\gamma^5}{2}\psi_0(x),
   \label{(Eq.85a)}
\end{eqnarray}
\begin{eqnarray}
      \bar{\textbf{j}}_{R}(t,\textbf{q})=\int d^3xe^{i\textbf{q}\cdot \textbf{x}}\bar{\psi}_0(x)\boldsymbol{\gamma}\frac{1+\gamma^5}{2}\psi_0(x)
   \label{(Eq.86a)},
\end{eqnarray}
where $\gamma^5=i\gamma^0\gamma^1\gamma^2\gamma^3$. $\bar{N}_{L}(t,\textbf{q})$ and $\bar{N}_{R}(t,\textbf{q})$ are the average number densities of left handed particles and right handed particles in momentum space at time $t$,
\begin{eqnarray}
      \bar{N}_{L}(t,\textbf{q})=\int d^3xe^{i\textbf{q}\cdot \textbf{x}}\psi_{0L}^\dag(x)\psi_{0L}(x)
   \label{(Eq.87a)},
\end{eqnarray}
\begin{eqnarray}
      \bar{N}_{R}(t,\textbf{q})=\int d^3xe^{i\textbf{q}\cdot \textbf{x}}\psi_{0R}^\dag(x)\psi_{0R}(x),
   \label{(Eq.88a)}
\end{eqnarray}
where
\begin{eqnarray}
      \psi_{0L}(x)=\frac{1-\gamma^5}{2}\psi_0(x)
   \label{(Eq.89a)},
\end{eqnarray}
\begin{eqnarray}
      \psi_{0R}(x)=\frac{1+\gamma^5}{2}\psi_0(x)),
   \label{(Eq.90a)}
\end{eqnarray}

The above discussion give the relation between the aforementioned four quantities and the characteristic quantities of the system. However, the rest ones $\tilde{\psi}^T_0(q)\gamma^0\tilde{\psi}^*_0(q)$, $\tilde{\psi}^T_0(q)\gamma^0\gamma^\mu\tilde{\psi}^*_0(q)$, $\tilde{\psi}^T_0(q)\gamma^0 S^{\mu\nu}\tilde{\psi}^*_0(q)$, and $\tilde{\psi}^T_0(q)\gamma^0 S^{\nu\rho}\gamma^\mu\tilde{\psi}^*_0(q)$ can not be directly determined by performing the tilde-conjugation operation on the first four since it will bring some undesired minus and complex conjugation signs. To overcome this difficult, we expand the classical fields $\bm{\psi}_0(x)$ and $\bar{\bm{\psi}}_0(x)$ as follows
\begin{eqnarray}
   &~&\bm{\psi}_0(x)= \left(
      \begin{array}{c}
        \bm{\psi}_0^{(1)}(x)\\
        \bm{\psi}_0^{(2)}(x)
      \end{array}
    \right)\;=\left(                 % 左括号
      \begin{array}{c}   %该矩阵一共2列，每一列都居中放置
        \psi_0(x) \\  %第一行元素
        i\tilde{\psi}_0^*(x) \\  % 第二行元素
      \end{array}
   \right)\nonumber\\
   &=&\int \frac{d^3p}{(2\pi)^3}\frac{1}{\sqrt{2E_p}}\sum_{s=1,2}[\left(                 % 左括号
      \begin{array}{c}   %该矩阵一共2列，每一列都居中放置
        \bar{c}^s(t,\textbf{p}) \\  %第一行元素
        i\tilde{\bar{c}}^s(t,\textbf{p})^* \\  % 第二行元素
      \end{array}
   \right)u^s(\textbf{p})e^{-ip\cdot x}\nonumber\\
            &~&~~~~~~~~~~~~+\left(                 %左括号
      \begin{array}{c}   %该矩阵一共2列，每一列都居中放置
        \bar{d}^s(t,\textbf{p})^* \\  %第一行元素
        i\tilde{\bar{d}}^s(t,\textbf{p}) \\  %第二行元素
      \end{array}
   \right)v^s(\textbf{p})e^{ip\cdot x}],
   \label{(Eq.62a)}
\end{eqnarray}
\begin{eqnarray}
  &~&\bar{\bm{\psi}}_0(x)=\left(
      \begin{array}{c}
        \bar{\bm{\psi}}_0^{(1)}(x)\\
        \bar{\bm{\psi}}_0^{(2)}(x)
      \end{array}
    \right)^T\;=\left(                 % 左括号
      \begin{array}{c}   %该矩阵一共2列，每一列都居中放置
        \bar{\psi}_0(x) \\  %第一行元素
        -i\tilde{\psi}_0^T(x)\gamma^0 \\  % 第二行元素
      \end{array}
   \right)^T\nonumber\\
   &=&\int \frac{d^3p}{(2\pi)^3}\frac{1}{\sqrt{2E_p}}\sum_{s=1,2}[\left(                 %左括号
      \begin{array}{c}   %该矩阵一共2列，每一列都居中放置
        \bar{c}^s(t,\textbf{p})^* \\  %第一行元素
        -i\tilde{\bar{c}}^s(t,\textbf{p}) \\  % 第二行元素
      \end{array}
   \right)^\mathrm{T}\bar{u}^s(\textbf{p})e^{-ip\cdot x}\nonumber\\
            &~&~~~~~~~~~~~~+\left(                 %左括号
      \begin{array}{c}   %该矩阵一共2列，每一列都居中放置
        \bar{d}^s(t,\textbf{p}) \\  %第一行元素
        -i\tilde{\bar{d}}^s(t,\textbf{p})^* \\  %第二行元素
      \end{array}
   \right)^\mathrm{T}\bar{v}^s(\textbf{p})e^{ip\cdot x}],
   \label{(Eq.63a)}
\end{eqnarray}
The coefficients, i.e., $\bar{c}^s(t,\textbf{p})$, $\bar{d}^s(t,\textbf{p})$, $\tilde{\bar{c}}^s(t,\textbf{p})$, and $\tilde{\bar{d}}^s(t,\textbf{p})$, should be defined such that the path integral quantization agrees with the canonical one. It suggests the following two requirements:

(\rmnum{1}) The Fourier's expansion of $\bm{\psi}^{(1)}(x)$ coincides with that of Dirac field in the usual QFT.

(\rmnum{2}) The operation rules of the tilde symbol, describing by Eq. ({\ref{(Eq.3a)}}) to Eq. ({\ref{(Eq.7a)}}), can also be extended to apply to these coefficients.

\noindent Then, we define $\bar{c}^s(t,\textbf{p})$ and $\bar{d}^s(t,\textbf{p})$ as Grassmann numbers satisfying the anticommutative relations
\begin{eqnarray}
     \begin{split}
       \{\bar{c}^s(t,\textbf{p}),\bar{c}^r(t',\textbf{p}')\}=0,~\{\bar{c}^s(t,\textbf{p}),\bar{c}^r(t',\textbf{p}')^*\}=0,\\% 此处需要换行，两个公式上下排列
       \{\bar{d}^s(t,\textbf{p}),\bar{d}^r(t',\textbf{p}')\}=0,~\{\bar{d}^s(t,\textbf{p}),\bar{d}^r(t',\textbf{p}')^*\}=0,\\
       \{\bar{c}^s(t,\textbf{p}),\bar{d}^r(t',\textbf{p}')\}=0,~\{\bar{c}^s(t,\textbf{p}),\bar{d}^r(t',\textbf{p}')^*\}=0.
     \end{split}
   \label{(Eq.64a)}
\end{eqnarray}
$\tilde{\bar{c}}^s(t,\textbf{p})$ and $\tilde{\bar{d}}^s(t,\textbf{p})$ are also defined as Grassmann numbers, and they satisfy the same anticommutative relations
\begin{eqnarray}
     \begin{split}
       \{\tilde{\bar{c}}^s(t,\textbf{p}),\tilde{\bar{c}}^r(t',\textbf{p}')\}=0,~\{\tilde{\bar{c}}^s(t,\textbf{p}),\tilde{\bar{c}}^r(t',\textbf{p}')^*\}=0,\\% 此处需要换行，两个公式上下排列
       \{\tilde{\bar{d}}^s(t,\textbf{p}),\tilde{\bar{d}}^r(t',\textbf{p}')\}=0,~\{\tilde{\bar{d}}^s(t,\textbf{p}),\tilde{\bar{d}}^r(t',\textbf{p}')^*\}=0,\\
       \{\tilde{\bar{c}}^s(t,\textbf{p}),\tilde{\bar{d}}^r(t',\textbf{p}')\}=0,~\{\tilde{\bar{c}}^s(t,\textbf{p}),\tilde{\bar{d}}^r(t',\textbf{p}')^*\}=0.
     \end{split}
   \label{(Eq.64h)}
\end{eqnarray}
Furthermore, we define
\begin{eqnarray}
    \{\bar{c}^s(t,\textbf{p}),\tilde{\bar{c}}^r(t,\textbf{p})\}=0,
   \label{(Eq.64c)}
\end{eqnarray}
\begin{eqnarray}
    \{\bar{d}^s(t,\textbf{p}),\tilde{\bar{d}}^r(t,\textbf{p})\}=0,
   \label{(Eq.64b)}
\end{eqnarray}
\begin{eqnarray}
    \{\bar{c}^s(t,\textbf{p}),\tilde{\bar{d}}^r(t,\textbf{p})\}=0.
   \label{(Eq.64d)}
\end{eqnarray}
To check the rationalities of these definitions, one can use Eq. ({\ref{(Eq.3a)}}) and ({\ref{(Eq.5a)}}) to obtain the Expression ({\ref{(Eq.62a)}}). Besides, Eq. ({\ref{(Eq.64h)}}) can be derived from Eq. ({\ref{(Eq.4a)}}). In addition, it is easily seen from Eq. ({\ref{(Eq.64a)}}) to Eq. ({\ref{(Eq.64d)}}) that the coefficients satisfy Eq. ({\ref{(Eq.6a)}}) and Eq. ({\ref{(Eq.7a)}}).

Considering
\begin{eqnarray}
    \left(\!\!
      \begin{array}{c}
        c_{\bm{p}}^s\\
        \bar{c}^s(t,\bm{p})
      \end{array}\!\!
    \right)\;=\int d^3x\frac{1}{\sqrt{2E_{\bm{p}}}}u^s(\bm{p})^\dag e^{ip\cdot x}
     \left(\!\!
      \begin{array}{c}
        \Psi(x)\\
        \psi_0(x)
      \end{array}\!\!
    \right),
\label{(Eq.64e)}
\end{eqnarray}
\begin{eqnarray}
    &~&\!\!\!\!\!\!\!\!\!\!\!\!\!\!\!\!\!\!\!\!
    \left(\!\!
      \begin{array}{c}
        d_{\bm{p}}^{s}\\
        \bar{d}^s(t,\bm{p})
      \end{array}\!\!
    \right)\;\!=\!\int d^3x
     \left(\!\!
      \begin{array}{c}
        \bar{\Psi}(x)\\
        \bar{\psi}_0(x)
      \end{array}\!\!
    \right)\!
    \frac{1}{\sqrt{2E_{\bm{p}}}}\gamma^0v^s(\bm{p}) e^{-ip\cdot x},
\label{(Eq.64j)}
\end{eqnarray}
and with the help of Eq. ({\ref{(Eq.64a)}}), we have
\begin{eqnarray}
      &~&N^{(+)}(t,\textbf{p})=\sum_{s=1,2}\langle\Omega(\beta)|c_{\bm{p}}^{s\dag}c_{\bm{p}}^s|\Omega(\beta)\rangle\nonumber\\
      &\sim&\sum_{s=1,2}\bar{c}^s(t,\textbf{p})^*\bar{c}^s(t,\textbf{p})=\bar{N}^{(+)}(t,\textbf{p}),
    \label{(Eq.70a)}
\end{eqnarray}
\begin{eqnarray}
      &~&N^{(-)}(t,\textbf{p})=\sum_{s=1,2}\langle\Omega(\beta)|d_{\bm{p}}^{s\dag}d_{\bm{p}}^s|\Omega(\beta)\rangle\nonumber\\ &\sim&\sum_{s=1,2}\bar{d}^s(t,\textbf{p})^*\bar{d}^s(t,\textbf{p})=\bar{N}^{(-)}(t,\textbf{p})
    \label{(Eq.71a)},
\end{eqnarray}
where $N^{(+)}(t,\textbf{p})$ and $N^{(-)}(t,\textbf{p})$ are the particle number density at given momentum $\textbf{p}$ and time $t$ for $e_\psi^-$ and $e_\psi^+$, and $\bar{N}^{(+)}(t,\textbf{p})$ and $\bar{N}^{(-)}(t,\textbf{p})$ are the corresponding ones in classical limit. From subsection C in Sec. \Rmnum{3}, they are equal to the number density of electrons and positrons, respectively.

Similarly, from the Fourier's transformation
\begin{eqnarray}
    &~&\!\!\!\!\!\!\!\!\!\!\!\!\!\!\!\!\!\!\!\!\!\!\!\!
    \left(\!\!
      \begin{array}{c}
        \tilde{c}_{\bm{p}}^{s}\\
        \tilde{\bar{c}}^s(t,\bm{p})
      \end{array}\!\!
    \right)\;\!\!=\!\int d^3x
     \left(\!\!
      \begin{array}{c}
       \bar{\tilde{\Psi}}(x)\\
        \tilde{\psi}^T_0(x)\gamma^0
      \end{array}\!\!
    \right)\!
    \frac{1}{\sqrt{2E_{\bm{p}}}}\gamma^0u^s(\bm{p}) e^{ip\cdot x},
\label{(Eq.64e)}
\end{eqnarray}
\begin{eqnarray}
    &~&\!\!\!\!\!\!\!\!\!\!\!\!\!\!\!\!\!\!\!\!\!\!\!\!
    \left(\!\!
      \begin{array}{c}
        \tilde{d}_{\bm{p}}^{s}\\
        \tilde{\bar{d}}^s(t,\bm{p})
      \end{array}\!\!
    \right)\;=\int d^3x\frac{1}{\sqrt{2E_{\bm{p}}}}u^s(\bm{p})^\dag e^{ip\cdot x}
     \left(\!\!
      \begin{array}{c}
        (\tilde{\Psi}^\dag(x))^T\\
        \tilde{\psi}^*_0(x)
      \end{array}\!\!
    \right),
  \label{(Eq.64e)}
\end{eqnarray}
and with the help of Eq. ({\ref{(Eq.59f)}}), we obtain
\begin{eqnarray}
      &~&N^{(+)}(t,\textbf{p})=\sum_{s=1,2}\langle\Omega(\beta)|\tilde{c}_{\bm{p}}^{s}\tilde{c}_{\bm{p}}^{s\dag}|\Omega(\beta)\rangle\nonumber\\
      &\sim&\sum_{s=1,2}\tilde{\bar{c}}^s(t,\textbf{p})\tilde{\bar{c}}^s(t,\textbf{p})^*=\bar{N}^{(+)}(t,\textbf{p}),
    \label{(Eq.72a)}
\end{eqnarray}
\begin{eqnarray}
      &~&N^{(-)}(t,\textbf{p})=\sum_{s=1,2}\langle\Omega(\beta)|\tilde{d}_{\bm{p}}^{s}\tilde{d}_{\bm{p}}^{s\dag}|\Omega(\beta)\rangle\nonumber\\ &\sim&\sum_{s=1,2}\tilde{\bar{d}}^s(t,\textbf{p})\tilde{\bar{d}}^s(t,\textbf{p})^*=\bar{N}^{(-)}(t,\textbf{p})
    \label{(Eq.73a)},
\end{eqnarray}
Inserting Expression ({\ref{(Eq.62a)}}), ({\ref{(Eq.63a)}}), ({\ref{(Eq.72a)}}) and ({\ref{(Eq.73a)}}) into the quantities $\tilde{\psi}^T_0(q)\gamma^0\tilde{\psi}^*_0(q)$, $\tilde{\psi}^T_0(q)\gamma^0\gamma^\mu\tilde{\psi}^*_0(q)$, $\tilde{\psi}^T_0(q)\gamma^0 S^{\mu\nu}\tilde{\psi}^*_0(q)$, and $\tilde{\psi}^T_0(q)\gamma^0 S^{\nu\rho}\gamma^\mu\tilde{\psi}^*_0(q)$, we obtain
\begin{eqnarray}
      \bar{\psi}_0(q)\psi_0(q)=\tilde{\psi}^T_0(q)\gamma^0\tilde{\psi}^*_0(q)
    \label{(Eq.73b)},
\end{eqnarray}
\begin{eqnarray}
      \bar{\psi}_0(q)\gamma^\mu\psi_0(q)=\tilde{\psi}^T_0(q)\gamma^0\gamma^\mu\tilde{\psi}^*_0(q)
    \label{(Eq.73c)},
\end{eqnarray}
\begin{eqnarray}
      \bar{\psi}_0(q)S^{ij}\psi_0(q)=\tilde{\psi}^T_0(q)S^{ij}\tilde{\psi}^*_0(q)
    \label{(Eq.73d)},
\end{eqnarray}
\begin{eqnarray}
      \bar{\psi}^T_0(q)S^{\nu\rho}\gamma^\mu\psi_0(q)=\tilde{\psi}^T_0(q)\gamma^0 S^{\nu\rho}\gamma^\mu\tilde{\psi}^*_0(q)
    \label{(Eq.73e)},
\end{eqnarray}

We make the following remarks on these results. First, although no classical electromagnetic field appears in the expression (\ref{(Eq.53a)}), it can still affect $\Pi_{T\mathrm{bac},2}(k)$. For instance, the classical electric field drives the motion of the electrons and positrons to form a current, thus affecting $\Pi_{T\mathrm{bac},2}(k)$ through the terms $\bar{\psi}_0(q)\gamma^\mu\psi_0(q)$ and $\tilde{\psi}^T_0(q)\gamma^0\gamma^\mu\tilde{\psi}^*_0(q)$, as shown in Eq. ({\ref{(Eq.76a)}}). Second, as will be seen in Expression (\ref{(Eq.103a)}), the terms involving $S^{\mu\nu}$ vanished. It indicates that the system is not affected by spin, to the approximation of $e^2$ order. Last, in this approximation, the following five quantities directly determine the background polarization tensor: total energy of Dirac particles, distributions of left and right handed particle in momentum space, and left- and right-handed current densities.

Note that Expression ({\ref{(Eq.53a)}}) is an integral on momentum, we give the following formulas in momentum space for later use. The Fourier transform of $\bm{\psi}_0(x)$ and $\bar{\bm{\psi}}_0(x)$ are
\begin{eqnarray}
     \bm{\psi}_0(q)&=&\int d^4xe^{iq\cdot x}\bm{\psi}_0(x)\nonumber\\
              &=&\frac{1}{\sqrt{2E_{\bm{q}}}}\sum_{s=1,2}[\left(                 %左括号
      \begin{array}{c}   %该矩阵一共2列，每一列都居中放置
        \bar{\mathrm{c}}^s(q^0-E_{\bm{q}},\textbf{q})\\  % 第一行元素
        i\tilde{\bar{\mathrm{c}}}^s(q^0-E_{\bm{q}},\textbf{p})^* \\  %第二行元素
      \end{array}
   \right)u^s(\textbf{q})\nonumber\\
              &~&~~~+\left(                 %左括号
      \begin{array}{c}   %该矩阵一共2列，每一列都居中放置
        \bar{\mathrm{d}}^s(q^0+E_{\bm{q}},-\textbf{q})^*\\  %第一行元素
        i\tilde{\bar{\mathrm{d}}}^s(q^0+E_{\bm{q}},-\textbf{q}) \\  %第二行元素
      \end{array}
   \right)v^s(-\textbf{q})],
   \label{(Eq.67a)}
\end{eqnarray}
\begin{eqnarray}
     \bar{\bm{\psi}}_0(q)&=&\int d^4xe^{iq\cdot x}\bar{\bm{\psi}}_0(x)\nonumber\\
              &=&\frac{1}{\sqrt{2E_{\bm{q}}}}\sum_{s=1,2}[\left(                 %左括号
      \begin{array}{c}   %该矩阵一共2列，每一列都居中放置
        \bar{\mathrm{c}}^s(q^0-E_{\bm{q}},\textbf{q})^*\\  %第一行元素
        i\tilde{\bar{\mathrm{c}}}^s(q^0-E_{\bm{q}},\textbf{q}) \\  %第二行元素
      \end{array}
   \right)\bar{u}^s(\textbf{q})\nonumber\\
              &~&~~~+\left(                 %左括号
      \begin{array}{c}   %该矩阵一共2列，每一列都居中放置
        \bar{\mathrm{d}}^s(q^0+E_{\bm{q}},-\textbf{q})\\  %第一行元素
        i\tilde{\bar{\mathrm{d}}}^s(q^0+E_{\bm{q}},-\textbf{p})^* \\  %第二行元素
      \end{array}
   \right)\bar{v}^s(-\textbf{q})]
   \label{(Eq.67b)},
\end{eqnarray}
where the Fourier transform of the coefficients are denoted by roman $\bar{\mathrm{c}}^s$ and $\bar{\mathrm{d}}^s$, and determined as follows
\begin{eqnarray}
     \bar{\mathrm{c}}^s(q^0-E_{\bm{q}},\textbf{q})=\int dte^{i(q^0-E_{\bm{q}})t}\bar{c}^s(t,\textbf{q}),
   \label{(Eq.68a)}
\end{eqnarray}
\begin{eqnarray}
     \tilde{\bar{\mathrm{c}}}^s(q^0-E_{\bm{q}},\textbf{q})^*=\int dte^{i(q^0-E_{\bm{q}})t}\tilde{\bar{c}}^s(t,\textbf{q})^*,
   \label{(Eq.68b)}
\end{eqnarray}
\begin{eqnarray}
     \bar{\mathrm{d}}^s(q^0+E_{\bm{q}},-\textbf{q})=\int dte^{i(q^0-E_{\bm{q}})t}d^s(t,-\textbf{q}),
   \label{(Eq.69a)}
\end{eqnarray}
\begin{eqnarray}
     \tilde{\bar{\mathrm{d}}}^s(q^0+E_{\bm{q}},-\textbf{q})^*=\int dte^{i(q^0-E_{\bm{q}})t}\tilde{\bar{d}}^s(t,-\textbf{q})^*.
   \label{(Eq.69b)}
\end{eqnarray}

\subsubsection{\label{sec:level2}Useful conclusions and assumptions}

We make the following two conclusions and four assumptions useful for deriving $\Pi_{T\mathrm{bac}}^{\mu\nu}(k)$ in this subsection.

Conclusion (\rmnum{1}): There are an equal number of left- and right-handed $e_\psi$ in this plasma, i.e.,
\begin{eqnarray}
      \bar{N}_{L}(\textbf{q})= \bar{N}_{R}(\textbf{q})=\frac{1}{2}\bar{N}(\textbf{q})
   \label{(Eq.94c)}.
\end{eqnarray}
It can be easily obtained from the commutation rules of gamma matrix and the equations ({\ref{(Eq.89a)}}) and ({\ref{(Eq.90a)}}).

Conclusion (\rmnum{2}):
\begin{eqnarray}
      \bar{j}^\mu_{L}(\textbf{q})= \bar{j}^\mu_{R}(\textbf{q})
   \label{(Eq.94d)}.
\end{eqnarray}
The reason is as follows. Note that $\mathcal{L}_{\mathrm{eff}}$ is $CPT$ invariance (charge conjugation, parity and time reversal symmetry). From Eq. ({\ref{(Eq.85a)}}) and ({\ref{(Eq.86a)}}), under the CPT transform, we have
\begin{eqnarray}
      CPT:~~~~~~~\bm{j}_{0L}\longleftrightarrow \bm{j}_{0R}
   \label{(Eq.94f)}.
\end{eqnarray}
It leads to the equation ({\ref{(Eq.94d)}}), otherwise the $CPT$ symmetry of the system will be violated. Inserting Eq.(\ref{(Eq.92a)})-(\ref{(Eq.93b)}), (\ref{(Eq.94c)}) and (\ref{(Eq.94d)}) into  (\ref{(Eq.77a)})-(\ref{(Eq.84a)}), we obtain
\begin{eqnarray}
      \bar{\psi}_0(q)S^{\nu\rho}\gamma^\mu\psi_0(q)=\tilde{\psi}^T_0(q)S^{\nu\rho}\gamma^\mu\tilde{\psi}^*_0(q)=0
   \label{(Eq.94e)}.
\end{eqnarray}

For the sake of simplicity, we make the four assumptions:

Assumption (\rmnum{1}): We consider the small amplitude oscillation in this paper.

Assumption (\rmnum{2}): The system is assumed in static state, which is a common assumption in the usual study of wave propagation in plasmas. It indicates
\begin{eqnarray}
      \bar{N}(t,\textbf{q})=\bar{N}(\textbf{q})
   \label{(Eq.91a)},
\end{eqnarray}
and
\begin{eqnarray}
      \bar{\textbf{j}}(t,\textbf{q})=0.
   \label{(Eq.92a)}
\end{eqnarray}

Assumption (\rmnum{3}): The plasma system is assumed as neutral. It implies
\begin{eqnarray}
      \bar{N}^{(+)}(\textbf{q})= \bar{N}^{(-)}(\textbf{q})=\frac{1}{2}\bar{N}(\textbf{q})
   \label{(Eq.93a)},
\end{eqnarray}
and
\begin{eqnarray}
      \bar{j}^0(\textbf{q})=0.
   \label{(Eq.94a)}
\end{eqnarray}
From Eq. (\ref{(Eq.76a)}) and (\ref{(Eq.92a)})-(\ref{(Eq.94a)}), we obtain
\begin{eqnarray}
      \bar{\psi}^T_0(q)\gamma^\mu\psi_0(q)=\tilde{\psi}_0(q)\gamma^0\gamma^\mu\tilde{\psi}^*_0(q)=0
   \label{(Eq.93b)}.
\end{eqnarray}
From Eq.(\ref{(Eq.72a)}), Eq.(\ref{(Eq.73a)}) and Eq.(\ref{(Eq.93a)}), we have
\begin{eqnarray}
      \bar{c}^1(t,\textbf{q})=\frac{1}{2}e^{i\phi_1(t)}\sqrt{\bar{N}(\textbf{q})}
   \label{(Eq.95a)},
\end{eqnarray}
\begin{eqnarray}
      \bar{c}^2(t,\textbf{q})=\frac{1}{2}e^{i\phi_2(t)}\sqrt{\bar{N}(\textbf{q})}
   \label{(Eq.96a)},
\end{eqnarray}
\begin{eqnarray}
      \bar{d}^1(t,\textbf{q})=\frac{1}{2}e^{i\chi_1(t)}\sqrt{\bar{N}(\textbf{q})}
   \label{(Eq.97a)},
\end{eqnarray}
\begin{eqnarray}
      \bar{d}^2(t,\textbf{q})=\frac{1}{2}e^{i\chi_2(t)}\sqrt{\bar{N}(\textbf{q})}
   \label{(Eq.98a)},
\end{eqnarray}
where $\phi_1(t)$, $\phi_2(t)$, $\chi_1(t)$, and $\chi_2(t)$ are four arbitrary functions depending on $t$. For simplicity, we set
\begin{eqnarray}
      \phi_1(t)=\phi_2(t)=\chi_1(t)=\chi_2(t)=0.
   \label{(Eq.98b)}
\end{eqnarray}
Then, according to Eq.(\ref{(Eq.68a)})-(\ref{(Eq.69b)}), we have
\begin{eqnarray}
      \bar{\mathrm{c}}^s(q^0-E_{\bm{q}},\textbf{q})=\pi\delta(q^0-E_{\bm{q}})\sqrt{\bar{N}(\textbf{q})}
   \label{(Eq.99a)},
\end{eqnarray}
\begin{eqnarray}
      \bar{\mathrm{d}}^s(q^0+E_{\bm{q}},\textbf{q})=\pi\delta(q^0+E_{\bm{q}})\sqrt{\bar{N}(\textbf{q})}
   \label{(Eq.100a)}.
\end{eqnarray}

Assumption (\rmnum{4}): We assume that the plasmas are in the ideal gas state. In the classical case, it indicates that the average kinetic energy of these particles is much larger than the potential energy. In the degenerate case, it is satisfied if the Fermi energy $E_F$ far exceeds the average interaction potential energy. It is easily satisfied in the high density circumstance.
%For the non-relativistic approximation, $\bar{N}(\bm{x})\gg (\frac{e^2m}{2\pi})^3\sim 10^{19}$cm$^{-3}$.
This assumption helps us to simplify Eq.(\ref{(Eq.75a)}). Considering
\begin{eqnarray}
     &~&\int\frac{d^4q}{(2\pi)^4}\psi^\dag_0(q)\psi_0(q)\nonumber\\
     &=&\int d^4q\sum_{s=1,2}[\bar{\mathrm{c}}^s(q^0-E_{\bm{q}},\textbf{q})\bar{\mathrm{c}}^s(q^0-E_{\bm{q}},\bm{q})^*\nonumber\\
     &~&~~~~~~~~~~~~~~~-\bar{\mathrm{d}}^s(q^0+E_{\bm{q}},-\textbf{q})\bar{\mathrm{d}}^s(q^0+E_{\bm{q}},-\textbf{q})^*]\nonumber\\
     &=&\int dt\frac{d^3q}{(2\pi)^3}\sum_{s=1,2}[\bar{c}^s(t,\textbf{q})\bar{c}^s(t,\bm{q})^*\nonumber\\
     &~&~~~~~~~~~~~~~~~~~~~~~~~~~~-\bar{d}^s(t,\textbf{q})\bar{d}^s(t,\bm{q})^*]
   \label{(Eq.101b)},
\end{eqnarray}
and with the help of Expression ({\ref{(Eq.68a)}}) and ({\ref{(Eq.68b)}}), we have
\begin{eqnarray}
     &~&\!\!\!\!\!\!\!\!\!\!\!\!\!\!\!\!\!\!\!\!\!
       \int dt\bar{c}^s(t,\textbf{q})\bar{c}^s(t,\bm{q})^*\nonumber\\
     &~&=\int \frac{dq}{2\pi}\bar{c}^s(q^0-E_{\bm{q}},\textbf{q})\bar{c}^s(q^0-E_{\bm{q}},\bm{q})^*,
   \label{(Eq.101c)}
\end{eqnarray}
\begin{eqnarray}
     &~&\!\!\!\!\!\!\!\!\!\!\!\!\!\!\!\!\!\!\!\!\!
     \int dt\bar{d}^s(t,\textbf{q})\bar{d}^s(t,\bm{q})^*\nonumber\\
     &~&=\int \frac{dq}{2\pi}\bar{d}^s(q^0+E_{\bm{q}},-\textbf{q})\bar{c}^s(q^0+E_{\bm{q}},-\bm{q})^*.
   \label{(Eq.101c)}
\end{eqnarray}
Then, Eq.(\ref{(Eq.75a)}) reduces to
\begin{eqnarray}
     &~&\int\!\frac{d^4q}{(2\pi)^4}\bar{\psi}_0(q)\psi_0(q)\nonumber\\
     &=&\frac{1}{m}\!\int\! dt\frac{d^3q}{{(2\pi)^3}}\bar{E}_{\bm{q}}\!\sum_{s=1,2}[\bar{c}^s(t,\textbf{q})\bar{c}^s(t,\bm{q})^*\nonumber\\
     &~&~~~~~~~~~~~~~~~~~~~~~~~~~~~~~~~~+\bar{d}^s(t,\textbf{q})\bar{d}^s(t,\bm{q})^*]\nonumber\\
     &=&\frac{1}{m}\!\int\! \frac{d^4q}{{(2\pi)^4}}\bar{E}_{\bm{q}}\!\sum_{s=1,2}[\bar{c}^s(q^0-E_{\bm{q}},\textbf{q})\bar{c}^s(q^0-E_{\bm{q}},\bm{q})^*\nonumber\\
     &~&~~~~~~~~~+\bar{d}^s(q^0+E_{\bm{q}},-\textbf{q})\bar{c}^s(q^0+E_{\bm{q}},-\bm{q})^*]
   \label{(Eq.101a)}.
\end{eqnarray}
From Eq.(\ref{(Eq.99a)}), (\ref{(Eq.100a)}) and (\ref{(Eq.101a)}), we get a more general formula
\begin{eqnarray}
     &~&\int \frac{d^4q}{(2\pi)^4}\bar{\psi}_0(q)f(q)\psi_0(q)\nonumber\\
     &=&2\pi\delta (0)\int \frac{d^3q}{(2\pi)^3}\frac{\bar{E}_{\bm{q}}}{m}[\bar{N}^{(+)}_0(\textbf{q})f(\textbf{q})|_{q^0=E_{\bm{q}}}\nonumber\\
     &~&~~~~~~~~~~~~~+\bar{N}^{(-)}_0(-\textbf{q})f(\textbf{q})|_{q^0=-E_{\bm{q}}}]
   \label{(Eq.102a)},
\end{eqnarray}
where $f(\textbf{q})$ is an arbitrary function of $\textbf{q}$.

\subsubsection{\label{sec:level2}Background polarization tensor}

Inserting Eq. ({\ref{(Eq.94c)}}), ({\ref{(Eq.94e)}}), ({\ref{(Eq.93b)}}) and ({\ref{(Eq.102a)}}) into the expression (\ref{(Eq.53a)}), we obtain
\begin{widetext}
\begin{eqnarray}
  &~&\!\!\!\!\!\!\!\!\!\!\!\!\!\!\!\!\!\!\!\!\!\!\!\!\!\!\!\!\!\!\!\!
      \sum_{\rho,\lambda=1,2}i\Pi^{(\rho\lambda)\mu\nu}_{Tbac,2}(k)(2\pi)^4\delta^{(4)}(0)\nonumber\\
  &~&=-4ie^22\pi\delta(0)\!\int\!\frac{d^3q}{(2\pi)^3}E_{\bm{q}}\Big\{N^{(+)}(\textbf{q})\Big\{\frac{(k^2+q^2-m^2)g^{\mu\nu}+4iq\cdot kS^{\mu\nu}}{[(k+q)^2-m^2+i\epsilon][(k-q)^2-m^2+i\epsilon]}\nonumber\\
  &~&~~~~~~~~~~~~~+\pi(g^{\mu\nu}-2iS^{\mu\nu})\{\delta[(q-k)^2-m^2]+\delta[(q+k)^2-m^2]\}\Big\}\Big|_{q^0=E_{\bm{q}}}\nonumber\\
  &~&~~~~~~~~~~~~~+N^{(-)}(-\textbf{q})\Big\{\frac{(k^2+q^2-m^2)g^{\mu\nu}+4iq\cdot kS^{\mu\nu}}{[(k+q)^2-m^2+i\epsilon][(k-q)^2-m^2+i\epsilon]}\nonumber\\
  &~&~~~~~~~~~~~~~+\pi(g^{\mu\nu}-2iS^{\mu\nu})\{\delta[(q-k)^2-m^2]+\delta[(q+k)^2-m^2]\}\Big\}\Big|_{q^0=-E_{\bm{q}}}\Big\}\nonumber\\
  &~&=-\frac{4ie^2}{V}(2\pi)^4\delta^{(4)}(0)\int\frac{d^3q}{(2\pi)^3}
     E_{\bm{q}}N_0(\textbf{q})\frac{k^2g^{\mu\nu}}{(k^2-2k\cdot q+i\epsilon)(k^2+2k\cdot q+i\epsilon)}\Big|_{q^0=E_{\bm{q}}},
 \label{(Eq.103a)}
\end{eqnarray}
\end{widetext}
where
\begin{eqnarray}
     V=\int d^3x=(2\pi)^3\delta^{(3)}(0)
   \label{(Eq.104a)}
\end{eqnarray}
is the volume of space the plasma occupied. The two terms involving $S^{\mu\nu}$ cancel each other out, which is easily seen by considering $\bar{N}^{(+)}(\bm{q})=\bar{N}^{(-)}(\bm{q})$, and replacing $\textbf{q}$ by $-\textbf{q}$ in the second term in the brace in the integral. For convenience, we write
\begin{eqnarray}
  \sum_{\rho,\lambda=1,2}i\Pi^{(\rho\lambda)\mu\nu}_{T\mathrm{bac},2}(k)=ig^{\mu\nu}k^2B(k)
 \label{(Eq.104b)}
\end{eqnarray}
%We make an important remark on this expression. First, A heuristic view of the appearance of $(2\pi)^4\delta^{(4)}(0)$ in the background polarization tensor is that it comes from the S-matrix elements. Imagine the physical process for the scattering of a fluctuation vector boson by the effective medium. It is well known that the S-matrix can always be written as
%\begin{eqnarray}
%     S_{fi}&=&_{out}\langle k|S|k \rangle_{in}\nonumber\\
%           &=&\mathbbm{1}_{fi}+iM_{fi}(2\pi)^4\delta^{(4)}(0)\nonumber\\
%           &\approx&\mathbbm{1}_{fi}+i\varepsilon_\mu(k)\varepsilon^*_\nu(k)\Pi_2^{\mu\nu}(k)(2\pi)^4\delta^{(4)}(0).
%   \label{(Eq.105a)}
%\end{eqnarray}
%The delta function comes from the integral of vertex in Feynman diagram representing the momentum conservation. It also explains why $\phi_1(t)$, $\phi_2(t)$, $\chi_1(t)$ and $\chi_2(t)$ in Eq.(\ref{(Eq.95a)})-Eq.(\ref{(Eq.98a)}) can be chosen as zero. In fact, there is no need for these constants to be unique determined. They are just being chosen for ensuring the appearance of $(2\pi)^4\delta^{(4)}(0)$ in the background polarization tensor. Second,
As pointed out before, the expression (\ref{(Eq.103a)}) violates Ward identity $k_\mu\Pi_T^{\mu\nu}(k)=0$, which is, however, acceptable. The reason is given in Appendix B.

For simplicity, we choose a coordinate system such that $k^\mu=(\omega,0,0,|\textbf{k}|)$. From the assumption (\rmnum{4}), the charged particles obey Fermi-Dirac distribution, reads
\begin{eqnarray}
     N_0^{(+)}(\textbf{q})=N_0^{(-)}(\textbf{q})=\frac{1}{e^{\beta(E_{\bm{q}}-\mu)}+1},
   \label{(Eq.106a)}
\end{eqnarray}
where $\mu$ is the chemical potential. Plugging it into the expression ({\ref{(Eq.103a)}}), then $\Pi^{\mu\nu}_{T\mathrm{bac},2}(k)$ is going to be discussed in both of THE low-energy and high-energy approximations below.

The term of low-energy approximation is understood as that $k^2$ is well below the threshold for real $e^-e^+$ production. As will be seen in the next subsection, it means
\begin{eqnarray}
     k^2\ll m^2.
   \label{(Eq.107b)}
\end{eqnarray}
Besides, to compare with the previous works, we restrict the discussion within the non-relativistic approximation and long wavelength limit
\begin{eqnarray}
     |\textbf{q}|\ll m,~~~|\bm{k}|\ll m,~~~\omega\ll m.
   \label{(Eq.107a)}
\end{eqnarray}
Furthermore, the plasma is assumed in the degenerate state which gives
\begin{eqnarray}
  \mu'/T\gg 1,
 \label{(Eq.107c)}
\end{eqnarray}
where $\mu'=\mu-m$.

Then the real and imaginary part of $B(k)$ are calculated separately. In the condition ({\ref{(Eq.107a)}}), most particles are with momentum $|\bm{q}|$ far away from the pole of the integral ({\ref{(Eq.103a)}}). The distribution ({\ref{(Eq.106a)}}) is almost zero at this point. It indicates that one can always neglect this singularity in computing $\mathrm{Re}B(k)$. Subsequently, we have
\begin{eqnarray}
  &~&\!\!\!\!\!\!\!\!\!\!\!\!\!\!\!\!\!\!\!\!\!\!
      \mathrm{Re}B(k)\approx-\frac{e^2m}{2\pi^2}\frac{1}{(\omega^2-|\textbf{k}|^2)^2-4m^2\omega^2}\nonumber\\
   &~&~~\times\int_0^\infty d|\bm{q}|\frac{1}{e^{\beta(\textbf{q}^2/2m-\mu')}+1}\{|\textbf{q}|^2\nonumber\\
   &~&~~+\frac{(\omega^2-|\textbf{k}|^2)^2+4m^2\omega^2-8m^2|\textbf{k}|^2}
  {2m^2[(\omega^2-|\textbf{k}|^2)^2-4m^2\omega^2]}|\textbf{q}|^4\}.
\label{(Eq.108a)}
\end{eqnarray}
It involves the integral of the form $\int_0^\infty d\varepsilon f(\varepsilon)[e^{\beta(\varepsilon-\mu')}+1]^{-1}$, where $f(\varepsilon)$ is the function such that the integral converges. In the condition ({\ref{(Eq.107c)}}), we have the formula
\begin{eqnarray}
     &~&\int_0^\infty d\varepsilon \frac{f(\varepsilon)}{e^{\beta(\mu'-\varepsilon)}+1}\nonumber\\
     &=&\int_0^{\mu'} d\varepsilon f(\varepsilon)+2T^2f^\prime(\mu')\int_0^\infty dx\frac{x}{e^x+1}+\cdot\cdot\cdot .~~~~~
   \label{(Eq.109a)}
\end{eqnarray}
Subsequently,
\begin{eqnarray}
     \int_0^\infty d\varepsilon \frac{\varepsilon^\frac{1}{2}}{e^{\beta(\mu'-\varepsilon)}+1}\approx\frac{2}{3}\mu'^{\frac{3}{2}}+\frac{T^2}{2\sqrt{\mu'}}\zeta(2),
   \label{(Eq.110a)}
\end{eqnarray}
\begin{eqnarray}
     \int_0^\infty d\varepsilon \frac{\varepsilon^\frac{3}{2}}{e^{\beta(\mu'-\varepsilon)}+1}\approx\frac{2}{5}\mu'^{\frac{5}{2}}+\frac{3}{2}T^2\sqrt{\mu'}\zeta(2).
   \label{(Eq.111a)}
\end{eqnarray}
Here, $\zeta(x)=\sum_{n=1}^\infty 1/n^x$ is the Riemann zeta function, which gives $\zeta(2)=\pi^2/6$. The chemical potential can be written as a power series in $T^2$ as
\begin{eqnarray}
     \mu'=\mu'_0[1-\frac{\pi^2}{12}(\frac{T}{\mu_0})^2+\cdot\cdot\cdot],
   \label{(Eq.112a)}
\end{eqnarray}
where $\mu'_0=E_F(T=0)=(3\pi^2n_0)^{\frac{2}{3}}/2m$ is Fermi energy at $T=0$, $n_0$ is the charged particle number per unit volume, satisfying
\begin{eqnarray}
     \int d^3xn_0=\int \frac{d^3q}{(2\pi)^3}N_0(\textbf{q})
   \label{(Eq.113c)}.
\end{eqnarray}
Then from the expression (\ref{(Eq.108a)}), to the order of $(T/\mu'_0)^2$, we obtain
\begin{eqnarray}
     \mathrm{Re}B(k)\!\approx&-&\!\!\frac{4m^2\omega_p^2}{(\omega^2-|\textbf{k}|^2)^2-4m^2\omega^2}\nonumber\\
     &\times&\!\!\{1+[\frac{3}{10m^2}(3\pi^2n_0)^{\frac{2}{3}}+\frac{\pi^2T^2}{2(3\pi^2n_0)^{\frac{2}{3}}}]\nonumber\\
     &\times&\!\! \frac{(\omega^2-|\textbf{k}|^2)^2+4m^2\omega^2-8\omega^2|\textbf{k}|^2}{(\omega^2-|\textbf{k}|^2)^2-4m^2\omega^2}\},
   \label{(Eq.113a)}
\end{eqnarray}
where $\omega_p^2=e^2n_0/m$ is the plasma frequency.

By using the relation
\begin{eqnarray}
  \frac{1}{x-x_0+i\epsilon}=\mathcal{P}\frac{1}{x-x_0}-i\pi \delta(x-x_0)
 \label{(Eq.107d)}
\end{eqnarray}
with $\mathcal{P}$ denoting that the principle value is taken in integrations, in the spacelike region ($k^2<0$), the integral ({\ref{(Eq.103a)}}) gives
\begin{eqnarray}
     &~&\!\!\!\!\!\!\!\!\!\!\!\!
        \mathrm{Im}B(k)=\frac{2\alpha}{|\bm{k}|k^2}\!\int_{\frac{1}{2}(|\bm{k}|-\omega\sqrt{1-\frac{4m^2}{k^2}})}^\infty d|\bm{q}|\frac{|\bm{q}|\sqrt{|\bm{q}|^2+m^2}}{e^{\beta(\sqrt{|\bm{q}|^2+m^2}-\mu)}+1}.\nonumber\\
   \label{(Eq.114g)}
\end{eqnarray}
In the timelike region ($k^2>0$), the integral ({\ref{(Eq.103a)}}) gives
\begin{eqnarray}
     &~&\!\!\!\!\!\!\!\!\!\!\!\!\!\!\!\!\!\!
        \mathrm{Im}B(k)=\frac{\alpha}{|\bm{k}|k^2}\theta(k^2-4m^2)\int_{-\frac{1}{2}(|\bm{k}|-\omega\sqrt{1-\frac{4m^2}{k^2}})}^{\frac{1}{2}(|\bm{k}|+\omega\sqrt{1-\frac{4m^2}{k^2}})} d|\bm{q}|\nonumber\\
     &~&~~~~~~~~~~~~~
        \times\frac{|\bm{q}|\sqrt{|\bm{q}|^2+m^2}}{e^{\beta(\sqrt{|\bm{q}|^2+m^2}-\mu)}+1},
   \label{(Eq.114c)}
\end{eqnarray}
where
\begin{eqnarray}
 \theta(x-x_0)=
		\begin{cases}
		  1 & \quad \mbox{for}\quad x\geqslant x_0,\ \\
		  0 & \quad \mbox{for}\quad x< x_0.
	\end{cases}
 \label{(Eq.107e)}
\end{eqnarray}
In the low energy approximation ({\ref{(Eq.107b)}}), the integrals ({\ref{(Eq.114g)}}) and ({\ref{(Eq.114c)}}) give
\begin{eqnarray}
  \lim_{-k^2/m^2\rightarrow 0}\mathrm{Im}B(k)\big|_{k^2<0}=0
 \label{(Eq.114f)}
\end{eqnarray}
and
\begin{eqnarray}
  \mathrm{Im}B(k)\big|_{k^2>0}=0,
 \label{(Eq.114h)}
\end{eqnarray}
respectively.

The term of high-energy limit is understood as that $k^2$ far exceeds the threshold for $e^-e^+$ pair production, i.e.,
\begin{eqnarray}
     k^2\gg m^2.
   \label{(Eq.114a)}
\end{eqnarray}
Here fermions can be in both relativistic and non-relativistic state. For simplicity, we shall restrict our discussion on $\mathrm{Im}B(k)$ in this case. Since Pauli's exclusion principle inhibits the pair production with $k^2$ below $\mu_0^2$ (If the density is high enough that $\mu_0>m$) at zero temperature \cite{Chin}. In this region, the collective mode is stable. It is hard to obtain the analytic expression in the moderate region $k^2\sim\mu_0^2$ (although $k^2\gg m^2$) with nonzero temperature where the decay of collective mode is partially affected by the inhibition. A numerical study is needed. In this paper, we just focus on the region $(k^2-\mu_0^2)/T^2\gg 1$ where this inhibition effect is excluded. Then, from ({\ref{(Eq.114c)}}), we have
\begin{eqnarray}
     \lim_{\beta^2(k^2-\mu_0^2)\rightarrow\infty}\mathrm{Im}B(k)=0.
   \label{(Eq.114b)}
\end{eqnarray}

%\begin{widetext}
%\begin{eqnarray}
%  \sum_{\rho,\lambda=1,2}\!\!i\Pi^{\mu\nu}_{Tbac,2}(k)\!&=&\!ig^{\mu\nu}m\omega_p^2\{\frac{1}{4\mu}-\frac{1}{32\mu^3|\textbf{k}|}[(\omega-|\textbf{k}|)^3\ln(1-\frac{4\mu^2}{(\omega-|\textbf{k}|)^2})
%  -(\omega+|\textbf{k}|)^3\ln(1-\frac{4\mu^2}{(\omega+|\textbf{k}|)^2})\nonumber\\
%  &~&-\frac{\mu^3}{3}\ln|\frac{\omega^2-(|\textbf{k}|+2\mu)^2}{\omega^2-(|\textbf{k}|-2\mu)^2}|]+\zeta(2)\frac{3m\omega_p^2T^2}{4\mu^3}[\frac{\mu^2}{(\omega^2-|\textbf{k}|^2)^2
%  }(6+\frac{8\mu^2(\omega^2-|\textbf{k}|^2+4\mu^2)}{[(\omega-|\textbf{k}|)^2-4\mu^2][(\omega+|\textbf{k}|)^2-4\mu^2]})\nonumber\\
%  &~&-\frac{4\mu^3}{|\textbf{k}|(\omega^2-|\textbf{k}|^2)^2}\ln|\frac{\omega^2-(|\textbf{k}|-2\mu)^2}{\omega^2-(|\textbf{k}|+2\mu)^2}|]\}.
%\label{(Eq.115a)}
%\end{eqnarray}

\subsection{\label{sec:level2}Calculation of $\sum_{\rho,\lambda=1,2}\Pi^{(\rho\lambda)\mu\nu}_{T\mathrm{vac},2}(k)$}

To the order of $e^2$, the temperature-dependent vacuum polarization tensor is reduced to
\begin{eqnarray}
     &~&\sum_{\rho,\lambda=1,2}i\Pi_{T2,\mathrm{vac}}^{(\rho\lambda)\mu\nu}(k)\nonumber\\
     &=&\sum_{\rho,\lambda=1,2}~~\ff{\vertexlabel^{\mu}\vertexlabel_{(\rho)}g~~~~\momentum{flSA}{p+k}\momentum{flSuV}{p}~~~~g\vertexlabel^{\nu}\vertexlabel_{(\lambda)}}\nonumber\\
     &~&\nonumber\\
     &=&e^2\sum_{\rho,\lambda=1,2}\int\frac{d^4p}{(2\pi)^4}\mathrm{tr}[(\Gamma_{(\rho\rho)})^\mu\nonumber\\
     &~&~~~~~~~~~~~~~\times S_{TF}^{(\rho\lambda)}(p+k)(\Gamma_{(\lambda\lambda)})^\nu S^{(\lambda\rho)}_{TF}(p)]\nonumber\\
     &=&e^2\!\!\int\!\!\frac{d^4p}{(2\pi)^4}\Big\{\frac{\mathrm{tr}[\gamma^\mu(\slashed{p}+\slashed{k}+m)\gamma^\nu(\slashed{p}+m)]}{[(p+k)^2-m^2+i\epsilon][p^2-m^2+i\epsilon]}\nonumber\\
     &~&~~~~~+\frac{\mathrm{tr}[\gamma^\mu(\slashed{p}+\slashed{k}+m)\gamma^\nu(\slashed{p}+m)]}{[(p+k)^2-m^2-i\epsilon][p^2-m^2-i\epsilon]}\Big\},
   \label{(Eq.115b)}
\end{eqnarray}
where we use $\mathrm{tr}()$ to denote Dirac traces. With the help of the formulas
\begin{displaymath}
  \delta(x-x_0)=\frac{\epsilon}{\pi[(x-x_0)^2+\epsilon^2]},
\end{displaymath}
and the relation ({\ref{(Eq.107d)}}), by using Feynman parameters integral method and standard dimensional regularization, Eq. ({\ref{(Eq.115b)}}) becomes
\begin{eqnarray}
     \sum_{\rho,\lambda=1,2}i\Pi_{T2,\mathrm{vac}}^{(\rho\lambda)\mu\nu}(k)=i(k^2g^{\mu\nu}-k^\mu k^\nu)\Pi_{T\mathrm{vac},2}(k^2),
   \label{(Eq.116c)}
\end{eqnarray}
where
\begin{widetext}
\begin{eqnarray}
     \Pi_{T\mathrm{vac},2}(k^2)
                     &=&(\mathrm{Re}+2i\mathrm{Im})\Big(-\frac{2\alpha}{\pi}\int_0^1dxx(1-x)\ln\frac{m^2}{m^2-x(1-x)k^2}\Big)\nonumber\\
                     &=&(\mathrm{Re}+2i\mathrm{Im})\Big\{\frac{2\alpha}{9\pi}[5+\frac{12m^2}{k^2}+\frac{6(k^4-2m^2k^2-8m^4)}{k^3\sqrt{4m^2-k^2}}\arctan\frac{k}{\sqrt{4m^2-k^2}}]\Big\}
   \label{(Eq.116a)}.
\end{eqnarray}
%\end{widetext}
The condition of $k^2>4m^2$ indicates the decay process of the boson propagation. In the low-energy limit, we have
\begin{eqnarray}
     \Pi_{T\mathrm{vac},2}(k^2)\approx\frac{2\alpha}{9\pi}(-1+\frac{3k^2}{4m^2})
   \label{(Eq.116b)}.
\end{eqnarray}
In the high-energy limit, we have the imaginary part of $\Pi_{T\mathrm{vac},2}(k^2)$ as follows
\begin{eqnarray}
     \mathrm{Im}\Pi_{T\mathrm{vac},2}(k^2+i\epsilon)=-\frac{2\alpha}{3}\sqrt{1-\frac{4m^2}{k^2}}(1+\frac{2m^2}{k^2})\approx-\frac{2\alpha}{3}
   \label{(Eq.116d)}.
\end{eqnarray}

\section{Dispersion Relations}

Substituting $g^{\mu\nu}k^2B(k)$ and $(k^2g^{\mu\nu}-k^\mu k^\nu)\Pi_{T\mathrm{vac},2}(k^2)$ into $D^{-1}_F(k)-i\!\sum_{\rho\lambda=1,2}\!\Pi_{T,2}^{(\rho\lambda)}(k)$, and using Feynman gauge ($\xi=1$), we have
%\begin{widetext}
\begin{eqnarray}
  \left(                 %左括号
      \begin{array}{cccc}   %该矩阵一共2 列，每一列都居中放置
        |\textbf{k}|^2[1-\Pi_{T\mathrm{vac},2}(k^2)]+k^2B(k)\!\!\!\!\!\!\!\!\!
         &  0
         &  0
         & \!\!\!\!\!\!\!\!\!\omega|\textbf{k}|[1-\Pi_{T\mathrm{vac},2}(k^2)] \\  % 第一行元素
        0 &  k^2[1-\Pi_{T\mathrm{vac},2}(k^2)-B(k)]& 0 & 0  \\  % 第二行元素
          0 & 0 & \!\!\!\!\!\!\!\!\! k^2[1-\Pi_{T\mathrm{vac},2}(k^2)-B(k)] & 0  \\
         \omega|\textbf{k}|[1-\Pi_{T\mathrm{vac},2}(k^2)] & 0 & 0 & \!\!\!\!\!\!\!\!\! \omega^2[1-\Pi_{T\mathrm{vac},2}(k^2)]-k^2B(k) \\
      \end{array}
   \right).
\label{(Eq.117a)}
\end{eqnarray}
\end{widetext}
Subsequently, Eq.(\ref{(Eq.39w)}) reduces to
\begin{eqnarray}
     1-\Pi_{T\mathrm{vac},2}(k^2)-B(k)=0.
   \label{(Eq.119a)}
\end{eqnarray}
\begin{eqnarray}
     k^2=0,
   \label{(Eq.118b)}
\end{eqnarray}
\begin{eqnarray}
     B(k)=0,
   \label{(Eq.118a)}
\end{eqnarray}
In principle, one can obtain the dispersion relations from these three equations. However, it asks for the exact evaluation of the integral (\ref{(Eq.103a)}), which is very difficult and is of less interest to the present works. In this paper, we explain the dispersion relations in both low-energy and high-energy limits and compare them with the previous studies.

\subsection{\label{sec:level2}Low energy limit}

\subsubsection{\label{sec:level2}Timelike longitudinal wave dispersion relation}

To the second order approximation, Eq.(\ref{(Eq.113a)}), (\ref{(Eq.116a)}) and (\ref{(Eq.119a)}) give the longitudinal wave dispersion relation as follows
\begin{eqnarray}
     \omega^2\!&=&\!\omega^2_p[1+\frac{\omega^2_p}{4m^2}(1-\frac{3q_F^2}{10m^2}+\frac{\pi^2T^2}{q_F^2})\nonumber\\
               &~&\!-\frac{3q_F^2}{10m^2}+\frac{\pi^2T^2}{q_F^2}-\frac{2\alpha}{9\pi}(1+\frac{\omega_p^2}{m^2})]\nonumber\\
               &~&\!+(\frac{3q_F^2}{5m^2}-\frac{\omega_p^2}{2m^2}+\frac{\pi^2T^2}{q_F^2})|\textbf{k}|^2+\frac{|\textbf{k}|^4}{4m^2}.
   \label{(Eq.120a)}
\end{eqnarray}
where, $\omega_p^2=e^2n_0/m$. The leading order approximation of this expression gives
\begin{eqnarray}
     \omega^2=\omega_p^2,
   \label{(Eq.121a)}
\end{eqnarray}
which is just the well known classical plasma frequency. In the zero temperature case, to the next to leading order approximation, the longitudinal dispersion relation is
\begin{eqnarray}
     \omega^2&=&\omega_p^2(1-\frac{3q_F^2}{10m^2}-\frac{2\alpha}{9\pi}+\frac{\omega_p^2}{4m^2})\nonumber\\
             &~&~~~~~~~~~+(\frac{3q_F^2}{5m^2}-\frac{\omega_p^2}{2m^2})|\textbf{k}|^2+\frac{|\textbf{k}|^4}{4m^2}.
   \label{(Eq.122a)}
\end{eqnarray}
We explain each term as follows. The last three terms in the first parentheses are the corrections to the plasma frequency. They can be explained as the corrections to the mass, effective charge, and effective number of charged particles, respectively. The term of $-3q_F^2/10m^2$ comes from the relativistic correction to mass. It can also be obtained by replacing $m^{-1}$ in $\omega_p^2=e^2n_0/m$ by $m^{-1}\langle\sqrt{1-v_T^2}\rangle\approx m^{-1}(1-\frac{\overline{v^2_T}}{2})$. Here, $\overline{v^2_T}$ is the statistical average of velocity squared. In the highly degenerate limit case at $T=0$, we have $\overline{v^2_T}=3q_F^2/5m^2$. It gives $\omega_p^2\approx (1-3q_F^2/10m^2)e^2n_0/m$ if only the correction due to the mass increasing is included. There is no such term in the scalar QED plasmas since the average velocity of bosons at zero temperature is vanished. The appearance of $-2\alpha/9\pi$ is caused by the virtual $e^-e^+$ pair production since it comes from $\Pi^{({\rho\lambda})\mu\nu}_{T\mathrm{vac},2}(k)$ if we retrace the calculation steps. A similar result is reported in the scalar plasmas \cite{Shi Fisch Qin}. The term $\frac{\omega_p^4}{4m^2}$ is attributed to virtual particle redistribution. The physical interpretation is as follows: Due to the vacuum polarization, Coulomb potential for individual $e^-$ or $e^+$ must be corrected as follows
\begin{eqnarray}
     V(r)=\mp\frac{\alpha}{r}(1+\frac{\alpha}{4\sqrt{\pi}}\frac{e^{-2mr}}{(mr)^{\frac{3}{2}}}+\cdot\cdot\cdot),
   \label{(Eq.122f)}
\end{eqnarray}
which is known as Uehling potential \cite{Peskin Schroeder}. It indicates that most virtual $e^-e^+$ pair appear in the region around the real electrons or positrons with volume $(z/2m)^3$ (with constant $z=O(1)$). Strictly speaking, the distribution of the virtual particles is determined not only by its center real charge but also by other charged particles. Besides, in the plasma oscillation process, one can not simply consider the movement of real fermions with the polarization cloud as a whole. The shape of the polarization cloud changes in time. Both of these two cases lead to deviation of $V(r)$ from Uehling potential. If $\kappa$, in Eq. ({\ref{(Eq.1a)}}), is not much bigger than $O(1)$, the real fermions begin to affect or even penetrate each other's vacuum polarization cloud. The distribution of the virtual $e^-e^+$ pairs will be significant changed. Subsequently, when $n_0=\frac{3}{4\pi}(2 m/\kappa)^3$ we have
\begin{eqnarray}
     \frac{\omega_p^2}{4m^2}=\frac{6\alpha}{\kappa^3}.
   \label{(Eq.122g)}
\end{eqnarray}
The correction is significant as expected. The increasing cut-off frequency is intuitively explained as the enhancing of restore force due to this virtual charges redistribution. To visualize it, we define the number density of ``effective created particle'', denoting by $\delta n_{\mathrm{eff}}$ as follows
\begin{eqnarray}
     \omega {'}_p^2=\frac{e^2(n_0+\delta n_{\mathrm{eff}})}{m}.
   \label{(Eq.122b)}
\end{eqnarray}
Ignoring other corrections in the expression ({\ref{(Eq.122a)}}), we have
\begin{eqnarray}
     \frac{\delta n_{\mathrm{eff}}}{n_0}=\frac{\omega^2_p}{4m^2}\approx 1.32\times 10^{-33}n_0\mathrm{cm}^3,
   \label{(Eq.122c)}
\end{eqnarray}
which represents the degree of cut-off frequency increasing. We must emphasize that this definition doesn't mean that there is any real on-shell particle production. Because in the non-relativistic region, $\mathrm{Im}B(k)=0$ for the timelike mode, as shown in Eq. ({\ref{(Eq.114h)}}). Besides, $n_0$ should be restricted below $10^{30}\mathrm{cm}^{-3}$ to satisfy the non-relativistic approximation. The general formula for any $n_0$ can be obtained from the integral ({\ref{(Eq.103a)}}) without any approximation. However, we can still infer from Eq. ({\ref{(Eq.122c)}}) that the cut-off frequency is significantly altered by dense plasma. One will observe about $0.1\%$ change in plasma frequency quare in plasma with density $10^{30}\mathrm{cm}^{-3}$. In this case $\kappa\approx 3.21$. It is reasonable to expect a further increase of cut-off frequency due to the virtual particle redistribution, if we extend the calculation to the relativistic scope. Further study is interesting. Especially in the extreme astrophysical environments, such as the outer crust of neutron star with typical density $10^4$g$\cdot$ cm$^{-3}\sim 10^{11}$g$\cdot$ cm$^{-3}$ \cite{Fantina Chamel Pearson Goriely} and white dwarfs with the density below the neutron drip, i.e., the mass density less than $10^{11}$g$\cdot$cm$^{-3}$ \cite{Fantoni}.
The effect has not been reported before to our knowledge. No similar term appears in any non-relativistic quantum \cite{Klimontovich Silin}\cite{Bohm Pines}\cite{Fetter Walecka}, semi-classical relativistic \cite{Kowalenko Frankel Hines} or the usual relativistic QFT \cite{Chin} studies of plasmas because RPA or the equivalent approximations are used in these studies. In RPA, the short range interparticle correlation is neglected. It means that the contribution of virtual $e^-_\psi e^+_\psi$ pair redistribution, in the nearby region of charged particles, is excluded (The non-relativistic theory has, of course, no such effect). It is not found in the study of scalar QED plasmas with the similar nontrivial background method by Shi, et al. \cite{Shi Fisch Qin} neither. Since they evaluate the self-energy with the help of the free particle wave function, as seen in Eq. (5) and Eq. (43) in their paper. Its essence is Hartree approximation in which the real interaction is replaced by some average one. There is no term concerning on the residue interaction that mainly contributed by the short range correlation. It leads to the fact that no such term appear in $\bar{\eta}^\mu$ (Eq. (11)) and $\Pi_{2,bk}^{\mu\nu}$ (Eq. (24)) in their paper. Subsequently, the effect of virtual $e^-e^+$ pair redistribution is excluded entirely. There is no such short-range effect missing in our theory since it is incorporated into the background field which can be exactly evaluated from our method. Before going further, we briefly summarize the influence of vacuum fluctuation on plasmas. The corrections to $\frac{2\alpha}{9\pi}$ and $\frac{\omega_p^2}{4m^2}$ indicate that the virtual $e^-$ and $e^+$ contribute to the oscillation of plasma in two ways: screening the bare charges and offering extra restore force on the other charged particles during oscillation.
%\begin{figure}[b]
%\includegraphics[width=0.8\columnwidth]{Fig1}% Here is how to import EPS art
%\caption{\label{Fig.1} The proportion of produced particles $\delta n_{\mathrm{eff}}/n_0$ as a function of $n_0$, according to Eq. (\ref{(Eq.122b)}). }
%\end{figure}
The term of $3q_F^2|\textbf{k}|^2/5m^2$ in the second parentheses of the Eq. (\ref{(Eq.122a)}) is commonly seen in the studies of non-relativistic degenerate plasma with various methods \cite{Bohm Pines}\cite{Kremp Schlanges Kraeft}. The term of $-\omega_p^2|\textbf{k}|^2/2m^2$ is found in the case of scalar QED plasmas \cite{Shi Fisch Qin}\cite{Kowalenko Frankel Hines}\cite{Hines Frankel}. The minus sign indicates a local minimum in the dispersion relation. This phenomenon is referred to, by Ref. \cite{Hines Frankel}, as negative dispersion due to the finite speed of light, and is attributed to a retardation effect in Bose plasmas. Also, this explanation can be extended to Fermi plasmas. If we neglect the effects of relativistic and vacuum polarization corrections, Eq.(\ref{(Eq.122a)}) reduces to the well-known longitudinal excitation spectrum for non-relativistic degenerate plasmas at zero temperature \cite{Bohm Pines}
\begin{eqnarray}
     \omega^2=\omega_p^2+\frac{3q_F^2}{5m^2}|\textbf{k}|^2+\frac{|\textbf{k}|^4}{4m^2}.
   \label{(Eq.122e)}
\end{eqnarray}

For the $T\neq 0$ case, as shown in Eq. (\ref{(Eq.120a)}), the plasma frequency decreases with the increase of temperature. A similar trend is found in classical plasmas \cite{Schlickeiser}. The temperature correction $\pi^2T^2/q_F^2$ in the coefficient before $|\bm{k}|^2$ agrees with the kinetic theory study on degenerate plasma \cite{Maafa}.

\subsubsection{\label{sec:level2}Lightlike transverse wave dispersion relation}

Another solution of Eq.(\ref{(Eq.118b)}) gives the lightlike dispersion relation of transverse mode
\begin{eqnarray}
     \omega^2=|\textbf{k}|^2.
   \label{(Eq.123a)}
\end{eqnarray}
Note that it is just the dispersion relation for free $\gamma_{_\mathcal{A}}$. As mentioned in Sec. \Rmnum{3}, the physical process can also be understood as a scattering of $\gamma_{_\mathcal{A}}$ by the effective medium. From the viewpoint of scattering theory, statistical speaking, in addition to the scattered $\gamma_{_\mathcal{A}}$, there are some other incident ones simply missing the target (plasma background) with the dispersion relation Eq. (\ref{(Eq.123a)}). However, it is just a purely mathematical scheme to divide the original fields into background fields and fluctuations. One can not separate them physically. The transverse mode represented by the relation ({\ref{(Eq.123a)}}) is not an observable object. To get the dispersion relation for the observable transverse wave, we note the followed two points: First, in principle, we can derive from the original total Lagrangian ({\ref{(Eq.11a)}}) the dispersion relation $\Omega(|\bm{k}|)$ for observable mode. It is because that the Lagrangians ({\ref{(Eq.19a)}}) and ({\ref{(Eq.11a)}}) are just two equivalent descriptions for the same system. Then $\omega(|\bm{k}|)$ and $\Omega(|\bm{k}|)$ are two different descriptions for the same transverse mode. However, the group velocity is an observable quantity, which implies
\begin{eqnarray}
     \frac{\partial\Omega}{\partial |\bm{k}|}=\frac{\partial\omega}{\partial |\bm{k}|}.
   \label{(Eq.123c)}
\end{eqnarray}
Subsequently
\begin{eqnarray}
     \Omega^2=\mathrm{constant}+|\textbf{k}|^2,
   \label{(Eq.123b)}
\end{eqnarray}
which differs from the relation ({\ref{(Eq.123a)}}) only by a constant term. Second, the transverse plasma mode can be considered as massive vector boson, with the mass equals to the cut-off plasma frequency, which was first pointed out by Anderson \cite{Anderson}. Considering that Eq. ({\ref{(Eq.123b)}}) implies a free propagating collective mode, from the mass-energy relation, to the first order approximation, we have the well-known relation
\begin{eqnarray}
     \Omega^2=\omega_p^2+|\textbf{k}|^2.
   \label{(Eq.124a)}
\end{eqnarray}
If the relativistic quantum and statistical effects are included, the transverse wave dispersion relation can be obtained by introducing the cut-off frequency part of Eq.(\ref{(Eq.120a)}) into Eq. ({\ref{(Eq.123a)}}), as follows
\begin{eqnarray}
     &~&\!\!\!\!\!\!\!\!\!\!\!\Omega^2=\omega^2_p[1+\frac{\omega^2_p}{4m^2}(1-\frac{3q_F^2}{10m^2}
             +\frac{\pi^2T^2}{q_F^2})\nonumber\\
     &~&~-\frac{3q_F^2}{10m^2}+\frac{\pi^2T^2}{q_F^2}
             -\frac{2\alpha}{9\pi}(1+\frac{\omega_p^2}{2m^2})]+|\textbf{k}|^2.
   \label{(Eq.125a)}
\end{eqnarray}

\subsubsection{\label{sec:level2}Spacelike longitudinal (acoustic) wave dispersion relation}

If $ mT\ll q_F^2$, i.e., $\mu\gg T$, the solution of Eq.(\ref{(Eq.118a)}) is
\begin{eqnarray}
     \omega^2=\frac{3q_F^2}{5m^2}|\textbf{k}|^2+\frac{|\textbf{k}|^4}{4m^2}.
   \label{(Eq.126a)}
\end{eqnarray}
Here $\sqrt{3}q_F/\sqrt{5}m=\sqrt{3}v_F/\sqrt{5}$ is not the acoustic velocity of ideal degenerate Fermi particles, i.e., $v_F/\sqrt{3}$. From Landau\cite{Landau}, it means that Eq.(\ref{(Eq.126a)}) is just the dispersion relation of zero sound. In fact, the condition that the temperature is much less than Fermi energy indicates that collisions are unimportant, and thermodynamic equilibrium is not established in each volume element in the time scale of $1/\omega$. The ordinary hydrodynamic sound wave doesn't propagate. Such zero sound term can be also found in electron plasma \cite{Chin}. The velocity of zero sound is the same as that of the longitudinal wave. There is no term concerning on zero sound in the scalar QED plasma as expect. The second order correction $|\textbf{k}|^4/4m^2$ to the sound mode of spinor QED plasma is just the leading term for scalar QED plasma \cite{Hines Frankel}\cite{Shi Fisch Qin}. According to Eq. ({\ref{(Eq.114f)}}), the sound wave is stable in the low-energy limit.

\subsection{\label{sec:level2}High-energy limit}

From Eq. ({\ref{(Eq.114b)}}) and ({\ref{(Eq.116d)}}), we have
\begin{eqnarray}
     \lim_{\frac{k^2- \max(\mu_0^2,m^2)}{\max(T^2,m^2)}\rightarrow\infty}\!\!\!\!\!\!\mathrm{Im}\Pi_{\mathrm{eff},2}^{\mu\nu}(k)=-\frac{2\alpha}{3}(k^2g^{\mu\nu}-k^\mu k^\nu).
   \label{(Eq.128a)}
\end{eqnarray}
The high energy fluctuation vector bosons decay into the fermions, and thereafter the waves stop propagating. A similar phenomenon can be also found in the scalar QED plasmas\ \cite{Shi Fisch Qin}. It is a pure vacuum polarization effect. According to Eq. ({\ref{(Eq.114b)}}), the background doesn't contribute to decay process in the high-energy limit.

\section{Summary and remarks}

We develop a fully quantized relativistic theory of spinor QED plasmas at finite temperature by using TFD. By decomposing the Dirac field and electromagnetic potential into the nontrivial background fields and the quantum fluctuation fields, we obtain an equivalent model for the plasma, making up of new particles $e_{\psi}$ and $\gamma_{_\mathcal{A}}$. The background field is critical for calculations. We propose the ``classical limit method'' to derive it. To the order of $e^2$, the dispersion relations of the longitudinal, transverse and sound waves are discussed in both of the low-energy and high-energy limits. The lowest order approximation gives the standard plasma oscillation expression, i.e., Eq.(\ref{(Eq.121a)}). The second order approximation gives the well-known results of longitudinal and transverse wave dispersions, i.e., Eq.(\ref{(Eq.122a)}) and (\ref{(Eq.124a)}), for non-relativistic degenerate plasmas. In addition, the zero sound in the electron-positron plasma is obtained. Further corrections appear in the next order approximation. As shown in Eq.(\ref{(Eq.120a)}) and Eq.(\ref{(Eq.125a)}), they include the relativistic correction to the mass, vacuum polarization correction to the effective charge, negative dispersion relation due to the finite light velocity, and temperature influence on the system. Besides, we find another vacuum fluctuation correction, i.e., virtual charge redistribution, to the plasma frequency, which has not been reported before to our knowledge. In the high-energy limit, the wave propagating stops caused by the strong vacuum polarization.

Next, we make the following remarks:

(1) Different from the method proposed by Yuan Shi, et al. \cite{Shi Fisch Qin} that they evaluate the background field by ground state wave function, which is hard to extend to the nonzero temperature and non-ideal cases, we develop the ``classical limit method'' to determine the nontrivial background fields for general many body system. Besides, our scheme gives conceptual clarity that each term involving the background field corresponds to a physical quantity, as shown in Eq.(\ref{(Eq.70b)})-Eq.(\ref{(Eq.84a)}).

(2) Although we assume that the background plasma charged particles obey Fermi-Dirac distribution in this paper, our method can be applied to the non-ideal case. The way is to replace all the macroscopic quantities (in determining the background fields) by the ones for the non-ideal system. Further efforts can be devoted to this subject which will exhibit a majority of interesting phenomenons.

(3) Unlike the usual QFT method for many-body system being restricted in the uniform and static circumstance, our theory can be applied to the slowly varying background cases. Since in our classical limit method, we connect each term including background fields with a corresponding physical quantity. There is no strict restriction on these quantities. One can give them any physically acceptable values changing in time and space on the macro scale, just keep valid of propagators of microscopic particles. Studies on these subjects will reveal more interesting effects.

(4) In this paper, we focus on the zero external field case. In the presence of external field, one should replace the free Dirac propagator $i/(\slashed{p}-m+i\epsilon)$ by $i/(\slashed{p}-e\bar{\slashed{A}}-m+i\epsilon)$, and $i/(\slashed{p}-m-i\epsilon)$ by $i/(\slashed{p}-e\bar{\slashed{A}}-m-i\epsilon)$. It will be much more complex in this case. Note that in the high intensity radiation case, two typical characteristic quantities, i.e., the classical parameter
 \begin{eqnarray}
  \xi=\frac{|e|\sqrt{-A^2}}{m}
\label{(Eq.1b)}
\end{eqnarray}
and quantum parameter
\begin{eqnarray}
  \chi=\frac{|e|\sqrt{(F_{\mu\nu}p^\nu)^2}}{m^3}
\label{(Eq.1c)}
\end{eqnarray}
play important roles \cite{Piazza Muller Hatsagortsyan Keitel}. The presence of intense electromagnetic field corresponding to $\xi\gg 1$ and $\chi\geqslant 1$ stimulates QED cascade \cite{Sokolov Naumova Nees}-\cite{Pukhov Nerush Kostyukov Shen Akli}. The usual perturbation method is of no use. It looks like the nontrivial background field method can overcome this obstacle, since it is an extension of ``Furry picture''. The strong field effect is included in the modified propagator. However, in this case, the system will be in the nonequilibrium state,  nonequilibrium TFD theory should be introduced. Besides, if the field strength exceeds the critical values ($F_{cr}^{\mu\nu}\sim e^2/m\sim 1.3\times 10^{16}$V$/$cm$\sim 4.4\times 10^{13}$G), the non-perturbative effect, such as Schwinger pair \cite{Schwinger}, appears. It can not be included in any perturbative theory.

(5) At last, we claim that our theory succeeds in obtaining the well-known quantum plasma modes in the low-energy limit, which strongly indicates its validity. Besides, our theory gives a more general scheme in dealing with many-body physics. The many-body effect is included in the background field which can be exactly evaluated by using the classical limit method, if the corresponding classical expectations, such as the particle distribution and the total energy etc., can be exactly determined. In the present work, the method overcomes the disadvantage of the approximations such as RPA and Hartree-Fock. Further, our scheme provides the possibility of dealing with non-static and nonuniform as well as static and uniform systems.
%Last, as emphasized by Yuan Shi, et al. \cite{Shi Fisch Qin}, it is necessary to develop a fully quantized theory for plasmas by using QFT. Quantum many body effects can be included without any confusion. Powerful relativistic quantum many body technology can be used in plasma study.  %Further, we introduce TFD into the quantum plasmas study which allows almost all the standard QFT techniques, including the background field method, being extended quite naturally to the temperature quantum many-body systems. Besides, new effects such as zero sound of electron-positron plasma and particle number increasing due to plasma oscillation are found which have not been reported before, as we known.%

\section{Acknowledgement}

This work was supported by the National Natural Science Foundation of
China under Grants No. 11547208 and No. 11547244, and the Foundation of Collaborative Innovation Team of Discipline Characteristics of Jianghan University under Fund No. 03100061.

\section{Data availability statement}

The data that supports the findings of this study are available within the article.

\appendix

\begin{widetext}
\section{Effective action for fluctuation vector boson in the presence of many-body system}

%The 2-point Green's function for fluctuation Dirac particle is derived as
%\begin{widetext}
%\begin{eqnarray}
%  &~&\langle\Omega(\beta)|T\bm{\psi}^{(\rho)}(x)\bar{\bm{\psi}}^{(\sigma)}(y)|\Omega(\beta)\rangle
%  =\sum_{\lambda,\tau=1,2}W_F^{(\rho\lambda)}(\beta;x)\langle\Omega|T\bm{\psi}^{(\lambda)}(x)\bar{\bm{\psi}}^{(\tau)}(y)|\Omega\rangle (W^{-1}_F(\beta;y))^{(\tau\sigma)}\nonumber\\
%  &=&\lim_{T\rightarrow \infty+i\varepsilon}\sum_{\lambda,\tau=1,2}\frac{W_F^{(\rho\lambda)}(\beta;x)\int D\bar{\bm{\psi}}D\bm{\psi} D\bm{\mathcal{A}}e^{i\int_T^T d^4x\hat{\mathcal{L}}[\bar{\bm{\psi}},\bm{\psi},\bm{\mathcal{A}}]}\bm{\psi}^{(\lambda)}(x)\bar{\bm{\psi}}^{(\tau)}(y)(W^{-1}_F(\beta,y))^{(\tau\sigma)}}{\int D\bar{\bm{\psi}}D\bm{\psi} D\bm{\mathcal{A}}e^{i\int_T^T d^4x\hat{\mathcal{L}}[\bar{\bm{\psi}},\bm{\psi},\bm{\mathcal{A}}]}}
%\label{(Eq.29a)},
%\end{eqnarray}
%\end{widetext}

By using the standard functional integral formulas over Grassmann variables
\begin{eqnarray}
  (\prod_i\int d\theta_i^*d\theta_i)\exp(-\sum_{ij}\theta_i^*B_{ij}\theta_j)=\det B,
  \label{(Eq.40b)}
\end{eqnarray}
\begin{eqnarray}
  (\prod_i\int d\theta_i^*d\theta_i)\theta_k\theta_l^*\exp(-\sum_{ij}\theta_i^*B_{ij}\theta_j)=(\det B)(B^{-1})_{kl},
  \label{(Eq.40c)}
\end{eqnarray}
\begin{eqnarray}
  (\prod_i\int d\theta_i^*d\theta_i)\theta_k\theta_l^*\theta_m\theta_n^*\exp(-\sum_{ij}\theta_i^*B_{ij}\theta_j)=(\det B)(B^{-1}_{kl}B^{-1}_{mn}-B^{-1}_{kn}B^{-1}_{ml}).
  \label{(Eq.40d)}
\end{eqnarray}
In the case of $\bar{A}=0$, we obtain the effective action $\Gamma_{\mathcal{A}}$ for $\gamma_{_\mathcal{A}}$, to the $e^2$ order approximation, as follows
\begin{eqnarray}
  e^{i\Gamma_\mathcal{A}}&=&\int D\bar{\psi}(x)D\psi(x)\exp(i\int d^4x\mathcal{L}_{\mathrm{e}\mathrm{f}\mathrm{f}}[\bar{\psi}(x),\psi(x),\mathcal{A}(x)])\nonumber\\
  &=&\det(i\slashed{D}-m\mathbbm{1}_{4\times 4})\exp\{\frac{i}{2}\int d^4x (\mathcal{A}_\mu(x))^\mathrm{T}(\partial^2g^{\mu\nu}-\partial^\mu \partial^\nu)\mathcal{A}_\nu(x)\}\nonumber\\
  &~&\times\{1-ie\int d^4x\mathrm{tr}[\gamma^\mu S_{F}(x-x)]\mathcal{A}_\mu(x)-\frac{e^2}{2}\int d^4xd^4y (\mathcal{A}_\mu(x))^\mathrm{T}\Pi^{(\mu\nu)}_{2}(x,y)\mathcal{A}_\nu(y)\},
\label{(Eq.30a)}
\end{eqnarray}
where
\begin{eqnarray}
    i\Pi_2^{\mu\nu}(x,y)&=&-ie^2\{\mathrm{tr}[-\gamma^\mu S_F(x,x)\gamma^\nu S_F(y,y)]-\gamma^\mu S_{F}(x,y)\gamma^\nu S_{F}(y,x)\nonumber\\
    &~&+\bar{\psi}_0^{(\alpha)}(x)\gamma^\mu S_{F}(x,y)\gamma^\nu\psi_0(y)+\bar{\psi}_0(y)\gamma^\mu S_{F}(y,x)\gamma^\nu\psi_0(x)\}.
  \label{(Eq.37a)}
\end{eqnarray}
\end{widetext}
The first term on the right side of the expression is
\begin{eqnarray}
   &~&~~~\ff{\vertexlabel^{\mu}\vertexlabel_{x} gc}~~\ff{cg\vertexlabel^{\nu}\vertexlabel_{y}}\nonumber\\
   &=&-e^2\mathrm{tr}[\gamma^\mu S_{F}(x,x)\gamma^\nu S_{F}(y,y)]
\label{(Eq.44b)}.
\end{eqnarray}
It represents the multiplication of two $\langle \Omega|\mathcal{A}_\mu|\Omega\rangle$. According to Lorentz's symmetry, this field expectation vanishes. The second term in the brace in Eq. ({\ref{(Eq.30a)}}) is linear in $\mathcal{A}_\mu$. It is responsible for the emission, absorption and scattering of $\gamma_{_\mathcal{A}}$ that not be concerned in this paper. Neglecting it, the Fourier transformation of $\Gamma_{\mathcal{A}}$ becomes, up to an unnecessary constant $-\ln\det(i\slashed{D}-m\mathbbm{1}_{4\times 4})$,
\begin{eqnarray}
   \Gamma_{\mathcal{A}}&\rightarrow& \int\frac{d^4k}{(2\pi)^4}\mathcal{A}_\mu(k)[-k^2g^{\mu\nu}+k^\mu k^\nu\nonumber\\
                     &~&~~~~~~~~~+\Pi^{\mu\nu}_{T2}(k)]\mathcal{A}_\nu(-k).
   \label{(Eq.49a)}
\end{eqnarray}
where
\begin{eqnarray}
   \mathcal{A}_\mu(k)=\int d^4xe^{ik\cdot x}\mathcal{A}_\mu(x).
  \label{(Eq.51a)}
\end{eqnarray}
Here, Feynman gauge ($\xi=1$) is used. Apply variational approach to the expression ({\ref{(Eq.49a)}}), we obtain the classical equation for $\mathcal{A}_\mu$ in the absence of many body system.

In general, in the presence of the many-body system, the effective action $\Gamma_{\mathcal{A}}$ for $\gamma_{_\mathcal{A}}$ can be expanded, to the $e^2$ order, as follows
\begin{eqnarray}
   &~&\!\!\!\!\!\!\!\!\!\!\!\!\Gamma_{\mathcal{A}}=\int\frac{d^4k}{(2\pi)^4} \mathcal{A}_\mu(k)[L_0^{\mu\nu}(k)+e^2L^{\mu\nu}_2(k)]\mathcal{A}_\nu(-k) \nonumber\\
   &~&~~~~~~~~~~~+e\int\frac{d^4k}{(2\pi)^4}L^\mu_1(k)\mathcal{A}_\mu(k),
  \label{(Eq.51b)}
\end{eqnarray}
where $L_0^{\mu\nu}(k)$, $L_1^\mu(k)$, and $L_2^{\mu\nu}(k)$ are functions of $k^\mu$. Considering that no thermal effect is taken into account for photons, free $\Gamma_{\mathcal{A}}$ should reduce to normal action for free $\gamma_{_\mathcal{A}}$, which implies
\begin{eqnarray}
   L_0^{\mu\nu}(k)=\frac{1}{2}(-k^2g^{\mu\nu}+k^\mu k^\nu).
  \label{(Eq.51c)}
\end{eqnarray}
As stated before, the linear term is ignored in the present study. From subsection E in Sec. \Rmnum{3}, we have
\begin{eqnarray}
   e^2L_2^{\mu\nu}(k)=i\Pi_{\mathrm{eff}}^{\mu\nu}(k)=i\!\sum_{\rho\lambda=1,2}\Pi_{T}^{(\rho\lambda)\mu\nu}(k).
  \label{(Eq.51d)}
\end{eqnarray}
According to these considerations, from the expression ({\ref{(Eq.51b)}}), the effective classical equation for $\mathcal{A}$ is
\begin{eqnarray}
   [k^2g^{\mu\nu}-k^\mu k^\nu-i\!\sum_{\rho\lambda=1,2}\Pi_{T}^{(\rho\lambda)\mu\nu}(k)]\mathcal{A}_\mu(k)=0.
  \label{(Eq.51d)}
\end{eqnarray}
The necessary condition for the existence of a nontrivial solution is Eq. ({\ref{(Eq.39w)}})
%It can be understood in two ways. One is that from classical point of view, one can derive a classical equation of motion of $\mathcal{A}^{(\alpha)}$, i.e. $([-k^2g^{\mu\nu}+k^\mu k^\nu+\Pi^{\mu\nu}_2(k)]\sigma^3)\mathcal{A}^{(\alpha)}_\mu=0$. Nontrivial solution of it asks for Eq. ({\ref{(Eq.50a)}}). The other is
%that from dressed Green's function point of view, the mode of fluctuation gauge boson is determined by the pole of dressed propagator. It is determined by summing the set of ring diagrams that is well documented in many body physics, and leads to Eq. ({\ref{(Eq.50a)}}). The equivalence of this two understandings is pointed out by Chin \cite{Chin}.

\section{Possible violation of Ward identity}

The polarization tensor $\Pi^{\mu\nu}_{T}$ can have form differing from $(k^2g^{\mu\nu}-k^\mu k^\nu)\Pi^{(\rho\sigma)}_T(k)$. We explain it, from three different perspectives, as follows:
\paragraph{}
In TFD, we consider a process similar to ({\ref{(Eq.50l)}}), but with $\gamma_{_\mathcal{A}}(\beta)$ on-shell. The corresponding S-matrix is
\begin{eqnarray}
  S_{fi}=\mathbbm{1}+i\mathcal{M}'_{fi}(k)(2\pi)^4\delta^{(4)}(0)
 \label{(Eq.50g)}.
\end{eqnarray}
Applying LSZ reduction formula to TFD, we have
\begin{eqnarray}
  &~&\!\!\!\!\!\!\!\!\!\!i\mathcal{M}'_{fi}(k)=\sum_{\rho\lambda}\sum_{\mathrm{polarizations}}\varepsilon_\mu^{(\rho)}(\bm{k})\varepsilon^{(\lambda)}_{\nu}(\bm{k})^*\nonumber\\
  &~&~~~~~~~~~~~~~\times\Pi_{T}^{(\rho\lambda)\mu\nu}(k)
   (2\pi)^4\delta^{(4)}(0)
 \label{(Eq.39e)}
\end{eqnarray}
where $\varepsilon_\mu^{(\rho)}(\bm{k})$ stands for the physical polarization vector of free $\gamma_{_\mathcal{A}}(\beta)$. More precisely, they are defined as follows
\begin{eqnarray}
  \contraction{}{\mathcal{A}}{_\mu^{(1)}(\beta,x)|\bm{k}(\beta}{)}\mathcal{A}_\mu^{(1)}{(\beta,x)}|\bm{k}(\beta),r\rangle\otimes|\widetilde{0}_{\mathrm{eff}}\rangle=\varepsilon_\mu^{(1)r}(\bm{k})e^{-ik\cdot x}|0_{\mathrm{eff}}(\beta)\rangle
 \label{(Eq.39f)},
\end{eqnarray}
\begin{eqnarray}
  \contraction{}{\mathcal{A}}{_\mu^{(2)}{(\beta,x)}|0_{\mathrm{eff}}\rangle\otimes|\bm{k}(\beta}{)}\mathcal{A}_\mu^{(2)}{(\beta,x)}|0_{\mathrm{eff}}\rangle\otimes|\bm{k}(\widetilde{\beta),r}\rangle=\varepsilon_\mu^{(2)r}(\bm{k})^*e^{ik\cdot x}|0_{\mathrm{eff}}(\beta)\rangle
 \label{(Eq.39g)},
\end{eqnarray}
\begin{eqnarray}
  \langle\bm{k}(\contraction{\beta}{)}{,r|\otimes\langle\widetilde{0}_{\mathrm{eff}}|}{\mathcal{A}}\beta),r|\otimes\langle\widetilde{0}_{\mathrm{eff}}|\mathcal{A}_\mu^{(1)}(\beta,x)=\langle 0_{\mathrm{eff}}(\beta)|\varepsilon_\mu^{(1)r}(\bm{k})^*e^{ik\cdot x}
 \label{(Eq.39h)},
\end{eqnarray}
\begin{eqnarray}
  \langle 0_{\mathrm{eff}}|\otimes\langle\bm{k}(\contraction{\beta}{)}{,r|}{\mathcal{A}} \widetilde{\beta),r}|\mathcal{A}_\mu^{(2)}(\beta,x)=\langle 0_{\mathrm{eff}}(\beta)|\varepsilon_\mu^{(2)r}(\bm{k})e^{-ik\cdot x}
 \label{(Eq.39i)},
\end{eqnarray}
where $|\bm{k}(\beta),r\rangle$ is the base vector of $\gamma_{_\mathcal{A}}(\beta)$ with 4-momentum $k$ and polarization indices $r$, $|\bm{k}(\widetilde{\beta),r}\rangle$ is the tilde conjugation one. Besides, $\contraction{}{~}{~}{~}~~~$ denotes Wick contraction.

$\varepsilon^{(\rho)r}_\mu(\bm{k})$ will be replaced by $\varepsilon^{(\rho)r}_\mu(\bm{k})+ck_\mu$ ($c$ is an arbitrary constant) under the change of gauge. It immediately gives Ward identity if $\mathcal{M}'_{fi}(k)$ is unchanged. However $\gamma_\mathcal{A}$ is just a mathematical object rather than an observable one. It indicates that $\mathcal{M}'_{fi}(k)$ is not an observable quantity, for which the gauge invariance is not necessary, and Ward identity violation is acceptable. Further, as seen in Sec. \Rmnum{4} that
\begin{eqnarray}
  k_\mu\Pi^{(\rho\sigma)\mu\nu}_{Tvac,2}(k)=0
 \label{(Eq.39j)},
\end{eqnarray}
we find $k_\mu\Pi^{(\alpha)\mu\nu}_{Tbac,2}(k)$ does not necessarily vanish.
\paragraph{}
From the diagrammatic expressions ({\ref{(Eq.39c)}}), ({\ref{(Eq.26g)}}) and ({\ref{(Eq.26h)}}), we can regard the medium as an ``external medium particle" with off shell momentum $q$ and ``polarization vector" $\bm{\psi}_0(q)$. Here external fermions in the usual diagram
$\!\!\Diagram{
                                           & ![ulft]{fvA}{q} \\
           \vertexlabel^{\mu}~ ![bot]{gA}k & ![bot]{fV}{q-k} & ![lrt]{gA}{k} ~\vertexlabel^{\nu} \\
                                           &                  & ![lrt]{fvA}{q} \\
   }$
is replaced by this ``external fermion-like particle". From this perspective, instead of an on-shell fermion line, the vertex
$\dd{
       fd \\
       & g \\
       fu \\
}$
is connected with an off shell ``external fermion-like particle", which contradicts to the necessary condition of Ward identity.
\paragraph{}
The possible Ward identity violation can be verified by direct computation. Let $\bm{\psi}^{(\rho)}(x)$ transforms according to the expression ({\ref{(Eq.20a)}}), with $\mathcal{A}_\mu^{(\rho)}(x)$ invariance. To the first order of $\chi_\rho$, the total Lagrangian $\hat{\mathcal{L}}[\bar{\psi},\psi,\mathcal{A}]$ becomes
\begin{eqnarray}
  \hat{\mathcal{L}}\rightarrow \hat{\mathcal{L}}-\sum_{\rho=1,2}\!\!\varepsilon_\rho P_\rho(\partial_\mu\chi_\rho)(\bm{\mathcal{J}}^{(\rho) \mu}_{\mathrm{v}\mathrm{a}\mathrm{c}}+\bm{\mathcal{J}}^{(\rho) \mu}_{\mathrm{b}\mathrm{a}\mathrm{c}})
 \label{(Eq.39k)}.
\end{eqnarray}
Expanding the integral
\begin{displaymath}
  \frac{1}{Z}P_\sigma\int D\bar{\bm{\psi}}D\bm{\psi} D\bm{\mathcal{A}}e^{i\int d^4x\hat{\mathcal{L}}}\bm{\mathcal{J}}^{(\sigma) \nu}(y)
\end{displaymath}
to first order in $\chi_\rho$, and considering the measure of functional integral is invariant under the  transformation ({\ref{(Eq.20a)}}), we obtain the Schwinger-Dyson equation associated with current as follows
\begin{eqnarray}
  &~&\!\!\!\!\!P_\rho P_\lambda\langle\Omega_{\mathrm{eff}}(\beta)|T\partial_\mu \bm{\mathcal{J}}^{(\rho)\mu}(x)\bm{\mathcal{J}}^{(\lambda)\nu}(y)|_{\mathrm{eff}}(\beta)\rangle\nonumber\\
  &~&\!\!\!\!\!\!\!\!\!\!\!\!\!=\delta^{(4)}(x-y)\delta^{\rho\lambda}P_\lambda\langle\Omega_{\mathrm{eff}}(\beta)|T\triangle\bm{\mathcal{J}}^{(\lambda)\nu}(y)|\Omega_{\mathrm{eff}}(\beta)\rangle
 \label{(Eq.39l)},
\end{eqnarray}
where $\triangle$ means that the configuration of thermal doublets $\bm{\psi}$ in $\bm{\mathcal{J}}$ is replaced by its deformation
\begin{eqnarray}
  \triangle\bm{\psi}^{(\rho)}=i(\bm{\psi}^{(\rho)}_0+\bm{\psi}^{(\rho)})
 \label{(Eq.39m)}.
\end{eqnarray}
Substitute this relation and Eq. ({\ref{(Eq.25a)}}) into Eq. {(\ref{(Eq.39l)})}, we obtain
\begin{eqnarray}
   P_\rho P_\lambda\langle\Omega_{\mathrm{eff}}(\beta)|T\partial_\mu \bm{\mathcal{J}}_{vac}^{(\rho)\mu}(x)\bm{\mathcal{J}}_{vac}^{(\lambda)\nu}(y)|\Omega_{\mathrm{eff}}(\beta)\rangle=0\nonumber\\
 \label{(Eq.39n)}
\end{eqnarray}
and
\begin{eqnarray}
   &~&P_\rho P_\lambda\langle\Omega_{\mathrm{eff}}(\beta)|T\partial_\mu \bm{\mathcal{J}}_{bac}^{(\rho)\mu}(x)\bm{\mathcal{J}}_{bac}^{(\lambda)\nu}(y)|_{\mathrm{eff}}(\beta)\rangle\nonumber\\
   &=&\delta^{(4)}(x-y)\delta^{\rho\lambda}P_\lambda[\bar{\bm{\psi}}_0^{(\lambda)}(y)\gamma^\nu\langle\Omega_{\mathrm{eff}}(\beta)|\bm{\psi}^{(\lambda)}(y)|\Omega_{\mathrm{eff}}(\beta)\rangle\nonumber\\
   &~&~~~~-\langle\Omega_{\mathrm{eff}}(\beta)|\bm{\psi}^{(\lambda)}(y)|_{\mathrm{eff}}(\beta)\rangle\gamma^\nu\bm{\psi}^{(\lambda)}_0(y)]
 \label{(Eq.39o)}.
\end{eqnarray}
Compute their Fourier transform by integrating
\begin{displaymath}
\int d^4xe^{-ik\cdot x}\int d^4ye^{-ik'\cdot y},
\end{displaymath}
and with the help of Eq. ({\ref{(Eq.48b)}}), Eq. ({\ref{(Eq.39n)}}) and ({\ref{(Eq.39o)}}) become
\begin{eqnarray}
  k_\mu\Pi^{(\rho\lambda)\mu\nu}_{T\mathrm{vac}}(k)=0
 \label{(Eq.39p)}
\end{eqnarray}
and
\begin{eqnarray}
   &~&k_\mu\Pi^{(\rho\lambda)\mu\nu}_{T\mathrm{bac}}(k)\nonumber\\
   &=&\delta^{\rho\lambda}P_\lambda\int\frac{d^4q}{(2\pi)^4}[\bar{\bm{\psi}}_0^{(\lambda)}(q)\gamma^\nu\langle\Omega_{\mathrm{eff}}(\beta)|\bm{\psi}^{(\lambda)}(q)|\Omega_{\mathrm{eff}}(\beta)\rangle\nonumber\\
   &~&~~~~-\langle\Omega_{\mathrm{eff}}(\beta)|\bar{\bm{\psi}}^{(\lambda)}(q)|\Omega_{\mathrm{eff}}(\beta)\rangle\gamma^\nu\bm{\psi}^{(\lambda)}_0(q)]
 \label{(Eq.39q)}.
\end{eqnarray}
We just need to show the possibility of $k_\mu\Pi^{(\rho\lambda)\mu\nu}_{T\mathrm{bac}}(k)\neq 0$, as follows. There are only two ways for $k_\mu\Pi^{(\rho\lambda)\mu\nu}_{T\mathrm{bac}}(k)$ being vanished: (1) $\langle\Omega_{\mathrm{eff}}(\beta)|\bm{\psi}^{(\lambda)}(x)|\Omega_{\mathrm{eff}}(\beta)\rangle\equiv0$. (2) The two terms on the right side of Eq. ({\ref{(Eq.39q)}}) cancel each other out. The necessary condition for (1) is that $|\Omega_{\mathrm{eff}}(\beta)\rangle$ is Lorentz invariance. However, it is impossible since the physical quantities, such as the energy, momentum, etc., may have various values in different frames of reference. One can invalidate (2) by choosing a special $\bm{\psi}^{(\lambda)}_0(q)$. It can always be achieved because this background field is not uniquely determined according to Sec. \Rmnum{4}. Let a specific $\bm{\psi}^{(\lambda)}_0(q)$ satisfying (2), then set $\phi_1(t)=\chi_1(t)=\frac{\pi}{2}$ and $\phi_2(t)=\chi_2(t)=-\frac{\pi}{2}$ in the equations ({\ref{(Eq.95a)}}) to ({\ref{(Eq.98a)}}), we obtain a new background field $\bm{\psi}'^{(\tau)}_0(q)=i\sum_{\tau=1,2}\sigma^3_{\lambda\tau}\bm{\psi}^{(\tau)}_0(q)$. The right-hand side of Eq. ({\ref{(Eq.39q)}}) becomes
\begin{eqnarray}
   &~&\!\!\!\!\!\!\!\!\!\!\!\!\!\!\!\!\!\!\!\!\!-i\delta^{\rho\lambda}P_\lambda\int\frac{d^4q}{(2\pi)^4}\sum_{\tau=1,2}[\bar{\bm{\psi}}_0^{(\tau)}(q)
      \sigma^3_{\tau\lambda}\gamma^\nu\nonumber\\
   &~&\times\langle\Omega_{\mathrm{eff}}(\beta)|\bm{\psi}^{(\lambda)}(q)|\Omega_{\mathrm{eff}}(\beta)\rangle\nonumber\\
   &~&+\langle\Omega_{\mathrm{eff}}(\beta)|\bar{\bm{\psi}}^{(\lambda)}(q)
      |\Omega_{\mathrm{eff}}(\beta)\rangle\gamma^\nu\sigma^3_{\lambda\tau}\bm{\psi}^{(\tau)}_0(q)]
 \label{(Eq.39r)},
\end{eqnarray}
where $\sigma^3$ is the third component of the Pauli matrix. Obviously, this expression is nonzero since ``$-$'' in the integral in Eq. ({\ref{(Eq.39q)}}) is replaced by ``$+$'' here.

\end{document}